\documentclass{aa}  

\usepackage{graphicx}
\usepackage[varg]{txfonts}
\usepackage{subcaption}         
\usepackage{lscape}             
\usepackage{placeins}           
\usepackage[breaklinks]{hyperref}
\usepackage{orcidlink}
\usepackage{afterpage}
\usepackage[normalem]{ulem}
\begin{document}

   \title{Analysis of mass-transferring binary candidates in the Milky Way}


   \author{G. Garcia-Moreno\inst{1,2} \orcidlink{0009-0001-4098-7706}
           \and N. Blagorodnova\inst{1,2,3} \orcidlink{0000-0003-0901-1606}
           \and F. Anders\inst{1,2,3} \orcidlink{0000-0003-4524-9363}
           \and M. Weiler\inst{1,2,3} \orcidlink{0000-0002-3007-3927}
           \and H. Wichern\inst{4} \orcidlink{0009-0004-1442-619X}
           \and N. Britavskiy\inst{5,6} \orcidlink{0000-0003-3996-0175}
           \and S. de Wet\inst{4} \orcidlink{0000-0003-2449-1329}}

   \institute{Departament de Física Quàntica i Astrofísica (FQA), Universitat de Barcelona (UB),  c. Martí i Franquès, 1, 08028 Barcelona, Spain
         \and
             Institut de Ciències del Cosmos (ICCUB), Universitat de Barcelona (UB), c. Martí i Franquès, 1, 08028 Barcelona, Spain
                \and
                    Institut d'Estudis Espacials de Catalunya (IEEC), Edifici RDIT, Campus UPC, 08860 Castelldefels (Barcelona), Spain
                        \and
                            DTU Space, Technical University of Denmark, Elektrovej 327, 2800 Kgs. Lyngby, Denmark
                                \and
                                    Royal Observatory of Belgium, Avenue Circulaire, Ringlaan 3, 1180 Brussels, Belgium
                                        \and
                                            University of Liège, Allée du 6 Août 19c (B5C), 4000 Sart Tilman, Liège, Belgium
             }

   \date{Received ; accepted }



    \abstract{Mass transfer between stars in binary systems profoundly impacts their evolution, yet many aspects of this process---especially the stability, mass loss, and eventual fate of such systems---remain poorly understood. One promising avenue to constrain these processes is through the identification and characterisation of systems undergoing active mass transfer. Inspired by the slow brightening preceding stellar merger transients, we worked on a method to identify Galactic mass-transferring binaries in which the donor is a Hertzsprung gap star. We constructed an initial sample of Hertzsprung gap stars using the \textit{Gaia} EDR3 contribution \texttt{Starhorse} catalogue, and we identified candidate mass-transferring systems by selecting sources that exhibit Balmer emission features (as seen in the low-resolution \textit{Gaia} XP spectra), mid-infrared excess (from WISE photometry), and photometric variability (inferred from the error in the \textit{Gaia} $G$-band magnitude). This multi-criteria selection yielded a sample of 67 candidates, which we further analysed using complementary photometric and spectroscopic data, as well as information from cross-matched archival catalogues. Among our candidates, we identified at least nine eclipsing binaries and some sources that are potential binaries as well. Three sources in our sample are strong candidates for mass-transferring binaries with a yellow component, and three more are binaries with a Be star. Notably, three sources in our sample are strong candidates for hosting a compact companion, based on their ultraviolet or X-ray signatures. The main sources of contamination in our search are hot but highly reddened stars---primarily Oe and Be stars---as well as regular pulsating stars such as $\delta$ Scuti and Cepheid variables. As an additional outcome of this work, we present a refined new catalogue of 308 Hertzsprung gap stars, selected using improved extinction corrections and stricter emission-line criteria. This enhanced sample is expected to contain a significantly higher fraction of scientifically valuable mass-transferring binaries.
    }
       

   \maketitle
%

\section{Introduction}
\nolinenumbers

A large fraction of stars in the observable Universe are found in binary or multiple systems, with the likelihood of multiplicity increasing with stellar mass \citep{Sana, mink, Moe}. The evolution of most of these systems is significantly influenced by mass transfer between the components. How mass transfer unfolds and its eventual outcome depends strongly on the initial configuration of the binary and the nature of the mass transfer process. Broadly, two scenarios are considered: stable mass transfer, in which the mass exchange occurs without disrupting the system, and unstable mass transfer, where the accreting star is unable to absorb all the material from the donor, resulting in the formation of a common envelope that engulfs both stars \citep{kuiper, paczynski}.

Mass transfer is an accepted explanation for the creation of many astrophysical phenomena, such as the Algol-type binaries \citep{crowford}, the evolution of closely orbiting stellar binaries into X-ray binaries \citep{Xraybinaries}, cataclysmic variable stars \citep{kraft}, supernovae \citep{supernovaI, supernova2}, accretion-induced collapse \citep{Ivanova}, and gravitational wave sources \citep{gwsources}, among others. However, the details of mass transfer and the possible mass and angular momentum loss of the process, as well as the fate of the evolving stars, are not well established and modelled yet.

In recent years, the study of mass transfer has gained renewed attention due to its fundamental link to the formation of compact binaries and, ultimately, gravitational-wave sources. To constrain these processes, detailed studies of systems currently undergoing mass exchange are crucial. These efforts span a wide range of approaches: observational analyses of interacting systems such as TZ Dra \citep{TZDra}, AU Monocerotis \citep{AUMonocerotis}, and the contact binary V2840 Cygni \citep{V2840Cygni}; modelling of well-studied binaries like $\beta$ Lyrae A \citep{thick, thin}; and theoretical work aimed at understanding the underlying physics of mass transfer and common envelope evolution \citep{Jakub, CEPod2001, CEIvanova2001, CEPablo2021, CEDeMarco2023}.

The outcome of unstable mass transfer has recently been linked to a class of optical transients known as luminous red novae (LRNe), whose peak luminosities lie between those of novae and supernovae \citep{V838Mon1, Soker2003, V1309ScoMerger, Pejcha2014, Pejcha2016}. These transients are generally interpreted as the observational signatures of stellar coalescence events, providing rare but direct evidence of binary interaction and merger processes.

Over the last two decades, LRNe have been studied both in our Galaxy and in nearby galaxies.  Despite the proximity, the only five confirmed and studied stellar mergers in the Milky Way (MW) include V4332 Sagittarii (discovered in 1994; \citealt{V4332Sag}), V838 Monocerotis (2002; \citealt{V838Mon1, V838Mon2}), OGLE-2002-BLG-360 (2002; \citealt{OGLE-2002}), CK Vul (1670; \citealt{CKVul}), and most notably, V1309 Scorpii (2008, \citealt{V1309ScoDiscovery}), whose pre-outburst photometry captured the final inspiral of a contact binary system \citep{V1309ScoMerger}. Additionally, about two dozen extra-galactic LRNe were found (see \citealt{pastorello}, \citealt{Blagorodnova2021}, \citealt{Reguitti2025}, and references therein). These events share a number of defining characteristics: a rapid rise and redward evolution in colour, usually a double-peaked light curve followed by a plateau, and the presence of strong and narrow emission lines for hydrogen and low-ionisation elements.

Archival data taken prior to the outburst have shown that most confirmed LRN progenitors are predominantly yellow giants (YG) or yellow supergiants (YSG) with spectral types G, F or, to a lesser extent, A \citep{Blagorodnova2017, MacLeod2017, Blagorodnova2021, Cai2022, Pastorello2023}. These stars are evolving off the main sequence (MS) and crossing the Hertzsprung gap (HG) on their way to the red giant branch (RGB). This phase is characterised by a rapid increase in the stellar radius of the evolving star. Given the presence of a nearby companion, the donor can overfill its Roche lobe and initiate unstable mass transfer, finally resulting in a stellar merger and a transient. In the few cases where pre-outburst data exist, the progenitor shows a gradual, years-long brightening, interpreted as mass loss and the formation of an extended pseudo-photosphere before coalescence. Despite these key discoveries, LRN progenitors remain largely unexplored \citep{pastorello}.

Guided by the location of known LRN progenitors on the {Hertzsprung-Russell (HR) diagram, a recent study presented a method for identifying Galactic LRN candidates likely to outburst within the next 1--10 years \citep{Harry}. The selection of candidates was based on the position in the HR diagram and the variability of the sources. The study first identified a set of stars falling in a low stellar density area between the MS and the RGB from the \textit{Gaia} DR2 and \textit{Gaia} EDR3, and explored their variability in different time-domain surveys to select slowly brightening sources. From an initial set of approximately 10$^7$ initial sources, 21 LRN precursor candidates were selected. Most of them shared similar characteristics: the presence of H$\alpha$ and (sometimes) H$\beta$ emission lines together with the existence of an infrared (IR) excess. One of the limitations of that study was the lack of reliable estimates for Galactic extinction in \textit{Gaia} DR2 and EDR3, implying that the true colours of the sources were rather uncertain.

The release of extinction estimates in \textit{Gaia} DR3 and \textit{Gaia} related catalogues, such as \texttt{Starhorse} \citep{Starhorse}, opened the possibility to revise the previous work with more accurate colours and astrophysical parameters. In this paper, we make use of these catalogues to present a selection and characterisation of HG stars in the Milky Way (MW) with characteristics compatible with known LRN precursors and mass-transferring systems. We first make a tailored selection of candidates, followed by an analysis using archival and new follow-up to find binary systems during the unstable mass transfer phase, which could become LRNe in the near future. In addition, we aim to characterise the variability of stars in the HG to understand what other systems can mimic the behaviour expected for LRN precursor systems. Finding unstable mass-transferring binaries will be of great interest to the theoretical and observational communities studying the stability of mass transfer and its impact on the final fate of binary stars. 

\section{Sample selection} \label{section:sample_selection}
Our initial selection strategy made use of \textit{Gaia} DR3 \citep{theGaiaMission, GaiaDR3} data and the \textit{Gaia} EDR3 {\tt StarHorse} \citep{Starhorse} catalogue, in combination with mid-IR photometry from the \texttt{ALLWISE} catalogue \citep{allwise}. The selection steps are listed below:

\begin{enumerate}
    \item To select HG stars, we used \texttt{StarHorse}, which (among other astrophysical parameters) contains absolute magnitudes in the $G$ band and de-reddened $BP - RP$ colours. Based on {\it MESA Isochrones \& Stellar Tracks} \citep{mesa, mist}, we selected sources with absolute magnitudes $M_G < 1.5$, to avoid contamination from lower MS stars with masses $\lesssim 2 M_\odot$, and we also applied two cuts to select stars between the MS and the RGB. Finally, we removed a portion of the colour-magnitude diagram populated by Red Clump stars, known red giants in the horizontal branch. In Fig. \ref{selec_grid}a our initial parameter space is shown.
    \item To select sources with emission in H$\alpha$ or H$\beta$, we used the method developed by \citet{Weiler} to derive the equivalent width (EW) of the lines using the internally calibrated \textit{Gaia} XP spectra \citep{Carrasco2021, DeAngeli2023} (see Fig. \ref{selec_grid}b, which shows the XP spectrum of one of our sources alongside the corresponding follow-up spectrum). We first tested this method with the candidates in \cite{Harry}. The sources with H$\alpha$ emission had positive EW values (as expected), but some of these sources showed high errors and low statistical significance in our EW calculation, even though they had clear emission in both the \textit{Gaia} XP spectrum and a higher resolution follow-up spectrum. Therefore, we selected sources with a positive H$\alpha$ or H$\beta$ EW, without taking into account the EW error or the statistical significance.
    \item To select sources with mid-IR excess associated with warm dust emission (likely from a disk), we used the \texttt{ALLWISE} catalogue. We applied the following recommended quality cuts\footnote{\url{https://wise2.ipac.caltech.edu/docs/release/allwise/expsup/sec2_1a.html}}: \texttt{cc\_flags} $ = 0000$, making sure the sources are unaffected by known artifacts; \texttt{ext\_flag} $ = 0$, indicating that the sources' shape is consistent with a point-source, and \texttt{ph\_qual} $ = A, B$ or $C$, selecting sources with a flux signal-to-noise ratio better than 2. We selected sources with $W1 (3.4\,\mu m) - W4 (22\,\mu m) > 1$, based on the distribution of sources in the $W1 (3.4\,\mu m) - W4 (22\,\mu m)$ vs $W1 (3.4\,\mu m) - W2 (4.6\,\mu m)$ colour-colour diagram (see Fig. \ref{selec_grid}c).
    \item As the photometry listed in the main \textit{Gaia} DR3 catalogue is a weighted average of the individual measurements taken in different transits \citep{riello}, one expects that variable stars have higher photometric uncertainties (as seen in \citealt{DR2vari} or \citealt{DR3vari}). Thus, we placed our initial HG sample in the $\log_{10}(G_{err})$ vs $G$ plane (see Fig. \ref{selec_grid}d) and selected sources above the 90th percentile.
    \item As a final step, we removed the sources with a known type in the \texttt{SIMBAD} \citep{simbad} database, for which the variability was associated with pulsations rather than binarity (such as RR Lyrae or Cepheids). 
\end{enumerate}

\onecolumn
\begin{figure*}[p]
   \centering
   \includegraphics[scale=0.46]{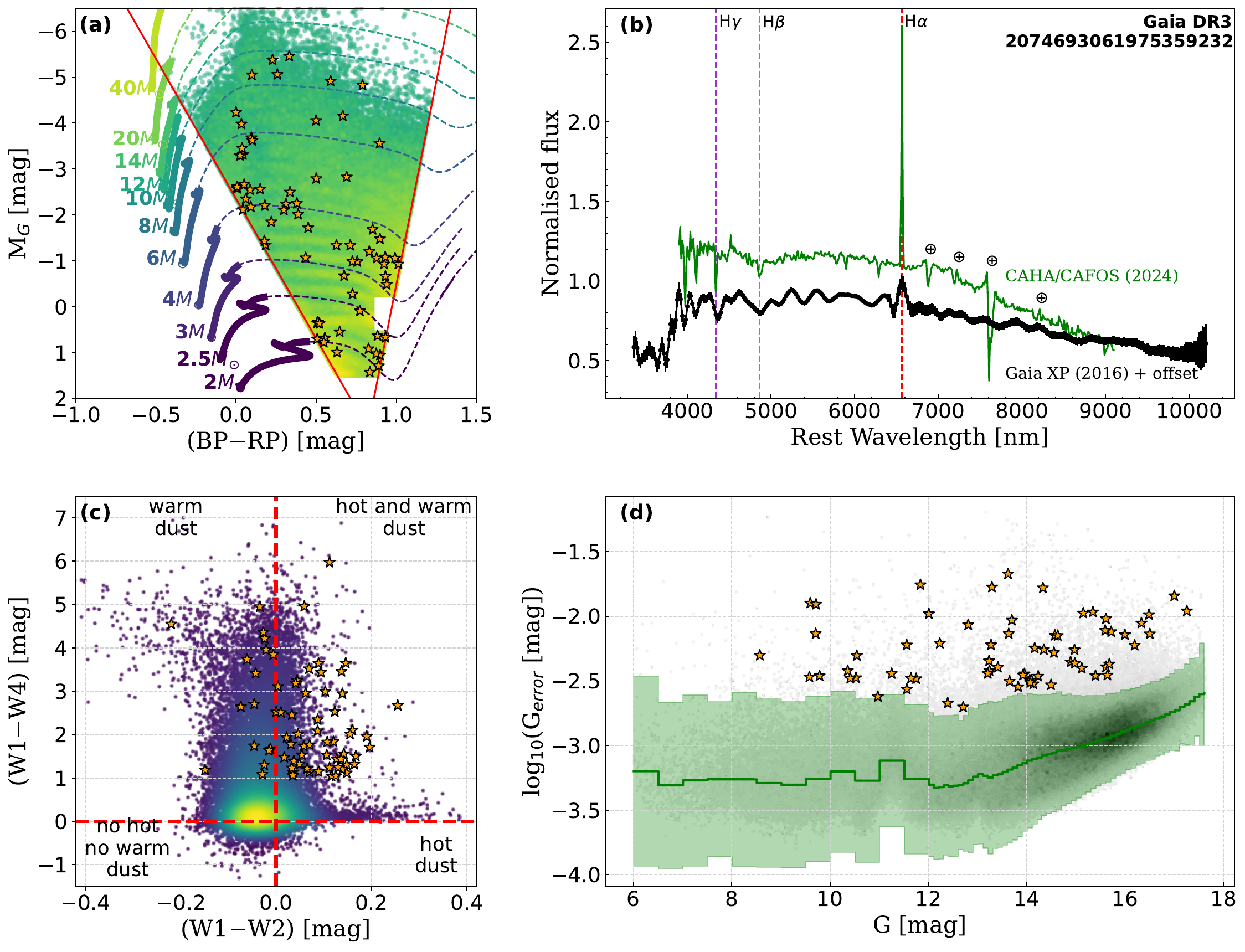} 
   \caption{Initial selection of our sample of mass-transferring candidates, based on a) the \textit{Gaia} EDR3 {\tt StarHorse} colour-magnitude diagram, b) the \textit{Gaia} XP spectra, c) the \texttt{ALLWISE} colour-colour diagram, and d) the \textit{Gaia} $G$-band variability. In \textit{a}, \textit{c}, and \textit{d}, we show the final sample of 67 candidates as orange stars. The red lines in \textit{a} show the cuts used to select stars between the MS and the RGB based on \textit{MESA Isochrones \& Stellar Tracks}. In \textit{b}, we show the \textit{Gaia} XP spectrum of one of the sources in our sample as an example, and we compare it with our follow-up spectrum taken with CAFOS (see Section \ref{od:spec}), and the $\oplus$ symbols represent telluric absorption (absorption by the Earth’s atmosphere). The red dashed lines in panel \textit{c} indicate the zero levels and divide the plot into four regions: warm dust, hot dust, both hot and warm dust, and no hot or warm dust. The shaded green region in \textit{d} corresponds to the 0.9 percentile of the initial HG sample.}
              \label{selec_grid}
\end{figure*}

\begin{figure*}[p]
\centering
\includegraphics[width=\textwidth]{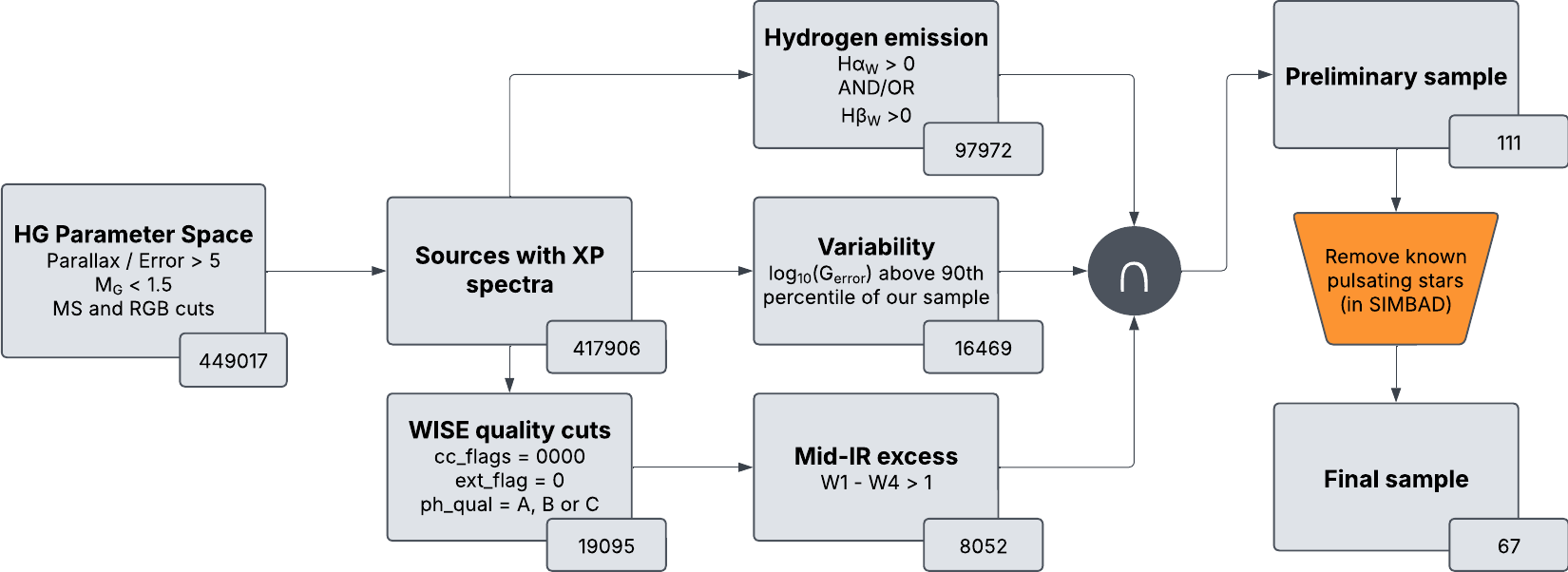}
    \caption{Overview of the selection process with the number of sources at each stage of the selection.}
    \label{diagram}
\end{figure*}
\twocolumn

\begin{table}[t]
\caption{Constraints and parameter space cuts made for the selection of our candidates. }
\label{table:selection}
\centering          
\renewcommand{\arraystretch}{1.1}
\begin{tabular}{cl}
\hline \hline
Selection step & Constraint \\
\hline
1 & parallax / parallax\_error $> 5$ \\ 
 & $M_G < 1.5$\,mag \\ 
 & $M_{BP} - M_{RP} > (21/128) M_G + (149/386)$ \\ 
 & $M_{BP} - M_{RP} < (-7/124) M_G + (603/620)$ \\ [0.6ex]

2 & H$\alpha_w > 0$\,nm OR H$\beta_w > 0$\,nm \\ [0.6ex]

3 & $W1 - W4 > 1$\,mag \\ [0.6ex] 

4 & $\log_{10}(G_{err})$ above 90th percentile \\

\hline
\end{tabular}
\tablefoot{The magnitudes used are the absolute magnitudes provided by \texttt{StarHorse}, and H$\alpha_w$ and H$\beta_w$ refer to the equivalent width of the respective lines derived using the internally calibrated \textit{Gaia} XP spectra.}
\end{table}

\begin{figure}[h!]
\centering
\includegraphics[width=\hsize]{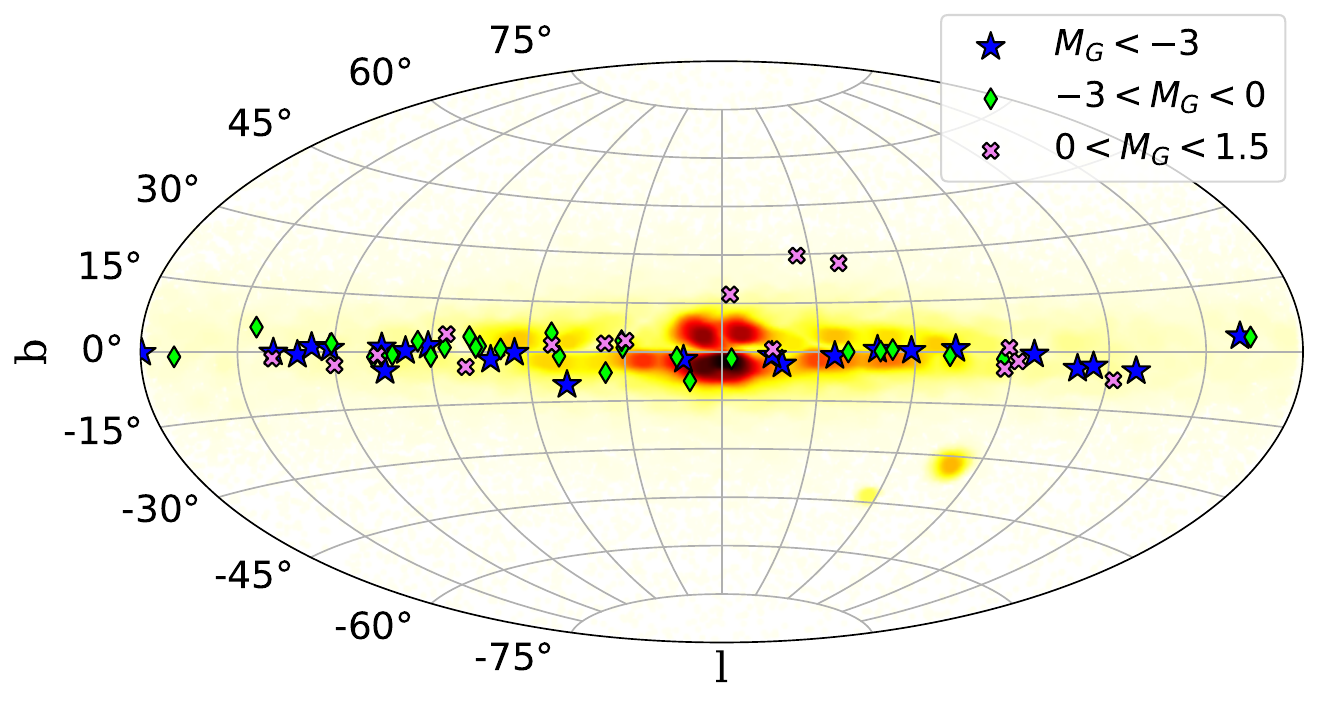}
    \caption{Sky map showing the position of the 67 sources in our initial sample with reference to the Galactic density of stars, constructed with a \textit{Gaia} DR3 random data set containing 100\,000 sources. Different ranges of absolute magnitude are shown with different markers.}
    \label{skymaps}
\end{figure}

We first applied step 1 to select the initial HG sample, then performed steps 2, 3, and 4 on this initial sample separately. Finally, step 5 was applied to the intersection of steps 2, 3, and 4, resulting in our final sample of 67 sources selected for further analysis. This process is summarised in Fig. \ref{diagram}, and a summary of the constraints used is presented in Table \ref{table:selection}. The 67 sources in the final sample are highlighted in Fig. \ref{selec_grid}, and their positions in the sky are shown in Fig. \ref{skymaps}. In Table \ref{longtable1:data} we gathered information on the targets in the final sample, such as their sky positions, distances and apparent magnitudes. In the following section, we describe the observational data used to characterise this sample.

\section{Observational data}
\subsection{Photometry}
In optical wavelengths, we retrieved light curves from the \textit{Zwicky Transient Facility} (ZTF) \citep{ztf}, the \textit{All Sky Automated Survey for SuperNovae} (ASAS-SN) \citep{ASAS-SN}, the \textit{Asteroid Terrestrial-impact Last Alert System} (ATLAS) \citep{ATLAS, ATLASVar}, and the \textit{Transiting Exoplanet Survey Satellite} (TESS) \citep{TESS}. In the near-infrared, we retrieved light curves from the \textit{NEO Wide-Field Infrared Survey Explorer} (NEOWISE) \citep{NEOWISE}. We cleaned all the photometric data using the recommended flags in the documentation of each catalogue. The TESS processed light curves were retrieved from catalogues that used different reduction pipelines. The main ones used in this work were the \textit{TESS Light Curves From Full Frame Images} (SPOC) \citep{SPOC} and the \textit{Quick-Look Pipeline} (QLP) \citep{QLP1, QLP2}. In addition, we also inspected TESS light curves from the \textit{TESS Asteroseismic Science Operations Center} (TASOC) \citep{TASOC}, the \textit{TESS FFI-Based Light Curves from the GSFC Team} (GSFC-ELEANOR-LITE) \citep{GFSC}, and the \textit{Cluster Difference Imaging Photometric Survey} (CDIPS) \citep{CDIPS}. In Fig. \ref{all_lcs_1} we show the light curves of all the sources in our sample, in 50-day bins.

\subsection{Spectroscopy} \label{od:spec}
Spectra were obtained using the following instruments:

\begin{itemize}
    \item The low-resolution instrument CAFOS \citep{cafos} at the \textit{Calar Alto Observatory} (CAHA) 2.2\,m telescope \citep{caha1, caha2}, with Grism G-200 ($R\sim300$, 4\,000--8\,500\,\AA{}). We reduced the data using a custom-developed \texttt{Python} pipeline, adapted by our team.
    \item The HERMES high-resolution spectrograph \citep{hermes} ($R=85\,000$, 3\,770--9\,000\,\AA{}) at the \textit{Mercator} telescope\footnote{\url{www.mercator.iac.es}} (Proposal ID: 5 (ROB service time), PI: Britavskiy, N.). These spectra were flux calibrated using the flux retrieved from the \textit{Gaia} DR3 low-resolution XP spectra.
    \item  The FIES instrument \citep{fies} at the the \textit{Nordic Optical Telescope} (NOT) \citep{NOT}, with the medium-resolution setup ($R=46\,000$, 3\,700--8\,300\,\AA{}) and with the low-resolution setup ($R=25\,000$, 3\,700--8\,300\,\AA{}) (Proposal ID: P69-854, PI: Wichern, H.; Proposal ID: P70-203, PI: Garcia-Moreno, G.).
    \item The ALFOSC instrument at the NOT, with Grism G4 ($R=710$ and $R=360$, 3\,200--9\,600\,\AA{}), Grism G7 ($R=650$, 3\,650--7\,110\,\AA{}), and Grism G19 ($R=1\,940$, 4\,400--6\,950\,\AA{}) (Proposal ID: P70-203, PI: Garcia-Moreno, G.; Proposal ID: P71-209, PI: Blagorodnova, N.). We reduced the data using the \texttt{Python} package \texttt{PypeIt} \citep{pypeit}.
    \item Low-resolution ($R=1\,800$, 3\,700--9\,000\,\AA{}) archival spectra from the \textit{Large Sky Area Multi-Object Fibre Spectroscopic Telescope} (LAMOST) \citep{lamost}.
    \item The low-resolution Mookodi spectrograph and imager mounted on the \textit{South African Astronomical Observatory} (SAAO) 1\,m Lesedi telescope \citep{Worters2016,Erasmus2024}, with a 4$^{\prime\prime}$ slit ($R{\approx}175$, 4\,000-–8\,000\,\AA{}). We reduced the data using a Mookodi-specific pipeline developed using the \texttt{ASPIRED} toolkit \citep{Lam2023}, which performs standard long slit reductions including trace fitting, spectral extraction, wavelength calibration and flux calibration.
\end{itemize}

In Fig. \ref{all_spec} we show single epoch spectra of the 26 candidates that we characterised spectroscopically, and the normalised flux of the H$\alpha$ and the H$\beta$ profiles in velocity space is shown in Fig. \ref{h_prof}. Details of the observations can be found in Table \ref{table:spec}.

\section{Analysis}
To verify the spectral type and binary nature of our candidates, we performed a light curve analysis and a spectroscopic analysis, in addition to a multi-wavelength characterisation using ultraviolet (UV) and X-ray catalogues from the literature. In this section, we describe the methodology of this analysis.

\onecolumn
\begin{figure*} [h!]
   \centering
   \includegraphics[scale=0.585]{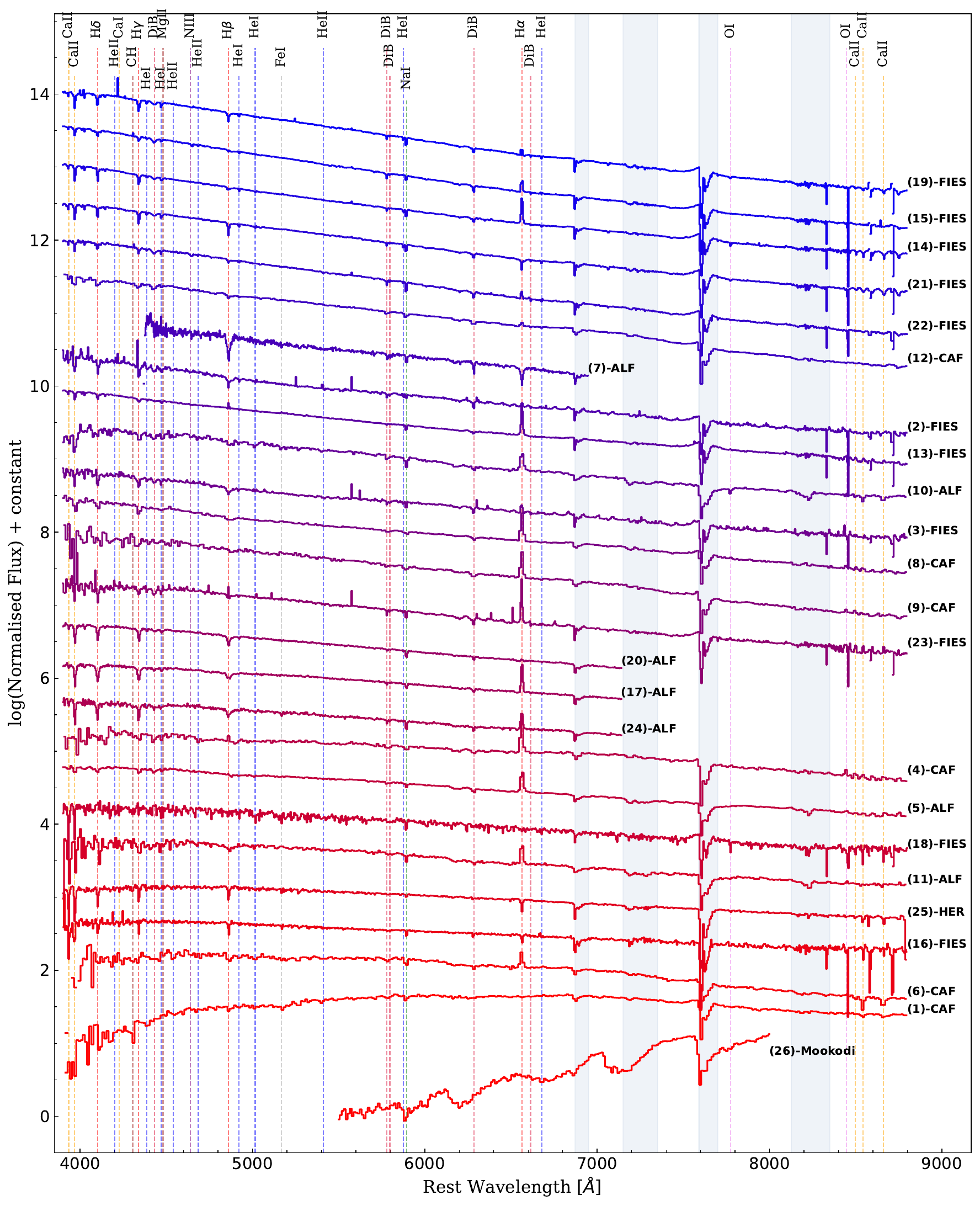} 
   \caption{Optical spectra for the subsample of spectroscopically characterised sources. Each spectrum, except for the Mookodi spectrum, is de-reddened for visualisation purposes using the best extinction values (see Sect. \ref{disc:extinc}, Tab. \ref{longtable1:data}), with \citet{Fitzpatrick99} dust extinction function and $R_V = 3.1$. The main emission and absorption lines are indicated. The areas with strong telluric absorption are shown by the shaded rectangles. On the right, the number in parenthesis indicates the index associated with each source (see Table \ref{table:spec}), and this index is followed by the instrument used to take each spectrum: CAFOS (CAF) from the \textit{Calar Alto Observatory} 2.2\,m telescope, HERMES from the \textit{Mercator} telescope, FIES and ALFOSC (ALF) from the \textit{Nordic Optical Telescope}, and Mookodi from the \textit{South African Astronomical Observatory} 1\,m Lesedi telescope.}
   \label{all_spec}
\end{figure*}

\begin{figure*} [h!]
   \centering
   \includegraphics[width=\textwidth]{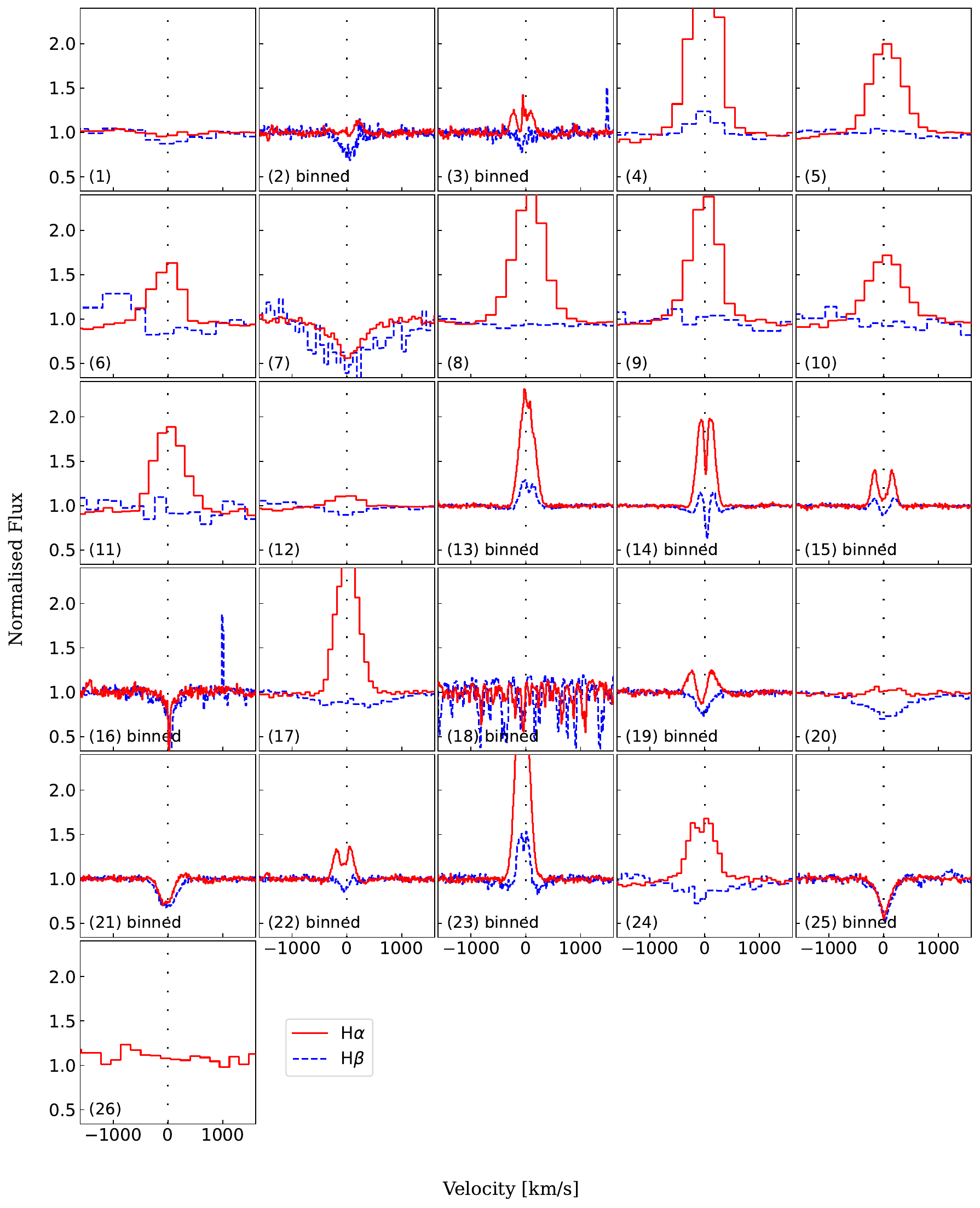}
   \caption{Flux normalised H$\alpha$ and H$\beta$ velocity profiles for the subsample of observed sources. All sources are shown on the same scale. The index corresponding to each source (see Table \ref{table:spec}) is shown in the bottom left corner of each spectrum. Medium- and high-resolution spectra have been binned using 10\,$km/s$ bins for improved visual representation.}
              \label{h_prof}
\end{figure*}
\twocolumn

\subsection{Light curve analysis}
\subsubsection{Visual inspection} \label{lc_visual_inspection}
We visually inspected the retrieved light curves of every source in our sample. We labelled the observed features in the light curves using the following categories:
\begin{itemize}
    \item Different kinds of variability (Diff): sources showing mixed, distinct variability patterns, including different oscillation modes or a combination of periodic and non-periodic variations, for example. 
    \item Outbursts (Outb): sources showing sudden brightening once or more than once, which could be linked to outbursts.
    \item Slow-rising in brightness (SR): sources that are slowly increasing in brightness. This increase can be observed in the optical bands, in the IR bands, or both. We only give this label to sources that have been increasing in brightness during the observed time frame or are brightening at later times and do not show any signs of a previous brightening.
\end{itemize}
An example lightcurve for each of these three labels is shown in Fig. \ref{figure:diff_outb_sr_example}.

\begin{figure}
\centering
\includegraphics[width=\hsize]{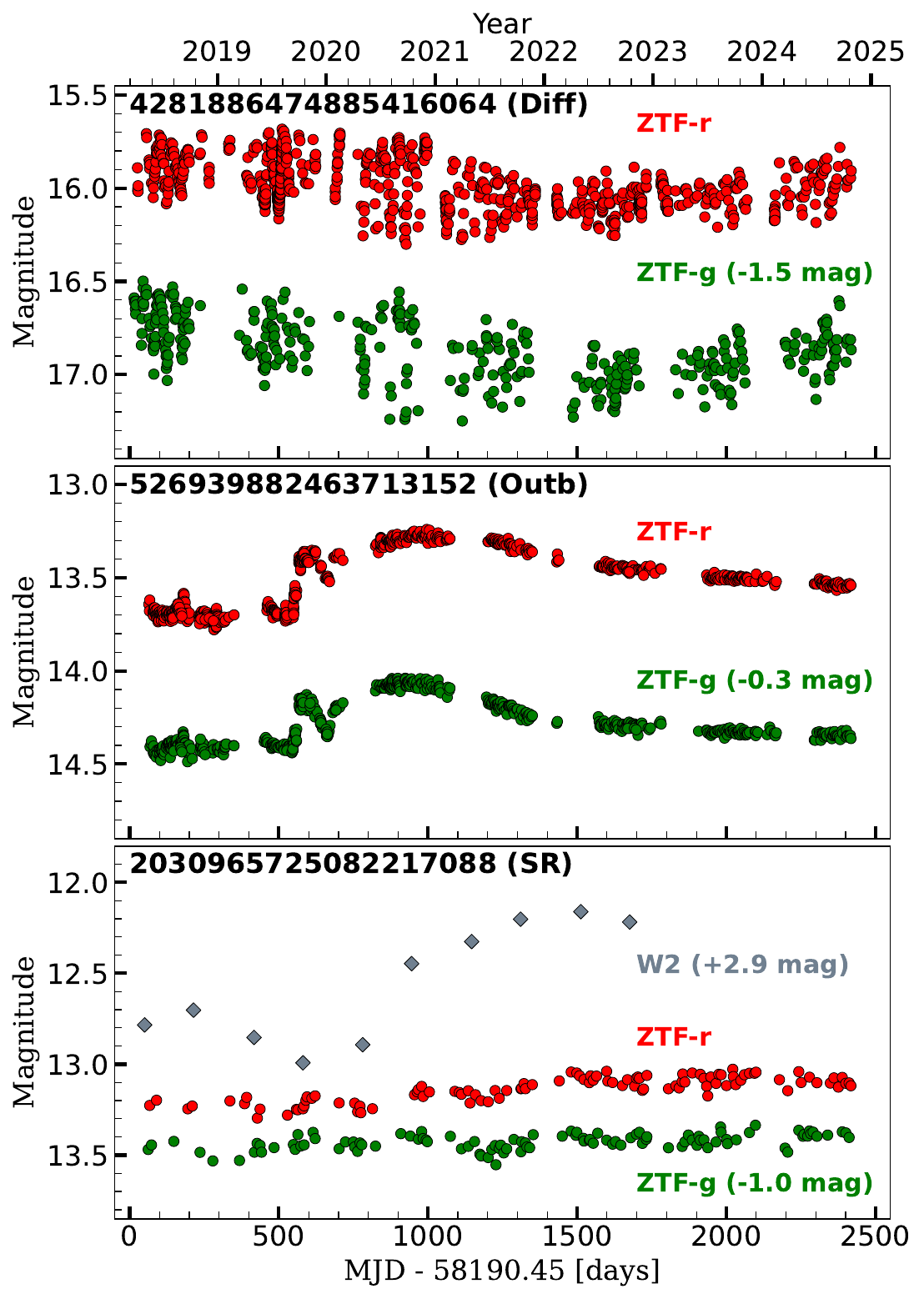} 
    \caption{Light curves of three sources in our sample illustrating the types of variability defined in this work (see Sect. \ref{lc_visual_inspection}). Circles represent the ZTF g band (green) and r band (red), while grey diamonds correspond to the NEOWISE W2 band. For clarity, the g-band and the W2-band light curves are vertically offset in all three panels, and the SR example is binned to better show its long-term evolution. The reference Modified Julian Date (MJD) corresponds to the start of the ZTF survey.}
    \label{figure:diff_outb_sr_example}
\end{figure}

\subsubsection{Periodicity analysis}
We performed a periodicity search using \texttt{FINKER} \citep{finker}, a nonparametric kernel regression code, applied to the ZTF, the ASAS-SN, and the ATLAS light curves. Frequencies from 0.001\,d$^{-1}$ (1\,000 days) to 100\,d$^{-1}$ (14.4 minutes) with steps of 0.0004\,d$^{-1}$ were tested for all sources. Alternative frequency grids were also explored, but they did not yield improved results. We performed an additional search for periods as short as five minutes for some sources where no successful longer period was found. \texttt{FINKER} allows uncertainty estimation in frequency determination via bootstrapping, but in our case, the errors given by this method were insignificant, and we do not report them. Lomb-Scargle periodograms \citep{Lomb, Scargle} were used for TESS light curves, which helped us find the shorter periods in our sample. Lomb-Scargle periodograms were also applied to the ZTF, the ASAS-SN, and the ATLAS light curves, but no significantly better results when compared with \texttt{FINKER} were obtained. In all attempts, we removed data points that were more than 5$\sigma$ away from the median magnitude of all the observations. Outlier removal was performed carefully to avoid excluding short-duration eclipses. Finally, all periods were tested by visually inspecting the folded light curves.

We labelled the sources based on their periodic variability as sinusoidal (Sin), non-sinusoidal pulsations (Puls), or eclipsing (ECL). Sources were labelled as eclipsing only when the eclipses could be clearly identified. Consequently, contact binaries were classified as sinusoidal variables, since in most cases their eclipses are indistinguishable from the sinusoidal variability of pulsators.


\subsection{Spectroscopic analysis}
To verify the spectral type of our stars and their location in the HG, we carried out a spectroscopic follow-up campaign targeting the most promising sources with periodic or peculiar light curves. While higher-resolution spectra would allow a more detailed study of specific spectral features, the low resolution obtained for some systems was adequate for our primary objectives: assessing whether the stars are "yellow" and verifying the presence of Balmer emission.

\subsubsection{Spectroscopic classification}
Our sources were given a spectral class by visually inspecting their spectra, and comparing them with theoretical spectra and spectra from standard stars, following \citet{princeton}. We only report temperature types (O-type, B-type, etc.) in our classification, as our only aim is to determine if the spectral type is compatible with being a yellow star. Additionally, to obtain a complementary spectral classification, we used \texttt{NutMaat} \citep{nutmaat2024}, a \texttt{Python} implementation of the \texttt{MKClass} code \citep{mcclass2014}. These classifications were also compared with the spectral classification provided by \textit{Gaia} DR3.

\subsubsection{Hydrogen line profiles}
We inspected the spectral region around H$\alpha$ and H$\beta$ and labelled each source, given the properties of their hydrogen profiles:
\begin{itemize}
    \item P-Cygni profile (PCyg)
    \item Emission or absorption in H$\alpha$ (H$\alpha$ abs/emi): from the 26 sources with new spectra, this label tells which show emission or absorption in H$\alpha$, also indicating if it has a two-peak (2p) profile.
\end{itemize}

\subsubsection{Extinction estimation} \label{analysis:extinction}
For the spectra with the high and medium resolution, we estimated the extinction from the absorption caused by the diffuse interstellar bands (DIBs) at 5\,780\,\AA{} and 6\,614\,\AA{} following \cite{dibs}, where they derive an empirical relation between the line-of-sight extinction and the EW of these two DIBs. To convert line-of-sight extinctions to visual extinctions, we used $R_V = A_V / E(B-V) = 3.1$. The resulting $A_V$ values were then converted to extinctions in the \textit{Gaia} passbands using the \texttt{gaia\_edr3\_photutils} package\footnote{\url{https://github.com/fjaellet/gaia_edr3_photutils/}}. To use this method, we needed precise measurements of the EWs of the DIBs, which we could not get with our low-resolution spectra.

\subsection{Characterisation with auxiliary catalogues}
We cross-matched our sample with the UV catalogue \textit{GALEX} \citep{galex} (with a search radius of 10 arcseconds), and X-ray catalogues from \textit{Swift} \citep{swift} (with a search radius of 10 arcseconds), \textit{Chandra} \citep{chandra} (with a search radius of 50 arcseconds), \textit{eROSITA} \citep{erosita} (with a search radius of 10 arcseconds), and \textit{4XMM} \citep{XMM} (with a search radius of 12 arcseconds). In all cases, the search radius was chosen to be enough to contain the positional error of the catalogues. We labelled the sources with X-ray (Hard X-ray) or UV (Far UV) detections.

To test for known variable stars, we also cross-matched our sample with a catalogue containing a cross-match of over 8 million \textit{Gaia} sources with variable objects in 152 different catalogues from the literature \citep[or GVXM from now on]{GaiaVarXmatch}. This catalogue provides over 100 variability (sub)types and periods, if available. 


\section{Results} 

\subsection{Light curve analysis}
We observed variability in the light curves of all 67 sources in our sample, but not all show signs of periodicity. Among all the sources, 19 show multiple types of variability, the majority of them likely associated with more than one pulsation mode (e.g., sources 2164630463117114496 or 4281886474885416064); ten show a fast decrease in magnitude (e.g., sources 526939882463713152 or 2027563492489195520), and 11 sources present a slow-increase in brightness over several hundreds or thousands of days (e.g., source 4263591911398361472). For some of the latter, the gradual brightening may be linked to long-term variability; future light curve data will help determine whether this is the case.

The folded light curves of the 36 candidates for which we determined a period are shown in Fig. \ref{grid_lcs1} and Fig. \ref{grid_lcs2}. Periods of one day or less were found for 19 candidates, periods between one and eight days for ten candidates, and periods longer than ten days for the remaining seven, with the longest period being $\approx$106.76 days. Among these 36 sources with periodic variability, 26 show sinusoidal variability, nine show eclipses, and we identify pulsations for one (see Fig. \ref{pie_charts}).

We note that periods detected exclusively in TESS light curves should be interpreted with caution. Given the large plate scale of TESS (21$^{\prime\prime}$\,pixel$^{-1}$), light-curve contamination from nearby bright variables is possible when sources lie within the same pixel. This is a known issue discussed, for example, by \citet{Pedersen2023_TESS_confusion} in the context of fast yellow pulsating supergiants. In our case, the short period of $\approx0.171\,$days initially found for source \textit{5962956195185292288} using TESS data was later traced to an eclipsing binary located $\approx42^{\prime\prime}$ away, identified as \textit{Gaia} DR3 \textit{5962956229545027712}.

\begin{figure}
\centering
\includegraphics[scale=0.47]{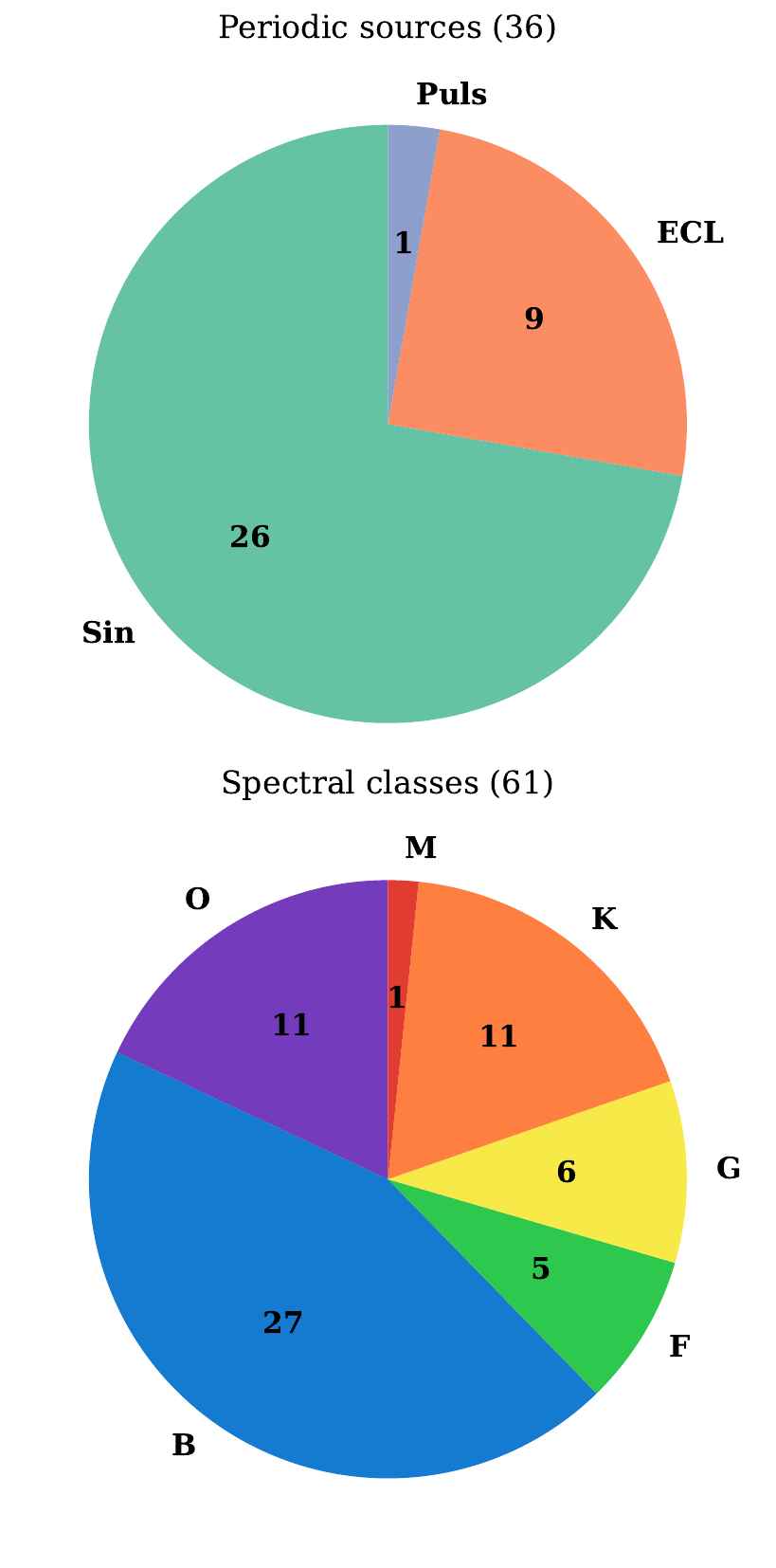} 
    \caption{Top: pie chart with the types of periodic variables identified in our sample. Bottom: pie chart with the spectral classes in our sample, combining the results of our classification and the classes reported in the \textit{Gaia} DR3.}
    \label{pie_charts}
\end{figure}

\subsection{Spectroscopic analysis}
The estimated spectral types we report from our visual inspection can be grouped into three main groups: hotter stars (O- and B-type), yellow stars (F- and G-type), and colder stars (K- or M-type). Out of the 26 candidates with spectra, we identified 17 as hotter stars, most of them Be stars. On the other side, we identified one source as a K-type star, and one as an M-type star. Overall, only four sources were finally assigned a spectral type compatible with being a YG or YSG: three F-type stars and one G-type star. The spectra obtained for three of the sources showed no clear features (apart from the main hydrogen lines), and we did not assign a spectral class to them. The \texttt{NutMaat} classifications rated as \textit{good} or better were consistent with our spectroscopic results, yielding the same temperature types as those we derived.

A comparison of our spectral classification with the one available in the \textit{Gaia} DR3 showed consistent results for 18 stars. For two, \textit{Gaia} identified them as G-type stars, whereas we classified them as B-type based on the presence of \ion{He}{i} absorption lines. Another source was classified as an O-type star by \textit{Gaia}, but our analysis indicates that its spectrum matches a B-type better. Additionally, \textit{Gaia} had spectral types for 38 sources in our sample for which we lacked spectra. These 38 spectral types are distributed as follows: 8 O-type, 13 B-type, 2 F-type, 5 G-type, and 10 K-type stars. All the spectral types identified are summarised in Fig. \ref{pie_charts}.

We note that our spectral classification was based on single stellar spectra. In the case of binaries, this approach is reasonable when one component dominates the system’s light, which we consider to be the case for our sources. Indeed, most of our spectra showed no evidence of composite features, suggesting that the contribution of a secondary component is negligible. However, this effect could explain the peculiar spectra observed in the three candidates for which no spectral type was assigned (sources \textit{2061252975440642816}, \textit{2164630463117114496}, and \textit{2175699216614191360}).

From the H$\alpha$ and H$\beta$ profiles shown in Fig. \ref{h_prof}, we identified H$\alpha$ emission in 19 sources (ten of which had a two-peak profile), H$\alpha$ absorption in five sources, and no discernible features in two of them. Five sources had H$\beta$ emission in addition to H$\alpha$ emission. Additionally, in Fig. \ref{h_prof_gaia_xp} we present a comparison between the H$\alpha$ profiles obtained from the \textit{Gaia} XP low-resolution spectra and our follow-up spectra, degraded to match the XP resolution. In most cases, the profiles show good agreement, while in others they differ significantly, suggesting that the H$\alpha$ emission may be transitory or variable over time. 

Following the method described in Sect. \ref{analysis:extinction}, we estimated the extinction for each source, resulting in $A_V$ values reported in Table \ref{table:av_dibs}. Updated extinction values have enabled us to revisit the locations of nine sources in the \textit{Gaia} colour-magnitude diagram, as shown in Fig. \ref{dibs}. The discussion of these results and a comparison of our extinction estimates with those from two external catalogues are discussed in Sections \ref{disc:extinc} and \ref{disc:shboost}.

\subsection{Characterisation with auxiliary catalogues} \label{res:multi_wavelength}
Our cross-match with UV and X-ray catalogues allowed the identification of three sources with X-ray emission, and one source with UV excess. The source with the UV detection in \textit{GALEX} was \textit{6123873398383875456}, which presented UV excess in the far UV channel (1350--1750\,\AA{}{}). This detection matched the optical counterpart at 0.44 arcseconds (within positional errors and with no other optical counterpart nearby).

The X-ray detections came from \textit{4XMM} and \textit{Swift}: one detection in \textit{Swift}'s \textit{2SXPS} catalogue matching source \textit{4054010697162430592} at 0.93 arcseconds (within positional errors and with no other optical counterpart nearby), and two detections in \textit{4XMM}-DR14, matching source \textit{5866345647515558400} at 1.1 arcseconds and source \textit{2083649030845658624} at 1.9 arcseconds, both within positional errors. However, \textit{2083649030845658624} had other optical counterparts within a few arcseconds (also within positional errors), so it is not clear if the X-ray emission is from our source.

\begin{table}[t]
\caption{Results of our extinction estimates using diffuse interstellar bands (DIBs).} 
\label{table:av_dibs}      
\centering
\begin{tabular}{clll}
\hline \hline
Index & Gaia DR3 ID & $A_V$ & $A_{V,err}$ \\ 
& & (mag) & (mag) \\
\hline 
2 & 187219239343050880 & 4.06 & 1.13 \\
3 & 2006088484204609408 & 3.72 & 1.02 \\
13 & 3355776901779440384 & 1.18 & 0.30 \\
14 & 3369399099232812160 & 2.47 & 0.65 \\
15 & 3444168325163139840 & 4.00 & 1.10 \\
19 & 461193695624775424 & 3.56 & 1.00 \\
21 & 508419369310190976 & 3.21 & 0.87 \\
22 & 512721444765993472 & 4.82 & 1.37 \\
23 & 526939882463713152 & 4.80 & 1.33 \\
\hline
\end{tabular}
\tablefoot{The reported $A_V$ values represent the average of the estimates derived from the 5\,780\,\AA{} and 6\,614\,\AA{} DIBs. Indices correspond to those reported in Table \ref{table:spec}. For details on the derivation of the $A_V$ values, see Sect \ref{analysis:extinction}}
\end{table}

\begin{figure}
\centering
\includegraphics[width=\hsize]{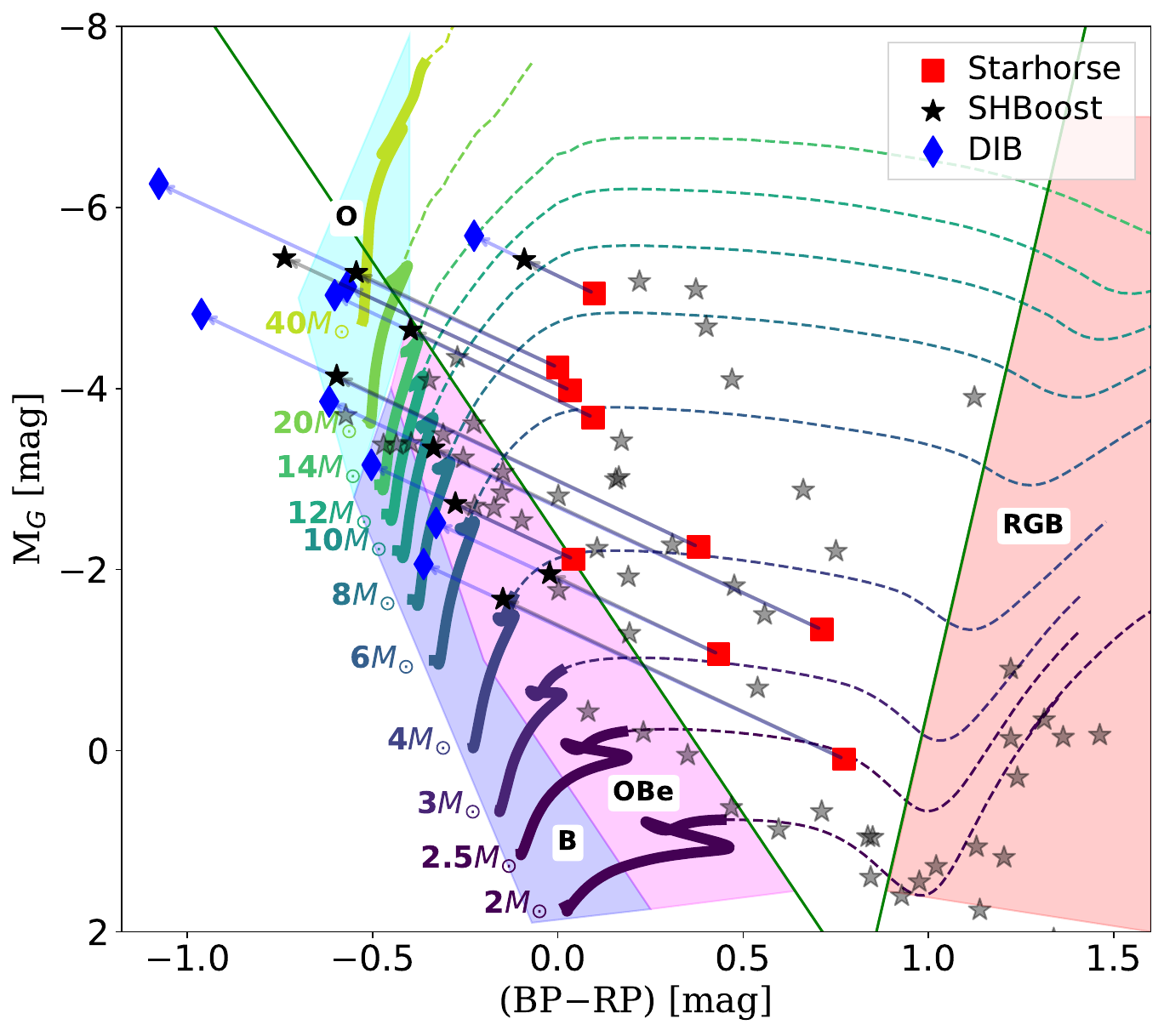}
    \caption{Positions of the nine sources for which we re-estimated extinction (see Sect. \ref{analysis:extinction}) in the colour-magnitude diagram. Red squares indicate the positions derived using \texttt{StarHorse}, while blue diamonds show the positions computed using our extinction estimates based on diffuse interstellar bands, averaging the results from the 5\,780\,\AA{} and 6\,614\,\AA{} lines. For comparison, we also show the positions of all sources in our sample using the \texttt{SHBoost} catalogue (black stars; see Sect. \ref{disc:extinc}). The schematic boxes denote the typical locations of different stellar types.}
    \label{dibs}
\end{figure}

Our search for periods reported in the literature provided 18 catalogued periods. Comparing these with the periods derived in this work, we found that eight sources had consistent values. For one eclipsing binary, our analysis yielded half the reported period. We were able to reproduce the reported period for four sources, while for the remaining five, the periods published in GVXM could not be confirmed with our data. The unconfirmed periods correspond to relatively long timescales, which may not be reproduced with our data either due to the limited observational baseline or because the literature periods themselves are uncertain, as the cross-matched values in GVXM are not independently verified.

The main results of our photometric and spectroscopic analysis are summarized in Fig. \ref{pie_charts}, and all the results of our analysis are provided in Table \ref{longtable1:results}, where we present the spectral types identified in this work together with the types listed by \textit{Gaia}, the main properties in the light curves and the spectra, the classes from the literature (from the GVXM and \texttt{SIMBAD}), and the periods found in this work and the literature. Literature periods that could not be confirmed with our data are marked with an asterisk in the last column of these tables.

\section{Discussion}

The goal of our selection was to identify mass-transferring binaries in the HG. However, our sample includes a variety of spectral types, most of which are contaminants. In this section, we first discuss how line-of-sight extinction affected our selection process, and then examine the nature of our candidates.

\subsection{The effect of extinction} \label{disc:extinc}

The most significant limitation of our method for selecting our sample of HG stars was the lack of reliable extinction estimates in the literature. As a result, for more than half of the sources in our sample, the extinction was underestimated, implying that the MS O- and B-type sources appeared fainter and redder. Consequently, many of our candidates were stars such as Oe or Be stars, which share some characteristics with our objects of interest but are generally not expected to lie in the HG \citet{BeHrd}. Therefore, obtaining precise extinction estimates is essential for reliably selecting sources within the narrow region of the colour-magnitude diagram occupied by the HG. 

Luckily, during the preparation of this work, several new astrophysical parameter catalogues derived from {\it Gaia} XP spectra became available (e.g., \citealt{Zhang2023, Fallows2024, Huson2025}). Of particular interest for us was the new \texttt{SHBoost} catalogue \citep{SHBoost}, which contains stellar parameters (effective temperature, surface gravity, mass, metallicity) and line-of-sight extinctions computed using \textit{Gaia} DR3 XP spectra, derived through supervised machine learning trained on a large high-quality dataset comprising the {\tt StarHorse} results for spectroscopic stellar surveys \citep{Queiroz2023} as well as complementary datasets of hot stars, white dwarfs, hot subdwarfs, among others. Particularly relevant for this work is that {\tt SHBoost} contains more reliable characterisation (including extinction) for high-mass stars than the {\tt StarHorse} catalogue, based on {\it Gaia} EDR3.

\subsection{Revisiting our first selection with SHBoost} \label{disc:shboost}

To assess the position of our sources in the colour-magnitude diagram with better extinction values, we cross-matched our sample with the \texttt{SHBoost} catalogue to obtain more accurate absolute magnitudes and colours. A comparison with the extinction corrected magnitudes from DIB spectroscopy is shown in Fig. \ref{dibs}, where we highlight two key observations: first, many of the Oe and Be stars now lie on the MS or just beyond the terminal-age MS, while the cooler stars (K- and M-types) are located on the RGB; second, all but one of the extinction values derived from our spectroscopic analysis are higher than those reported in \texttt{SHBoost}, with $A_V$ values larger by approximately 0.7\,mag on average. However, the values remain comparable within uncertainties, and we therefore consider the \texttt{SHBoost} estimates to be reliable. Figure \ref{dibs} also shows that from the initial 67 candidates selected based on the EDR3 {\tt StarHorse} catalogue, only 25 remain in our HG parameter space. Given these results, we can now discuss the true nature of the sources in our sample.

\subsection{Nature of our candidates} \label{disc:nature}

\subsubsection{Oe and Be stars}
Oe and Be stars are rapidly rotating, moderately massive stars (M $\sim 3.6$--$20$\,M$_\odot$) with Balmer emission associated with a gaseous envelope, and with an equatorial disk likely formed through pulsations or by radiatively-driven winds \citep{BeStars, Be-review}. Moreover, these stars show IR excess caused by free--free and free--bound emission from the gas in the disk \citep{Gehr}. Oe and Be stars are also known to exhibit different kinds of photometric variability \citep{Be-puls, Be-var}. 

Among the O- and B-type stars identified in our sample, 15 are strong candidates for Oe or Be stars based on the presence of Balmer emission. In some cases, Oe and Be stars also display \ion{Ca}{ii} triplet and Paschen emission (e.g., \citealt{CaEmissionBe, Banerjee2021}), an emission we identify in, for example, sources \textit{2013187240507011456} and \textit{2074693061975359232}. Only two B-type stars in our sample do not show Balmer emission in our follow-up spectra. As noted by \citet{Negueruela}, the emission lines in Oe and Be stars are variable and can transition from emission to absorption as the star temporarily exits the Be phase. This may explain the case of source \textit{508419369310190976}, for which H$\alpha$ emission is visible in the \textit{Gaia} XP spectrum (see Fig. \ref{h_prof_gaia_xp}), but it is absent in our follow-up spectrum.

Additionally, for the 44 sources for which we could not assign a spectral classification, either because we did not have follow-up spectra or because the spectra did not show enough features, 21 were classified as O- or B-type stars by \textit{Gaia}. Based on the selection criteria of our search and the results obtained, there is a strong reason to believe that most, if not all, of these sources are likely Oe or Be stars. 

\subsubsection{Regular variables}
Regular variables cannot be discarded as possible contaminants in our search. We discuss source \textit{6123873398383875456} in Sect. \ref{disc:gap_mass_transfer}, which we think is most likely a close binary misclassified as an RRc, but we found two other sources classified as regular pulsator stars in the literature: \textit{460686648965862528}, classified as a $\delta$ Scuti by \citet{deltaScuZTF}, and \textit{4076568861833452160}, classified as a Cepheid by \citet{asas_vari}. These kinds of variables reside in the instability strip, which crosses the HG, so it is expected to find them in our parameter space. Examples of different kinds of variables placed in the \textit{Gaia} HR diagram can be found in \citet{GaiaVarXmatch}.

Mid-IR excess is also expected in Cepheid variables \citep{CEP-IR, CEP-IR-theo}, and Balmer emission has been observed in the type II Cepheid W Virginis \citep{cep-emission}. Pre-MS $\delta$ Scuti variable stars have been observed before \citep{deltascuPMS}, and strong IR-excess and emission are expected in this type of star, linked to active accretion processes. Therefore, the selection of these variables with our search criteria is possible, and a new identification method will be needed to trace and remove them from our sample, especially if we want to distinguish them from close binaries, which can have similar light curves.

\subsubsection{Binaries with compact companions}
As described in Sect. \ref{res:multi_wavelength}, we identify one source with far UV excess and three sources with hard X-ray emission. From the three sources with X-ray detections, it is not clear if the X-ray detection of source \textit{2083649030845658624} corresponds to its optical counterpart, whereas sources \textit{4054010697162430592} (reported as a single-lined spectroscopic binary in \citealt{hardXray-sb}) and \textit{5866345647515558400} are good candidates for X-ray binaries.

Interestingly, the source with far UV excess---source \textit{473575777103322496}---is identified as a Be star, has sinusoidal variability with a period of 0.45878 days (Fig. \ref{grid_lcs2}), and shows two peaks in the IR light curve, one between 2014 and 2016 and a second one between 2018 and 2020 (Fig. \ref{all_lcs_1}). We suggest this source is a Be star with a hot companion (a subdwarf OB star or a white dwarf, for example), which is potentially accreting mass during a second mass transfer stage (as discussed in \citet{Be_xray} for $\gamma$ Cas stars), based on the UV emission and the observed variability.

\begin{figure}[t]
\centering
\includegraphics[scale=0.36]{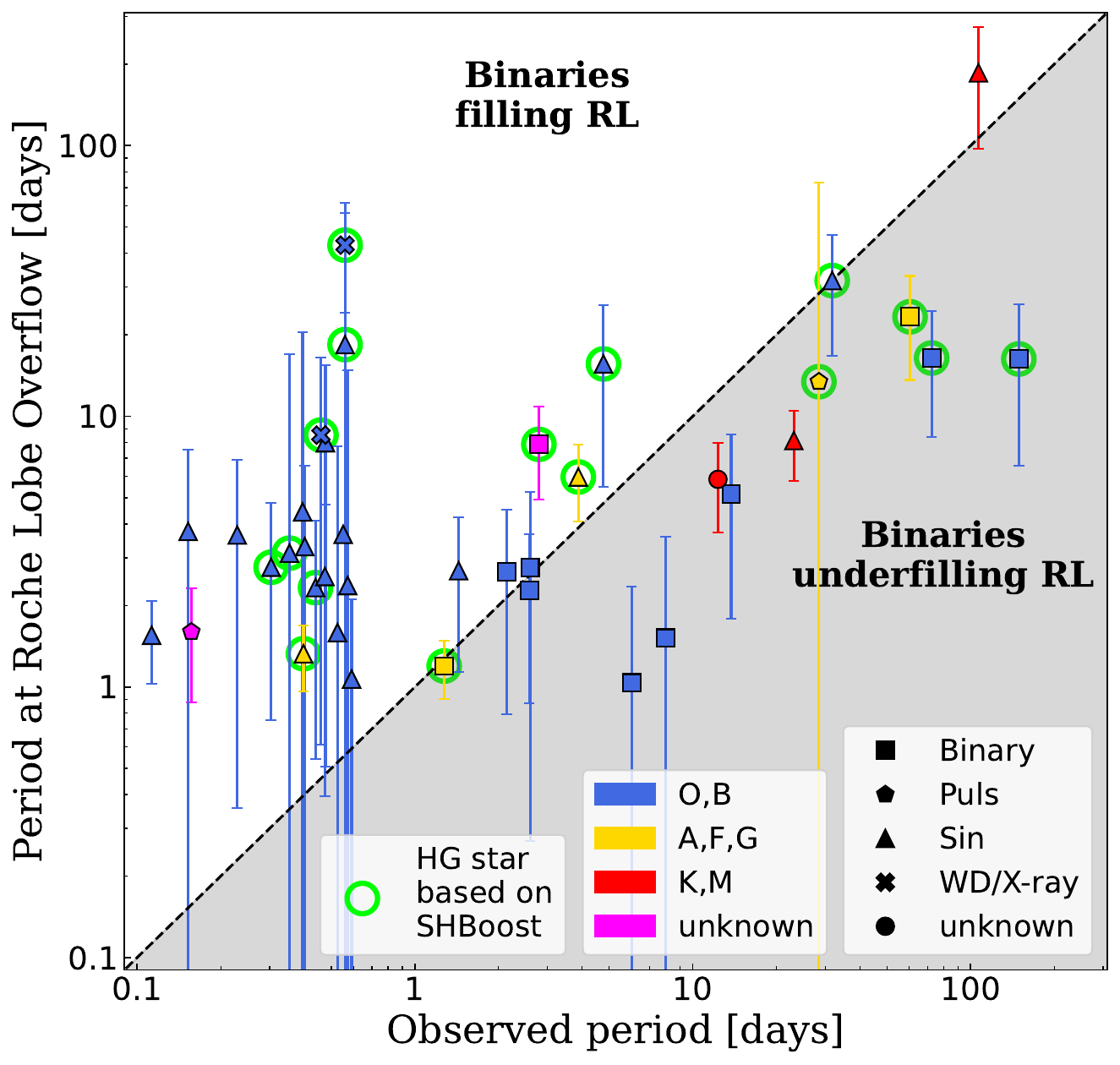} 
    \caption{Comparison of the observed period and the theoretical period of a binary in which one star fills its Roche lobe (see Sect. \ref{disc:gap_mass_transfer}). Different marker shapes correspond to different types of stars, and colours represent spectral classes (using our classification when available, and \textit{Gaia}'s otherwise). Sources that lie, based on \texttt{SHBoost} values, within our defined Hertzsprung gap parameter space are highlighted with circles.}
    \label{fig:periods}
\end{figure}

\subsubsection{Mass transferring systems} \label{disc:gap_mass_transfer}
One possible method for assessing whether our candidates were undergoing mass transfer in a binary system was to compare their observed orbital period with the theoretical period at which Roche lobe overflow would occur ($P_{RLO}$). Therefore, we take advantage of the improved parameters given by \texttt{SHBoost} to calculate this period,
\begin{equation}
    P_{RLO} \approx 0.35 \left(\frac{R^3}{M}\right)^{1/2} \left(\frac{2}{1+q}\right)^{0.2} \text{days}.
    \label{eq:period_orb}
\end{equation}

\begin{figure*}[h]
\centering
\includegraphics[width=\textwidth]{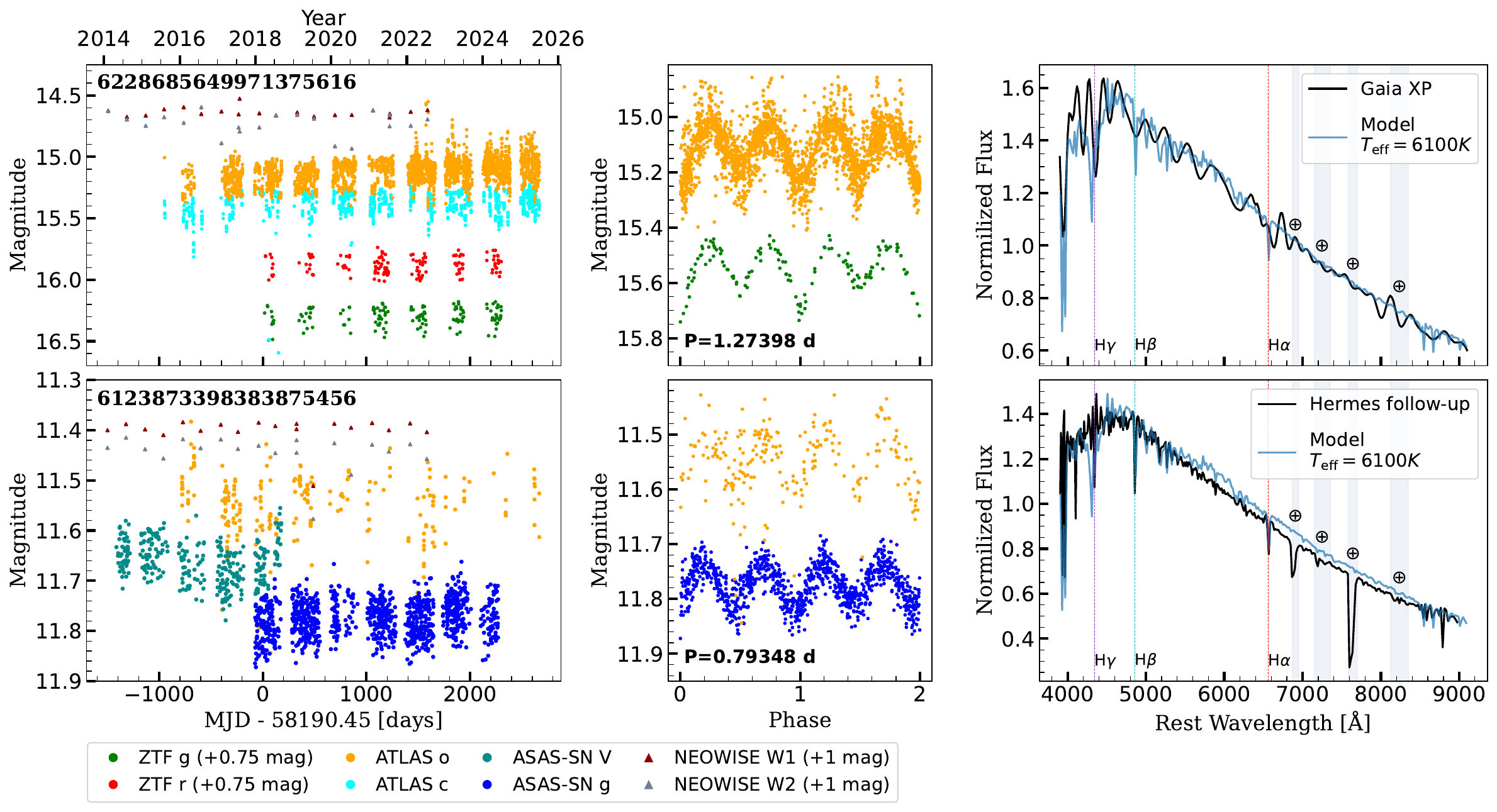}
    \caption{Light curve (left column), phase-folded light curve (middle column), and spectrum (right column) of our two best candidates for mass-transferring merger progenitors: \textit{6228685649971375616} (top row) and \textit{6123873398383875456} (bottom row). For both sources, a model spectrum from the Castelli–Kurucz Atlas \citep{CastelliANDKurucz}, with $T_\mathrm{eff}=6100\,$K and $\log g =3.75$, is shown for comparison. The ZTF and NEOWISE data in the light curves (left column) are offset for visual clarity, and the reference Modified Julian Date (MJD) in the light curve corresponds to the start of observations by ZTF. Although we report a sinusoidal period of 0.39674\,d for \textit{6123873398383875456}, the phase-folded light curve is shown using twice this period, consistent with our interpretation of the source as a contact eclipsing binary. The $\oplus$ symbols and shaded areas in the spectra (right column) represent telluric absorption (absorption by the Earth’s atmosphere).}
    \label{fig:6228685649971375616_lc}
\end{figure*}

The masses are provided directly by \texttt{SHBoost}, while the radii are computed using the $\log g$ values also given in \texttt{SHBoost}. We adopted a fixed mass ratio of $q = 0.5$ but accounting for the dependence on $q$ in the error bars by also computing the period at $q = 0.01$ (indicating that 99\% of the system's mass resides in the primary star) and at $q = 1$ (representing an equal-mass binary). The dependence of $P_{RLO}$ on $q$ is weak, and adopting different $q$ values does not significantly affect our results. It is to be noted that the masses in \texttt{SHBoost} are computed assuming single stars. In case the second component contributes to the total light, this may affect the final $P_{RLO}$, adding an additional uncertainty that we can not verify straightforwardly. 

We compared the $P_{RLO}$ to the periods found in this work, and we show this in Fig. \ref{fig:periods}, where we also mark different kinds of variables found among our candidates, their spectral class, and whether they remain in the HG using the extinctions from the \texttt{SHBoost} catalogue. Points above the diagonal correspond to sources with an observed period shorter than the period they would have if they were a binary and one of the components had filled its Roche lobe, whereas the opposite is true for points below the diagonal. Binaries in this region, with observed periods longer than $P_{RLO}$, are most likely detached binaries. Nonetheless, an interesting example of a source in this area---the G-type star \textit{2060841448854265216}---shows, in its folded light curve, eclipses akin to an Algol star with an accretion disk around the gainer (see \ref{grid_lcs1}). Algol binaries are systems where the mass gainer is the more massive star that is still in the MS, and the donor is the cooler, fainter and larger star \citep{Algol-binaries}. In this case, we could be seeing the less evolved star (or a combination of the two), which would mean $P_{RLO}$ is underestimated. Consequently, we would not rule out \textit{2060841448854265216} as a potential mass-transferring binary.

The most relevant sources for our study are those located near the line where the observed period equals $P_{RLO}$, as these represent the strongest candidates for ongoing mass transfer in binary systems. Among them, perhaps the most intriguing is source \textit{6228685649971375616}, a yellow HG star with a folded light curve characteristic of an interacting binary and an orbital period of 1.274 days. The light curves in different bands are shown in Fig. \ref{fig:6228685649971375616_lc}, together with the folded ATLAS o-band and ZTF g-band light curves. We also present the \textit{Gaia} XP spectrum and compare it with a model spectrum of a late F-type star ($T_{\mathrm{eff}} = 6100$\,K). The model was generated using \texttt{pysynphot} \citep{pysynphot} with the Castelli–Kurucz Atlas \citep{CastelliANDKurucz}, adopting $\log g = 3.75$, as reported by \texttt{SHBoost} (see Table \ref{longtable1:data}). The comparison shows a good agreement between the observed and model spectra, confirming that this is indeed a yellow source. Although no clear H$\alpha$ emission is detected in the \textit{Gaia} XP spectrum, the absence of a well-defined absorption feature suggests that the expected absorption is partially filled.

If this source is indeed evolving toward a merger (and a subsequent LRN), and assuming its progenitor behaves similarly to that of V1309 Sco \citep{V1309ScoMerger}, we would expect to observe three key signatures in its light-curve evolution: (1) a steady, monotonic brightening over several years (of order $\sim1$ mag in the $I$ band for V1309 Sco) prior to outburst; (2) a progressive decrease in the orbital period as the components spiral in; and (3) a transition in the folded light-curve morphology from an EW-type with two maxima per cycle to an asymmetric single maxima profile (see figures 1–3 in \citealt{V1309ScoMerger}). Such behaviour is not observed in \textit{6228685649971375616}. A slight increase in brightness is seen in the ATLAS bands (of about $\sim0.1$\,mag over approximately 10\,years), but this trend is not evident in other bands. Moreover, we detect no significant evolution in the folded light curve, which consistently shows two maxima and two minima per cycle across all epochs. We measured the period across different epochs, but any variations detected were smaller than the associated uncertainties. Investigating potential period evolution would require high-cadence observations over an extended timespan, which are not available for this source. If this object is indeed similar to the progenitor of V1309 Sco, we do not expect its outburst to occur in the near future.

While for the source discussed above we could clearly distinguish the two minima corresponding to the eclipses, this is not the case for some contact binaries. Contact binaries and regular pulsators can exhibit similarly sinusoidal light curves with short periods. Therefore, some of our candidates showing sinusoidal variability may in fact be close binary systems. One such case is \textit{6123873398383875456}, an F-type star identified in the ASAS-SN catalogue of variable stars \citep{asas_vari} as an RRc-type pulsator, although it is not included in the RR Lyrae table of \textit{Gaia} DR3. Notably, this source has a re-normalised unit weight error (RUWE) of nearly 6.5 in the \textit{Gaia} DR3 catalogue, a value well above the nominal threshold and often indicative of unresolved binarity \citep{ruwe-binary}. The folded light curve displays two symmetric minima with slightly different depths but the same periodicity. Although this behaviour could be attributed to period doubling in an RR Lyrae star, it could also correspond to the two eclipses of a contact binary. If the latter interpretation is correct, the star would lie just above the 1:1 line in Fig. \ref{fig:periods}, further supporting the hypothesis that it is a mass-transferring binary.

Similarly to \textit{6228685649971375616}, source \textit{6123873398383875456} does not show the behaviour expected for an LRN progenitor approaching outburst. In Fig. \ref{fig:6228685649971375616_lc}, we show its light curves across different bands, along with the folded ATLAS o-band and ASAS-SN g-band light curves. We also present the follow-up spectrum, which we compare to a model of a late F-type star (the same model described above).

The case of source \textit{6228685649971375616}, however, appears to be an exception rather than the rule, as we do not find any evidence of binarity in other sinusoidal sources. As discussed in the subsections above, regular variables and OBe stars can also be selected by our criteria. These likely account for the majority of stars with sinusoidal light curves located in the leftmost region of the $P_{RLO}$-observed period diagram. 

Interestingly, among all the O- and B-type stars in our sample, we identify at least seven in binary systems, including three Be stars: \textit{527155253604491392}, \textit{2006088484204609408}, and \textit{2166378312964576256}. The formation channel of Oe and Be stars remains a topic of debate, although several studies suggest that close binary interactions are an important channel for their formation (e.g., \citealt{BeBinary, Bebinary2014, PMBeStars}). These three Be stars in binary systems are therefore particularly compelling targets for investigating this formation channel. Two of them---\textit{527155253604491392} and \textit{2166378312964576256}---appear to be in wide binaries, possibly having acquired their Be characteristics through past binary interaction, as suggested by their location in Fig. \ref{fig:periods}. In contrast, \textit{2006088484204609408} may represent an actively interacting binary system, where the accreting star is currently evolving into a Be star.

\section{New sample of HG objects}

The analysis of our initial sample showed that uncertainties in the extinction were the main cause of including contaminants in our selection process. Given the appearance of new catalogues, we redid the same selection steps described in Sect. \ref{section:sample_selection}, but this time on the \texttt{SHBoost} input catalogue instead of {\tt StarHorse}. In this case, we only looked at H$\alpha$ EW values for the selection of hydrogen emitters, and we removed false positives by visual inspection of the calibrated XP spectra. We did not remove any further sources that might have a classification in \texttt{SIMBAD} or other catalogues. The other selection steps were kept the same. 

The filtering process provided a new preliminary sample composed of 444 sources. After visually inspecting the XP spectra of the preliminary sample, we finished with a new sample composed of 308 HG stars that fit our criteria. We will explore this new sample in future research. We present this new sample in a table made available in its entirety via CDS/VizieR.

\section{Summary and conclusions}

In this work, we aimed to identify Galactic mass-transferring binaries with quickly evolving donors in the HG with yellow spectral types that showed Balmer emission, mid-IR excess, and variability. After our initial selection, we obtained a sample of 67 candidates that we analysed using photometric data and spectroscopic data, along with multi-wavelength photometric archival data. We present a summary of the nature of our candidates (identified by their \textit{Gaia} DR3 IDs):

\begin{itemize}
    \item[$\bullet$] We computed periods for 36 sources, among which 26 showed sinusoidal variability, nine were eclipsing binaries, and one showed pulsations.
    \item[$\bullet$] The spectral types we identified from our follow-up spectra and the spectral types given by \textit{Gaia} were in agreement in 18 out of 21 cases. Combining our spectral types and the ones reported by \textit{Gaia}, the distribution of spectral classes in our sample was: 11 O-type, 27 B-type, 5 F-type, 6 G-type, 11 K-type, and 1 M-type star. 
    \item[$\bullet$] Among the O- and B-type stars in our sample, 15 are strong candidates for Oe or Be stars based on the presence of Balmer emission. These reddened Oe and Be stars contributed the most to the contaminants in our sample.
    \item[$\bullet$] Regular variables cannot be discarded as possible contaminants, and we identified two in our sample: a $\delta$ Scuti (\textit{460686648965862528}), and a Cepheid (\textit{4076568861833452160}). 
    \item[$\bullet$] Three sources in our sample showed associated hard X-ray emission. Two of them had good matches with their optical counterpart (sources \textit{4054010697162430592} and \textit{5866345647515558400}), and are strong candidates for X-ray binaries. Source \textit{2083649030845658624} had other optical counterparts within the positional errors of the X-ray detection.
    \item[$\bullet$] Source \textit{473575777103322496} shows far UV excess. This source was identified as a Be star, and we suggest it is in a binary system with a hot companion (subdwarf OB star or a white dwarf, for example), and it is potentially mass-transferring.
    \item[$\bullet$] The strongest candidates for mass-transferring binaries with a bright yellow primary are \textit{2060841448854265216}, a G-type star whose light curve resembles that of an Algol-type eclipsing binary with an accretion disk; \textit{6228685649971375616}, an F-type star in a close binary, and \textit{6123873398383875456}, an F-type contact binary that was previously misclassified as an RRc Lyrae star.
    \item[$\bullet$] Among all the O- and B-type stars in our sample, we identify seven in binary systems, including three Be stars: \textit{527155253604491392}, \textit{2006088484204609408}, and \textit{2166378312964576256}. These three Be stars in binary systems are particularly compelling targets for investigating the formation channel of Oe and Be stars through close binary interactions.
\end{itemize}

Most of the sources in our analysed sample are not yellow stars. Of the initial 67 candidates selected using the EDR3 \texttt{StarHorse} catalogue, only 25 remain within our HG parameter space when using the more reliable \texttt{SHBoost} extinction values. This underscores a key challenge in our original goal of identifying yellow HG stars as potential LRN progenitors: the large uncertainties in extinction estimates from \textit{Gaia} and \texttt{StarHorse}, which were central to defining our selection space. 

Comparing the observations of the two best candidates for stellar-merger progenitors in our sample---\textit{6228685649971375616} and \textit{6123873398383875456}---with those of the progenitor of V1309 Sco, the best-studied case of a stellar merger to date, suggests that if these systems are indeed potential merger progenitors, they are still far from the merger phase. Nevertheless, both remain interesting targets that deserve further investigation.

As an additional outcome of this work, we present a refined selection of 308 HG candidate stars using the improved extinction correction from the \texttt{SHBoost} catalogue. This new sample was derived using the same overall methodology but incorporates improved extinction estimates and visual inspection of the \textit{Gaia} XP spectra, ensuring that each selected object displays, at a minimum, clear H$\alpha$ emission. From this revised sample, we expect to identify a significantly larger fraction of scientifically valuable mass-transferring binary systems.

This new sample gives new opportunities to explore HG stars with interesting features. We encourage other teams to pursue follow-up observations of promising sources in this newly identified sample, and also of interesting sources in our initial sample. Additionally, forthcoming data releases, particularly \textit{Gaia} DR4, expected by the end of 2026, will substantially enhance the search for HG mass-transferring binaries by providing improved spectrophotometric and astrometric data, including BP/RP spectra for all sources down to $G\approx 19$\,mag, RVS spectra for sources brighter than $G \approx14.5$\,mag, and epoch photometry. These additions will offer a richer dataset for identifying and characterising evolved binaries and their emission features with greater precision.



\begin{acknowledgements} 
    G. G. M. would like to thank H. Tranin for assistance with the X-ray data, F. Stoppa for support with the use of \texttt{FINKER}, G. Iorio and S. Malhotra for valuable discussions, and D. Jones, V. Vuolteenaho, T. Pursimo, and M. Gort for their help during the observations and data reduction.

    G. G. M. and N. B. acknowledge to be funded by the European Union (ERC, CET-3PO, 101042610). Views and opinions expressed are however those of the author(s) only and do not necessarily reflect those of the European Union or the European Research Council Executive Agency. Neither the European Union nor the granting authority can be held responsible for them.

    F. A. acknowledges financial support from MCIN/AEI/10.13039/501100011033 through a RYC2021-031638-I grant co-funded by the European Union NextGenerationEU/PRTR.

    The authors acknowledge financial support from grant CEX2024-001451-M funded by MICIU/AEI/10.13039/501100011033. 

    This work has made use of data from the European Space Agency (ESA) mission {\it Gaia} \url{https://www.cosmos.esa.int/gaia}), processed by the {\it Gaia} Data Processing and Analysis Consortium (DPAC, \url{https://www.cosmos.esa.int/web/gaia/dpac/consortium}). Funding for the DPAC has been provided by national institutions, in particular the institutions participating in the {\it Gaia} Multilateral Agreement.

    This publication makes use of data products from the Wide-field Infrared Survey Explorer, which is a joint project of the University of California, Los Angeles, and the Jet Propulsion Laboratory/California Institute of Technology, and NEOWISE, which is a project of the Jet Propulsion Laboratory/California Institute of Technology. WISE and NEOWISE are funded by the National Aeronautics and Space Administration.

    Based on observations obtained with the Samuel Oschin Telescope 48-inch and the 60-inch Telescope at the Palomar Observatory as part of the Zwicky Transient Facility project. ZTF is supported by the National Science Foundation under Grants No. AST-1440341 and AST-2034437 and a collaboration including current partners Caltech, IPAC, the Oskar Klein Center at Stockholm University, the University of Maryland, University of California, Berkeley, the University of Wisconsin at Milwaukee, University of Warwick, Ruhr University, Cornell University, Northwestern University and Drexel University. Operations are conducted by COO, IPAC, and UW.

    This work has made use of data from the Asteroid Terrestrial-impact Last Alert System (ATLAS) project. The Asteroid Terrestrial-impact Last Alert System (ATLAS) project is primarily funded to search for near earth asteroids through NASA grants NN12AR55G, 80NSSC18K0284, and 80NSSC18K1575; byproducts of the NEO search include images and catalogs from the survey area. This work was partially funded by Kepler/K2 grant J1944/80NSSC19K0112 and HST GO-15889, and STFC grants ST/T000198/1 and ST/S006109/1. The ATLAS science products have been made possible through the contributions of the University of Hawaii Institute for Astronomy, the Queen’s University Belfast, the Space Telescope Science Institute, the South African Astronomical Observatory, and The Millennium Institute of Astrophysics (MAS), Chile.

    This paper includes data collected by the TESS mission. Funding for the TESS mission is provided by the NASA's Science Mission Directorate.

    Based on observations collected at Centro Astronómico Hispano en Andalucía (CAHA) at Calar Alto, operated jointly by Junta de Andalucía and Consejo Superior de Investigaciones Científicas (IAA-CSIC).

    Based on observations made with the Nordic Optical Telescope, owned in collaboration by the University of Turku and Aarhus University, and operated jointly by Aarhus University, the University of Turku and the University of Oslo, representing Denmark, Finland and Norway, the University of Iceland and Stockholm University at the Observatorio del Roque de los Muchachos, La Palma, Spain, of the Instituto de Astrofisica de Canarias. The NOT data were obtained under program IDs P69-854, P70-203, and P71-209.

    The data presented here were obtained [in part] with ALFOSC, which is provided by the Instituto de Astrofisica de Andalucia (IAA) under a joint agreement with the University of Copenhagen and NOT.

    This article is based on observations made in the Observatorios de Canarias del IAC with the Mercator telescope operated on the island of La Palma by KU Leuven in the Observatorio del Roque de los Muchachos.

    This research is based on observations made with the Galaxy Evolution Explorer, obtained from the MAST data archive at the Space Telescope Science Institute, which is operated by the Association of Universities for Research in Astronomy, Inc., under NASA contract NAS 5–26555.

    This research is based on observations made with the Neil Gehrels Swift Observatory, obtained from the MAST data archive at the Space Telescope Science Institute, which is operated by the Association of Universities for Research in Astronomy, Inc., under NASA contract NAS 5–26555.
    
    This research has made use of data obtained from the 4XMM XMM-Newton serendipitous source catalogue compiled by the XMM-Newton Survey Science Centre consortium.

    This research has made use of the VizieR catalogue access tool, CDS, Strasbourg, France (DOI: 10.26093/cds/vizier).
    
    This work made extensive use of \texttt{Python}, specially the packages \texttt{NumPy} \citep{numpy}, \texttt{pandas} \citep{pandas},  \texttt{astropy} \citep{astropy}, \texttt{matplotlib} \citep{matplotlib}, and \texttt{astroquery} \citep{astroquery}.
\end{acknowledgements}

%
%

\bibliographystyle{aa} 
\bibliography{aanda} 

@ARTICLE{Huson2025,
       author = {{Huson}, Dylan and {Cowan}, Indiana and {Sizemore}, Logan and {Kounkel}, Marina and {Hutchinson}, Brian},
        title = "{Gaia Net: Toward Robust Spectroscopic Parameters of Stars of all Evolutionary Stages}",
      journal = {\apj},
         year = 2025,
        month = may,
       volume = {984},
       number = {1},
          eid = {58},
        pages = {58},
          doi = {10.3847/1538-4357/adc2fa},
archivePrefix = {arXiv},
       eprint = {2503.02958},
 primaryClass = {astro-ph.SR},
       adsurl = {https://ui.adsabs.harvard.edu/abs/2025ApJ...984...58H}
}

@ARTICLE{Zhang2023,
       author = {{Zhang}, Xiangyu and {Green}, Gregory M. and {Rix}, Hans-Walter},
        title = "{Parameters of 220 million stars from Gaia BP/RP spectra}",
      journal = {\mnras},
         year = 2023,
        month = sep,
       volume = {524},
       number = {2},
        pages = {1855-1884},
          doi = {10.1093/mnras/stad1941},
archivePrefix = {arXiv},
       eprint = {2303.03420},
 primaryClass = {astro-ph.SR},
       adsurl = {https://ui.adsabs.harvard.edu/abs/2023MNRAS.524.1855Z}
}

@ARTICLE{Fallows2024,
       author = {{Fallows}, Connor P. and {Sanders}, Jason L.},
        title = "{Stellar atmospheric parameters from Gaia BP/RP spectra using uncertain neural networks}",
      journal = {\mnras},
         year = 2024,
        month = jun,
       volume = {531},
       number = {1},
        pages = {2126-2147},
          doi = {10.1093/mnras/stae1303},
archivePrefix = {arXiv},
       eprint = {2405.10699},
 primaryClass = {astro-ph.SR},
       adsurl = {https://ui.adsabs.harvard.edu/abs/2024MNRAS.531.2126F}
}

@ARTICLE{Queiroz2023,
       author = {{Queiroz}, A.~B.~A. and {Anders}, F. and {Chiappini}, C. and {Khalatyan}, A. and {Santiago}, B.~X. and {Nepal}, S. and {Steinmetz}, M. and {Gallart}, C. and {Valentini}, M. and {Dal Ponte}, M. and {Barbuy}, B. and {P{\'e}rez-Villegas}, A. and {Masseron}, T. and {Fern{\'a}ndez-Trincado}, J.~G. and {Khoperskov}, S. and {Minchev}, I. and {Fern{\'a}ndez-Alvar}, E. and {Lane}, R.~R. and {Nitschelm}, C.},
        title = "{StarHorse results for spectroscopic surveys and Gaia DR3: Chrono-chemical populations in the solar vicinity, the genuine thick disk, and young alpha-rich stars}",
      journal = {\aap},
         year = 2023,
        month = may,
       volume = {673},
          eid = {A155},
        pages = {A155},
          doi = {10.1051/0004-6361/202245399},
archivePrefix = {arXiv},
       eprint = {2303.09926},
 primaryClass = {astro-ph.GA},
       adsurl = {https://ui.adsabs.harvard.edu/abs/2023A&A...673A.155Q}
}

@ARTICLE{DeAngeli2023,
       author = {{De Angeli}, F. and {Weiler}, M. and {Montegriffo}, P. and {Evans}, D.~W. and {Riello}, M. and {Andrae}, R. and {Carrasco}, J.~M. and {Busso}, G. and {Burgess}, P.~W. and {Cacciari}, C. and {Davidson}, M. and {Harrison}, D.~L. and {Hodgkin}, S.~T. and {Jordi}, C. and {Osborne}, P.~J. and {Pancino}, E. and {Altavilla}, G. and {Barstow}, M.~A. and {Bailer-Jones}, C.~A.~L. and {Bellazzini}, M. and {Brown}, A.~G.~A. and {Castellani}, M. and {Cowell}, S. and {Delchambre}, L. and {De Luise}, F. and {Diener}, C. and {Fabricius}, C. and {Fouesneau}, M. and {Fr{\'e}mat}, Y. and {Gilmore}, G. and {Giuffrida}, G. and {Hambly}, N.~C. and {Hidalgo}, S. and {Holland}, G. and {Kostrzewa-Rutkowska}, Z. and {van Leeuwen}, F. and {Lobel}, A. and {Marinoni}, S. and {Miller}, N. and {Pagani}, C. and {Palaversa}, L. and {Piersimoni}, A.~M. and {Pulone}, L. and {Ragaini}, S. and {Rainer}, M. and {Richards}, P.~J. and {Rixon}, G.~T. and {Ruz-Mieres}, D. and {Sanna}, N. and {Sarro}, L.~M. and {Rowell}, N. and {Sordo}, R. and {Walton}, N.~A. and {Yoldas}, A.},
        title = "{Gaia Data Release 3. Processing and validation of BP/RP low-resolution spectral data}",
      journal = {\aap},
         year = 2023,
        month = jun,
       volume = {674},
          eid = {A2},
        pages = {A2},
          doi = {10.1051/0004-6361/202243680},
archivePrefix = {arXiv},
       eprint = {2206.06143},
 primaryClass = {astro-ph.IM},
       adsurl = {https://ui.adsabs.harvard.edu/abs/2023A&A...674A...2D}
}

@ARTICLE{Carrasco2021,
       author = {{Carrasco}, J.~M. and {Weiler}, M. and {Jordi}, C. and {Fabricius}, C. and {De Angeli}, F. and {Evans}, D.~W. and {van Leeuwen}, F. and {Riello}, M. and {Montegriffo}, P.},
        title = "{Internal calibration of Gaia BP/RP low-resolution spectra}",
      journal = {\aap},
         year = 2021,
        month = aug,
       volume = {652},
          eid = {A86},
        pages = {A86},
          doi = {10.1051/0004-6361/202141249},
archivePrefix = {arXiv},
       eprint = {2106.01752},
 primaryClass = {astro-ph.IM},
       adsurl = {https://ui.adsabs.harvard.edu/abs/2021A&A...652A..86C}
}

@ARTICLE{thick,
       author = {{Mourard}, D. and {Bro{\v{z}}}, M. and {Nemravov{\'a}}, J.~A. and {Harmanec}, P. and {Budaj}, J. and {Baron}, F. and {Monnier}, J.~D. and {Schaefer}, G.~H. and {Schmitt}, H. and {Tallon-Bosc}, I. and {Armstrong}, J.~T. and {Baines}, E.~K. and {Bonneau}, D. and {Bo{\v{z}}i{\'c}}, H. and {Clausse}, J.~M. and {Farrington}, C. and {Gies}, D. and {Jury{\v{s}}ek}, J. and {Kor{\v{c}}{\'a}kov{\'a}}, D. and {McAlister}, H. and {Meilland}, A. and {Nardetto}, N. and {Svoboda}, P. and {{\v{S}}lechta}, M. and {Wolf}, M. and {Zasche}, P.},
        title = "{Physical properties of {\ensuremath{\beta}} Lyrae A and its opaque accretion disk}",
      journal = {\aap},
         year = 2018,
        month = oct,
       volume = {618},
          eid = {A112},
        pages = {A112},
          doi = {10.1051/0004-6361/201832952},
archivePrefix = {arXiv},
       eprint = {1807.04789},
 primaryClass = {astro-ph.SR},
       adsurl = {https://ui.adsabs.harvard.edu/abs/2018A&A...618A.112M}
}

@ARTICLE{thin,
       author = {{Bro{\v{z}}}, M. and {Mourard}, D. and {Budaj}, J. and {Harmanec}, P. and {Schmitt}, H. and {Tallon-Bosc}, I. and {Bonneau}, D. and {Bo{\v{z}}i{\'c}}, H. and {Gies}, D. and {{\v{S}}lechta}, M.},
        title = "{Optically thin circumstellar medium in the {\ensuremath{\beta}} Lyr A system}",
      journal = {\aap},
         year = 2021,
        month = jan,
       volume = {645},
          eid = {A51},
        pages = {A51},
          doi = {10.1051/0004-6361/202039035},
archivePrefix = {arXiv},
       eprint = {2010.05541},
 primaryClass = {astro-ph.SR},
       adsurl = {https://ui.adsabs.harvard.edu/abs/2021A&A...645A..51B}
}

@ARTICLE{V1309ScoMerger,
       author = {{Tylenda}, R. and {Hajduk}, M. and {Kami{\'n}ski}, T. and {Udalski}, A. and {Soszy{\'n}ski}, I. and {Szyma{\'n}ski}, M.~K. and {Kubiak}, M. and {Pietrzy{\'n}ski}, G. and {Poleski}, R. and {Wyrzykowski}, {\L}. and {Ulaczyk}, K.},
        title = "{V1309 Scorpii: merger of a contact binary}",
      journal = {\aap},
         year = 2011,
        month = apr,
       volume = {528},
          eid = {A114},
        pages = {A114},
          doi = {10.1051/0004-6361/201016221},
archivePrefix = {arXiv},
       eprint = {1012.0163},
 primaryClass = {astro-ph.SR},
       adsurl = {https://ui.adsabs.harvard.edu/abs/2011A&A...528A.114T}
}

@ARTICLE{OGLE-2002,
       author = {{Tylenda}, R. and {Kami{\'n}ski}, T. and {Udalski}, A. and {Soszy{\'n}ski}, I. and {Poleski}, R. and {Szyma{\'n}ski}, M.~K. and {Kubiak}, M. and {Pietrzy{\'n}ski}, G. and {Koz{\l}owski}, S. and {Pietrukowicz}, P. and {Ulaczyk}, K. and {Wyrzykowski}, {\L}.},
        title = "{OGLE-2002-BLG-360: from a gravitational microlensing candidate to an overlooked red transient}",
      journal = {\aap},
         year = 2013,
        month = jul,
       volume = {555},
          eid = {A16},
        pages = {A16},
          doi = {10.1051/0004-6361/201321647},
archivePrefix = {arXiv},
       eprint = {1304.1694},
 primaryClass = {astro-ph.SR},
       adsurl = {https://ui.adsabs.harvard.edu/abs/2013A&A...555A..16T}
}

@ARTICLE{CKVul,
       author = {{Kato}, T.},
        title = "{CK Vul as a candidate eruptive stellar merging event}",
      journal = {\aap},
         year = 2003,
        month = feb,
       volume = {399},
        pages = {695-697},
          doi = {10.1051/0004-6361:20021808},
archivePrefix = {arXiv},
       eprint = {astro-ph/0211557},
 primaryClass = {astro-ph},
       adsurl = {https://ui.adsabs.harvard.edu/abs/2003A&A...399..695K}
}

@ARTICLE{Harry,
       author = {{Addison}, Harry and {Blagorodnova}, Nadejda and {Groot}, Paul J. and {Erasmus}, Nicolas and {Jones}, David and {Mogawana}, Orapeleng},
        title = "{Searching for the next Galactic Luminous red nova}",
      journal = {\mnras},
         year = 2022,
        month = dec,
       volume = {517},
       number = {2},
        pages = {1884-1900},
          doi = {10.1093/mnras/stac2685},
archivePrefix = {arXiv},
       eprint = {2206.07070},
 primaryClass = {astro-ph.SR},
       adsurl = {https://ui.adsabs.harvard.edu/abs/2022MNRAS.517.1884A}
}

@ARTICLE{Starhorse,
       author = {{Anders}, F. and {Khalatyan}, A. and {Queiroz}, A.~B.~A. and {Chiappini}, C. and {Ard{\`e}vol}, J. and {Casamiquela}, L. and {Figueras}, F. and {Jim{\'e}nez-Arranz}, {\'O}. and {Jordi}, C. and {Mongui{\'o}}, M. and {Romero-G{\'o}mez}, M. and {Altamirano}, D. and {Antoja}, T. and {Assaad}, R. and {Cantat-Gaudin}, T. and {Castro-Ginard}, A. and {Enke}, H. and {Girardi}, L. and {Guiglion}, G. and {Khan}, S. and {Luri}, X. and {Miglio}, A. and {Minchev}, I. and {Ramos}, P. and {Santiago}, B.~X. and {Steinmetz}, M.},
        title = "{Photo-astrometric distances, extinctions, and astrophysical parameters for Gaia EDR3 stars brighter than G = 18.5}",
      journal = {\aap},
         year = 2022,
        month = feb,
       volume = {658},
          eid = {A91},
        pages = {A91},
          doi = {10.1051/0004-6361/202142369},
archivePrefix = {arXiv},
       eprint = {2111.01860},
 primaryClass = {astro-ph.GA},
       adsurl = {https://ui.adsabs.harvard.edu/abs/2022A&A...658A..91A}
}

@ARTICLE{GaiaDR3,
       author = {{Gaia Collaboration} and {Vallenari}, A. and {Brown}, A.~G.~A. and {Prusti}, T. and {de Bruijne}, J.~H.~J. and {Arenou}, F. and {Babusiaux}, C. and {Biermann}, M. and {Creevey}, O.~L. and {Ducourant}, C. and {Evans}, D.~W. and {Eyer}, L. and {Guerra}, R. and {Hutton}, A. and {Jordi}, C. and {Klioner}, S.~A. and {Lammers}, U.~L. and {Lindegren}, L. and {Luri}, X. and {Mignard}, F. and {Panem}, C. and {Pourbaix}, D. and {Randich}, S. and {Sartoretti}, P. and {Soubiran}, C. and {Tanga}, P. and {Walton}, N.~A. and {Bailer-Jones}, C.~A.~L. and {Bastian}, U. and {Drimmel}, R. and {Jansen}, F. and {Katz}, D. and {Lattanzi}, M.~G. and {van Leeuwen}, F. and {Bakker}, J. and {Cacciari}, C. and {Casta{\~n}eda}, J. and {De Angeli}, F. and {Fabricius}, C. and {Fouesneau}, M. and {Fr{\'e}mat}, Y. and {Galluccio}, L. and {Guerrier}, A. and {Heiter}, U. and {Masana}, E. and {Messineo}, R. and {Mowlavi}, N. and {Nicolas}, C. and {Nienartowicz}, K. and {Pailler}, F. and {Panuzzo}, P. and {Riclet}, F. and {Roux}, W. and {Seabroke}, G.~M. and {Sordo}, R. and {Th{\'e}venin}, F. and {Gracia-Abril}, G. and {Portell}, J. and {Teyssier}, D. and {Altmann}, M. and {Andrae}, R. and {Audard}, M. and {Bellas-Velidis}, I. and {Benson}, K. and {Berthier}, J. and {Blomme}, R. and {Burgess}, P.~W. and {Busonero}, D. and {Busso}, G. and {C{\'a}novas}, H. and {Carry}, B. and {Cellino}, A. and {Cheek}, N. and {Clementini}, G. and {Damerdji}, Y. and {Davidson}, M. and {de Teodoro}, P. and {Nu{\~n}ez Campos}, M. and {Delchambre}, L. and {Dell'Oro}, A. and {Esquej}, P. and {Fern{\'a}ndez-Hern{\'a}ndez}, J. and {Fraile}, E. and {Garabato}, D. and {Garc{\'\i}a-Lario}, P. and {Gosset}, E. and {Haigron}, R. and {Halbwachs}, J. -L. and {Hambly}, N.~C. and {Harrison}, D.~L. and {Hern{\'a}ndez}, J. and {Hestroffer}, D. and {Hodgkin}, S.~T. and {Holl}, B. and {Jan{\ss}en}, K. and {Jevardat de Fombelle}, G. and {Jordan}, S. and {Krone-Martins}, A. and {Lanzafame}, A.~C. and {L{\"o}ffler}, W. and {Marchal}, O. and {Marrese}, P.~M. and {Moitinho}, A. and {Muinonen}, K. and {Osborne}, P. and {Pancino}, E. and {Pauwels}, T. and {Recio-Blanco}, A. and {Reyl{\'e}}, C. and {Riello}, M. and {Rimoldini}, L. and {Roegiers}, T. and {Rybizki}, J. and {Sarro}, L.~M. and {Siopis}, C. and {Smith}, M. and {Sozzetti}, A. and {Utrilla}, E. and {van Leeuwen}, M. and {Abbas}, U. and {{\'A}brah{\'a}m}, P. and {Abreu Aramburu}, A. and {Aerts}, C. and {Aguado}, J.~J. and {Ajaj}, M. and {Aldea-Montero}, F. and {Altavilla}, G. and {{\'A}lvarez}, M.~A. and {Alves}, J. and {Anders}, F. and {Anderson}, R.~I. and {Anglada Varela}, E. and {Antoja}, T. and {Baines}, D. and {Baker}, S.~G. and {Balaguer-N{\'u}{\~n}ez}, L. and {Balbinot}, E. and {Balog}, Z. and {Barache}, C. and {Barbato}, D. and {Barros}, M. and {Barstow}, M.~A. and {Bartolom{\'e}}, S. and {Bassilana}, J. -L. and {Bauchet}, N. and {Becciani}, U. and {Bellazzini}, M. and {Berihuete}, A. and {Bernet}, M. and {Bertone}, S. and {Bianchi}, L. and {Binnenfeld}, A. and {Blanco-Cuaresma}, S. and {Blazere}, A. and {Boch}, T. and {Bombrun}, A. and {Bossini}, D. and {Bouquillon}, S. and {Bragaglia}, A. and {Bramante}, L. and {Breedt}, E. and {Bressan}, A. and {Brouillet}, N. and {Brugaletta}, E. and {Bucciarelli}, B. and {Burlacu}, A. and {Butkevich}, A.~G. and {Buzzi}, R. and {Caffau}, E. and {Cancelliere}, R. and {Cantat-Gaudin}, T. and {Carballo}, R. and {Carlucci}, T. and {Carnerero}, M.~I. and {Carrasco}, J.~M. and {Casamiquela}, L. and {Castellani}, M. and {Castro-Ginard}, A. and {Chaoul}, L. and {Charlot}, P. and {Chemin}, L. and {Chiaramida}, V. and {Chiavassa}, A. and {Chornay}, N. and {Comoretto}, G. and {Contursi}, G. and {Cooper}, W.~J. and {Cornez}, T. and {Cowell}, S. and {Crifo}, F. and {Cropper}, M. and {Crosta}, M. and {Crowley}, C. and {Dafonte}, C. and {Dapergolas}, A. and {David}, M. and {David}, P. and {de Laverny}, P. and {De Luise}, F. and {De March}, R.},
        title = "{Gaia Data Release 3. Summary of the content and survey properties}",
      journal = {\aap},
         year = 2023,
        month = jun,
       volume = {674},
          eid = {A1},
        pages = {A1},
          doi = {10.1051/0004-6361/202243940},
archivePrefix = {arXiv},
       eprint = {2208.00211},
 primaryClass = {astro-ph.GA},
       adsurl = {https://ui.adsabs.harvard.edu/abs/2023A&A...674A...1G}
}

@ARTICLE{Weiler,
       author = {{Weiler}, M. and {Carrasco}, J.~M. and {Fabricius}, C. and {Jordi}, C.},
        title = "{Analysing spectral lines in Gaia low-resolution spectra}",
      journal = {\aap},
         year = 2023,
        month = mar,
       volume = {671},
          eid = {A52},
        pages = {A52},
          doi = {10.1051/0004-6361/202244764},
archivePrefix = {arXiv},
       eprint = {2211.06946},
 primaryClass = {astro-ph.IM},
       adsurl = {https://ui.adsabs.harvard.edu/abs/2023A&A...671A..52W}
}

@ARTICLE{deltaScuZTF,
       author = {{Chen}, Xiaodian and {Wang}, Shu and {Deng}, Licai and {de Grijs}, Richard and {Yang}, Ming and {Tian}, Hao},
        title = "{The Zwicky Transient Facility Catalog of Periodic Variable Stars}",
      journal = {\apjs},
         year = 2020,
        month = jul,
       volume = {249},
       number = {1},
          eid = {18},
        pages = {18},
          doi = {10.3847/1538-4365/ab9cae},
archivePrefix = {arXiv},
       eprint = {2005.08662},
 primaryClass = {astro-ph.SR},
       adsurl = {https://ui.adsabs.harvard.edu/abs/2020ApJS..249...18C}
}

@ARTICLE{ztf,
       author = {{Masci}, Frank J. and {Laher}, Russ R. and {Rusholme}, Ben and {Shupe}, David L. and {Groom}, Steven and {Surace}, Jason and {Jackson}, Edward and {Monkewitz}, Serge and {Beck}, Ron and {Flynn}, David and {Terek}, Scott and {Landry}, Walter and {Hacopians}, Eugean and {Desai}, Vandana and {Howell}, Justin and {Brooke}, Tim and {Imel}, David and {Wachter}, Stefanie and {Ye}, Quan-Zhi and {Lin}, Hsing-Wen and {Cenko}, S. Bradley and {Cunningham}, Virginia and {Rebbapragada}, Umaa and {Bue}, Brian and {Miller}, Adam A. and {Mahabal}, Ashish and {Bellm}, Eric C. and {Patterson}, Maria T. and {Juri{\'c}}, Mario and {Golkhou}, V. Zach and {Ofek}, Eran O. and {Walters}, Richard and {Graham}, Matthew and {Kasliwal}, Mansi M. and {Dekany}, Richard G. and {Kupfer}, Thomas and {Burdge}, Kevin and {Cannella}, Christopher B. and {Barlow}, Tom and {Van Sistine}, Angela and {Giomi}, Matteo and {Fremling}, Christoffer and {Blagorodnova}, Nadejda and {Levitan}, David and {Riddle}, Reed and {Smith}, Roger M. and {Helou}, George and {Prince}, Thomas A. and {Kulkarni}, Shrinivas R.},
        title = "{The Zwicky Transient Facility: Data Processing, Products, and Archive}",
      journal = {\pasp},
         year = 2019,
        month = jan,
       volume = {131},
       number = {995},
        pages = {018003},
          doi = {10.1088/1538-3873/aae8ac},
archivePrefix = {arXiv},
       eprint = {1902.01872},
 primaryClass = {astro-ph.IM},
       adsurl = {https://ui.adsabs.harvard.edu/abs/2019PASP..131a8003M}
}

@ARTICLE{finker,
       author = {{Stoppa}, F. and {Johnston}, C. and {Cator}, E. and {Nelemans}, G. and {Groot}, P.~J.},
        title = "{FINKER: Frequency Identification through Nonparametric KErnel Regression in astronomical time series}",
      journal = {\aap},
         year = 2024,
        month = jun,
       volume = {686},
          eid = {A158},
        pages = {A158},
          doi = {10.1051/0004-6361/202348848},
archivePrefix = {arXiv},
       eprint = {2312.05408},
 primaryClass = {astro-ph.IM},
       adsurl = {https://ui.adsabs.harvard.edu/abs/2024A&A...686A.158S}
}

@ARTICLE{ASAS-SN,
       author = {{Hart}, K. and {Shappee}, B.~J. and {Hey}, D. and {Kochanek}, C.~S. and {Stanek}, K.~Z. and {Lim}, L. and {Dobbs}, S. and {Tucker}, M. and {Jayasinghe}, T. and {Beacom}, J.~F. and {Boright}, T. and {Holoien}, T. and {Ong}, J.~M. Joel and {Prieto}, J.~L. and {Thompson}, T.~A. and {Will}, D.},
        title = "{ASAS-SN Sky Patrol V2.0}",
      journal = {arXiv e-prints},
         year = 2023,
        month = apr,
          eid = {arXiv:2304.03791},
        pages = {arXiv:2304.03791},
          doi = {10.48550/arXiv.2304.03791},
archivePrefix = {arXiv},
       eprint = {2304.03791},
 primaryClass = {astro-ph.IM},
       adsurl = {https://ui.adsabs.harvard.edu/abs/2023arXiv230403791H}
}

@ARTICLE{ATLASVar,
       author = {{Heinze}, A.~N. and {Tonry}, J.~L. and {Denneau}, L. and {Flewelling}, H. and {Stalder}, B. and {Rest}, A. and {Smith}, K.~W. and {Smartt}, S.~J. and {Weiland}, H.},
        title = "{A First Catalog of Variable Stars Measured by the Asteroid Terrestrial-impact Last Alert System (ATLAS)}",
      journal = {\aj},
         year = 2018,
        month = nov,
       volume = {156},
       number = {5},
          eid = {241},
        pages = {241},
          doi = {10.3847/1538-3881/aae47f},
archivePrefix = {arXiv},
       eprint = {1804.02132},
 primaryClass = {astro-ph.SR},
       adsurl = {https://ui.adsabs.harvard.edu/abs/2018AJ....156..241H}
}

@ARTICLE{ATLAS,
       author = {{Tonry}, J.~L. and {Denneau}, L. and {Heinze}, A.~N. and {Stalder}, B. and {Smith}, K.~W. and {Smartt}, S.~J. and {Stubbs}, C.~W. and {Weiland}, H.~J. and {Rest}, A.},
        title = "{ATLAS: A High-cadence All-sky Survey System}",
      journal = {\pasp},
         year = 2018,
        month = jun,
       volume = {130},
       number = {988},
        pages = {064505},
          doi = {10.1088/1538-3873/aabadf},
archivePrefix = {arXiv},
       eprint = {1802.00879},
 primaryClass = {astro-ph.IM},
       adsurl = {https://ui.adsabs.harvard.edu/abs/2018PASP..130f4505T}
}

@ARTICLE{allwise,
       author = {{Cutri}, R.~M. and {Wright}, E.~L. and {Conrow}, T. and {Fowler}, J.~W. and {Eisenhardt}, P.~R.~M. and {Grillmair}, C. and {Kirkpatrick}, J.~D. and {Masci}, F. and {McCallon}, H.~L. and {Wheelock}, S.~L. and {Fajardo-Acosta}, S. and {Yan}, L. and {Benford}, D. and {Harbut}, M. and {Jarrett}, T. and {Lake}, S. and {Leisawitz}, D. and {Ressler}, M.~E. and {Stanford}, S.~A. and {Tsai}, C. -W. and {Liu}, F. and {Helou}, G. and {Mainzer}, A. and {Gettngs}, D. and {Gonzalez}, A. and {Hoffman}, D. and {Marsh}, K.~A. and {Padgett}, D. and {Skrutskie}, M.~F. and {Beck}, R. and {Papin}, M. and {Wittman}, M.},
        title = "{VizieR Online Data Catalog: AllWISE Data Release (Cutri+ 2013)}",
      journal = {VizieR Online Data Catalog},
         year = 2021,
        month = feb,
          eid = {II/328},
        pages = {II/328},
       adsurl = {https://ui.adsabs.harvard.edu/abs/2014yCat.2328....0C}
}

@ARTICLE{NEOWISE,
       author = {{Mainzer}, A. and {Bauer}, J. and {Grav}, T. and {Masiero}, J. and {Cutri}, R.~M. and {Dailey}, J. and {Eisenhardt}, P. and {McMillan}, R.~S. and {Wright}, E. and {Walker}, R. and {Jedicke}, R. and {Spahr}, T. and {Tholen}, D. and {Alles}, R. and {Beck}, R. and {Brandenburg}, H. and {Conrow}, T. and {Evans}, T. and {Fowler}, J. and {Jarrett}, T. and {Marsh}, K. and {Masci}, F. and {McCallon}, H. and {Wheelock}, S. and {Wittman}, M. and {Wyatt}, P. and {DeBaun}, E. and {Elliott}, G. and {Elsbury}, D. and {Gautier}, T., IV and {Gomillion}, S. and {Leisawitz}, D. and {Maleszewski}, C. and {Micheli}, M. and {Wilkins}, A.},
        title = "{Preliminary Results from NEOWISE: An Enhancement to the Wide-field Infrared Survey Explorer for Solar System Science}",
      journal = {\apj},
         year = 2011,
        month = apr,
       volume = {731},
       number = {1},
          eid = {53},
        pages = {53},
          doi = {10.1088/0004-637X/731/1/53},
archivePrefix = {arXiv},
       eprint = {1102.1996},
 primaryClass = {astro-ph.EP},
       adsurl = {https://ui.adsabs.harvard.edu/abs/2011ApJ...731...53M}
}

@ARTICLE{TESS,
       author = {{Ricker}, George R. and {Winn}, Joshua N. and {Vanderspek}, Roland and {Latham}, David W. and {Bakos}, G{\'a}sp{\'a}r {\'A}. and {Bean}, Jacob L. and {Berta-Thompson}, Zachory K. and {Brown}, Timothy M. and {Buchhave}, Lars and {Butler}, Nathaniel R. and {Butler}, R. Paul and {Chaplin}, William J. and {Charbonneau}, David and {Christensen-Dalsgaard}, J{\o}rgen and {Clampin}, Mark and {Deming}, Drake and {Doty}, John and {De Lee}, Nathan and {Dressing}, Courtney and {Dunham}, Edward W. and {Endl}, Michael and {Fressin}, Francois and {Ge}, Jian and {Henning}, Thomas and {Holman}, Matthew J. and {Howard}, Andrew W. and {Ida}, Shigeru and {Jenkins}, Jon M. and {Jernigan}, Garrett and {Johnson}, John Asher and {Kaltenegger}, Lisa and {Kawai}, Nobuyuki and {Kjeldsen}, Hans and {Laughlin}, Gregory and {Levine}, Alan M. and {Lin}, Douglas and {Lissauer}, Jack J. and {MacQueen}, Phillip and {Marcy}, Geoffrey and {McCullough}, Peter R. and {Morton}, Timothy D. and {Narita}, Norio and {Paegert}, Martin and {Palle}, Enric and {Pepe}, Francesco and {Pepper}, Joshua and {Quirrenbach}, Andreas and {Rinehart}, Stephen A. and {Sasselov}, Dimitar and {Sato}, Bun'ei and {Seager}, Sara and {Sozzetti}, Alessandro and {Stassun}, Keivan G. and {Sullivan}, Peter and {Szentgyorgyi}, Andrew and {Torres}, Guillermo and {Udry}, Stephane and {Villasenor}, Joel},
        title = "{Transiting Exoplanet Survey Satellite (TESS)}",
      journal = {Journal of Astronomical Telescopes, Instruments, and Systems},
         year = 2015,
        month = jan,
       volume = {1},
          eid = {014003},
        pages = {014003},
          doi = {10.1117/1.JATIS.1.1.014003},
       adsurl = {https://ui.adsabs.harvard.edu/abs/2015JATIS...1a4003R}
}

@ARTICLE{astropy,
       author = {{Astropy Collaboration} and {Price-Whelan}, Adrian M. and {Lim}, Pey Lian and {Earl}, Nicholas and {Starkman}, Nathaniel and {Bradley}, Larry and {Shupe}, David L. and {Patil}, Aarya A. and {Corrales}, Lia and {Brasseur}, C.~E. and {N{\"o}the}, Maximilian and {Donath}, Axel and {Tollerud}, Erik and {Morris}, Brett M. and {Ginsburg}, Adam and {Vaher}, Eero and {Weaver}, Benjamin A. and {Tocknell}, James and {Jamieson}, William and {van Kerkwijk}, Marten H. and {Robitaille}, Thomas P. and {Merry}, Bruce and {Bachetti}, Matteo and {G{\"u}nther}, H. Moritz and {Aldcroft}, Thomas L. and {Alvarado-Montes}, Jaime A. and {Archibald}, Anne M. and {B{\'o}di}, Attila and {Bapat}, Shreyas and {Barentsen}, Geert and {Baz{\'a}n}, Juanjo and {Biswas}, Manish and {Boquien}, M{\'e}d{\'e}ric and {Burke}, D.~J. and {Cara}, Daria and {Cara}, Mihai and {Conroy}, Kyle E. and {Conseil}, Simon and {Craig}, Matthew W. and {Cross}, Robert M. and {Cruz}, Kelle L. and {D'Eugenio}, Francesco and {Dencheva}, Nadia and {Devillepoix}, Hadrien A.~R. and {Dietrich}, J{\"o}rg P. and {Eigenbrot}, Arthur Davis and {Erben}, Thomas and {Ferreira}, Leonardo and {Foreman-Mackey}, Daniel and {Fox}, Ryan and {Freij}, Nabil and {Garg}, Suyog and {Geda}, Robel and {Glattly}, Lauren and {Gondhalekar}, Yash and {Gordon}, Karl D. and {Grant}, David and {Greenfield}, Perry and {Groener}, Austen M. and {Guest}, Steve and {Gurovich}, Sebastian and {Handberg}, Rasmus and {Hart}, Akeem and {Hatfield-Dodds}, Zac and {Homeier}, Derek and {Hosseinzadeh}, Griffin and {Jenness}, Tim and {Jones}, Craig K. and {Joseph}, Prajwel and {Kalmbach}, J. Bryce and {Karamehmetoglu}, Emir and {Ka{\l}uszy{\'n}ski}, Miko{\l}aj and {Kelley}, Michael S.~P. and {Kern}, Nicholas and {Kerzendorf}, Wolfgang E. and {Koch}, Eric W. and {Kulumani}, Shankar and {Lee}, Antony and {Ly}, Chun and {Ma}, Zhiyuan and {MacBride}, Conor and {Maljaars}, Jakob M. and {Muna}, Demitri and {Murphy}, N.~A. and {Norman}, Henrik and {O'Steen}, Richard and {Oman}, Kyle A. and {Pacifici}, Camilla and {Pascual}, Sergio and {Pascual-Granado}, J. and {Patil}, Rohit R. and {Perren}, Gabriel I. and {Pickering}, Timothy E. and {Rastogi}, Tanuj and {Roulston}, Benjamin R. and {Ryan}, Daniel F. and {Rykoff}, Eli S. and {Sabater}, Jose and {Sakurikar}, Parikshit and {Salgado}, Jes{\'u}s and {Sanghi}, Aniket and {Saunders}, Nicholas and {Savchenko}, Volodymyr and {Schwardt}, Ludwig and {Seifert-Eckert}, Michael and {Shih}, Albert Y. and {Jain}, Anany Shrey and {Shukla}, Gyanendra and {Sick}, Jonathan and {Simpson}, Chris and {Singanamalla}, Sudheesh and {Singer}, Leo P. and {Singhal}, Jaladh and {Sinha}, Manodeep and {Sip{\H{o}}cz}, Brigitta M. and {Spitler}, Lee R. and {Stansby}, David and {Streicher}, Ole and {{\v{S}}umak}, Jani and {Swinbank}, John D. and {Taranu}, Dan S. and {Tewary}, Nikita and {Tremblay}, Grant R. and {de Val-Borro}, Miguel and {Van Kooten}, Samuel J. and {Vasovi{\'c}}, Zlatan and {Verma}, Shresth and {de Miranda Cardoso}, Jos{\'e} Vin{\'\i}cius and {Williams}, Peter K.~G. and {Wilson}, Tom J. and {Winkel}, Benjamin and {Wood-Vasey}, W.~M. and {Xue}, Rui and {Yoachim}, Peter and {Zhang}, Chen and {Zonca}, Andrea and {Astropy Project Contributors}},
        title = "{The Astropy Project: Sustaining and Growing a Community-oriented Open-source Project and the Latest Major Release (v5.0) of the Core Package}",
      journal = {\apj},
         year = 2022,
        month = aug,
       volume = {935},
       number = {2},
          eid = {167},
        pages = {167},
          doi = {10.3847/1538-4357/ac7c74},
archivePrefix = {arXiv},
       eprint = {2206.14220},
 primaryClass = {astro-ph.IM},
       adsurl = {https://ui.adsabs.harvard.edu/abs/2022ApJ...935..167A}
}

@ARTICLE{caha1,
       author = {{S{\'a}nchez}, S.~F. and {Aceituno}, J. and {Thiele}, U. and {P{\'e}rez-Ram{\'\i}rez}, D. and {Alves}, J.},
        title = "{The Night Sky at the Calar Alto Observatory}",
      journal = {\pasp},
         year = 2007,
        month = oct,
       volume = {119},
       number = {860},
        pages = {1186-1200},
          doi = {10.1086/522378},
archivePrefix = {arXiv},
       eprint = {0709.0813},
 primaryClass = {astro-ph},
       adsurl = {https://ui.adsabs.harvard.edu/abs/2007PASP..119.1186S}
}

@ARTICLE{caha2,
       author = {{S{\'a}nchez}, S.~F. and {Thiele}, U. and {Aceituno}, J. and {Cristobal}, D. and {Perea}, J. and {Alves}, J.},
        title = "{The Night Sky at the Calar Alto Observatory II: The Sky at the Near-infrared}",
      journal = {\pasp},
         year = 2008,
        month = nov,
       volume = {120},
       number = {873},
        pages = {1244},
          doi = {10.1086/593981},
archivePrefix = {arXiv},
       eprint = {0809.4988},
 primaryClass = {astro-ph},
       adsurl = {https://ui.adsabs.harvard.edu/abs/2008PASP..120.1244S}
}

@ARTICLE{cafos,
       author = {{Meisenheimer}, K.},
        title = "{Cafos 2.2 - Der Fokalreduktor des 2.2-m-Teleskops auf dem Calar-Alto.}",
      journal = {Sterne und Weltraum},
         year = 1994,
        month = jul,
       volume = {33},
       number = {7},
        pages = {516-522},
       adsurl = {https://ui.adsabs.harvard.edu/abs/1994S&W....33..516M}
}

@ARTICLE{fies,
       author = {{Telting}, J.~H. and {Avila}, G. and {Buchhave}, L. and {Frandsen}, S. and {Gandolfi}, D. and {Lindberg}, B. and {Stempels}, H.~C. and {Prins}, S. and {NOT staff}},
        title = "{FIES: The high-resolution Fiber-fed Echelle Spectrograph at the Nordic Optical Telescope}",
      journal = {Astronomische Nachrichten},
         year = 2014,
        month = jan,
       volume = {335},
       number = {1},
        pages = {41},
          doi = {10.1002/asna.201312007},
       adsurl = {https://ui.adsabs.harvard.edu/abs/2014AN....335...41T}
}

@INPROCEEDINGS{NOT,
       author = {{Djupvik}, Anlaug Amanda and {Andersen}, Johannes},
        title = "{The Nordic Optical Telescope}",
    booktitle = {Highlights of Spanish Astrophysics V},
         year = 2010,
       series = {Astrophysics and Space Science Proceedings},
       volume = {14},
        month = jan,
        pages = {211},
          doi = {10.1007/978-3-642-11250-8_21},
archivePrefix = {arXiv},
       eprint = {0901.4015},
 primaryClass = {astro-ph.IM},
       adsurl = {https://ui.adsabs.harvard.edu/abs/2010ASSP...14..211D}
}

@ARTICLE{hermes,
       author = {{Raskin}, G. and {van Winckel}, H. and {Hensberge}, H. and {Jorissen}, A. and {Lehmann}, H. and {Waelkens}, C. and {Avila}, G. and {de Cuyper}, J. -P. and {Degroote}, P. and {Dubosson}, R. and {Dumortier}, L. and {Fr{\'e}mat}, Y. and {Laux}, U. and {Michaud}, B. and {Morren}, J. and {Perez Padilla}, J. and {Pessemier}, W. and {Prins}, S. and {Smolders}, K. and {van Eck}, S. and {Winkler}, J.},
        title = "{HERMES: a high-resolution fibre-fed spectrograph for the Mercator telescope}",
      journal = {\aap},
         year = 2011,
        month = feb,
       volume = {526},
          eid = {A69},
        pages = {A69},
          doi = {10.1051/0004-6361/201015435},
archivePrefix = {arXiv},
       eprint = {1011.0258},
 primaryClass = {astro-ph.IM},
       adsurl = {https://ui.adsabs.harvard.edu/abs/2011A&A...526A..69R}
}

@ARTICLE{lamost,
       author = {{Cui}, Xiang-Qun and {Zhao}, Yong-Heng and {Chu}, Yao-Quan and {Li}, Guo-Ping and {Li}, Qi and {Zhang}, Li-Ping and {Su}, Hong-Jun and {Yao}, Zheng-Qiu and {Wang}, Ya-Nan and {Xing}, Xiao-Zheng and {Li}, Xin-Nan and {Zhu}, Yong-Tian and {Wang}, Gang and {Gu}, Bo-Zhong and {Luo}, A. -Li and {Xu}, Xin-Qi and {Zhang}, Zhen-Chao and {Liu}, Gen-Rong and {Zhang}, Hao-Tong and {Yang}, De-Hua and {Cao}, Shu-Yun and {Chen}, Hai-Yuan and {Chen}, Jian-Jun and {Chen}, Kun-Xin and {Chen}, Ying and {Chu}, Jia-Ru and {Feng}, Lei and {Gong}, Xue-Fei and {Hou}, Yong-Hui and {Hu}, Hong-Zhuan and {Hu}, Ning-Sheng and {Hu}, Zhong-Wen and {Jia}, Lei and {Jiang}, Fang-Hua and {Jiang}, Xiang and {Jiang}, Zi-Bo and {Jin}, Ge and {Li}, Ai-Hua and {Li}, Yan and {Li}, Ye-Ping and {Liu}, Guan-Qun and {Liu}, Zhi-Gang and {Lu}, Wen-Zhi and {Mao}, Yin-Dun and {Men}, Li and {Qi}, Yong-Jun and {Qi}, Zhao-Xiang and {Shi}, Huo-Ming and {Tang}, Zheng-Hong and {Tao}, Qing-Sheng and {Wang}, Da-Qi and {Wang}, Dan and {Wang}, Guo-Min and {Wang}, Hai and {Wang}, Jia-Ning and {Wang}, Jian and {Wang}, Jian-Ling and {Wang}, Jian-Ping and {Wang}, Lei and {Wang}, Shu-Qing and {Wang}, You and {Wang}, Yue-Fei and {Xu}, Ling-Zhe and {Xu}, Yan and {Yang}, Shi-Hai and {Yu}, Yong and {Yuan}, Hui and {Yuan}, Xiang-Yan and {Zhai}, Chao and {Zhang}, Jing and {Zhang}, Yan-Xia and {Zhang}, Yong and {Zhao}, Ming and {Zhou}, Fang and {Zhou}, Guo-Hua and {Zhu}, Jie and {Zou}, Si-Cheng},
        title = "{The Large Sky Area Multi-Object Fiber Spectroscopic Telescope (LAMOST)}",
      journal = {Research in Astronomy and Astrophysics},
         year = 2012,
        month = sep,
       volume = {12},
       number = {9},
        pages = {1197-1242},
          doi = {10.1088/1674-4527/12/9/003},
       adsurl = {https://ui.adsabs.harvard.edu/abs/2012RAA....12.1197C}
}

@ARTICLE{chandra,
       author = {{Evans}, Ian N. and {Primini}, Francis A. and {Glotfelty}, Kenny J. and {Anderson}, Craig S. and {Bonaventura}, Nina R. and {Chen}, Judy C. and {Davis}, John E. and {Doe}, Stephen M. and {Evans}, Janet D. and {Fabbiano}, Giuseppina and {Galle}, Elizabeth C. and {Gibbs}, Danny G., II and {Grier}, John D. and {Hain}, Roger M. and {Hall}, Diane M. and {Harbo}, Peter N. and {He}, Xiangqun Helen and {Houck}, John C. and {Karovska}, Margarita and {Kashyap}, Vinay L. and {Lauer}, Jennifer and {McCollough}, Michael L. and {McDowell}, Jonathan C. and {Miller}, Joseph B. and {Mitschang}, Arik W. and {Morgan}, Douglas L. and {Mossman}, Amy E. and {Nichols}, Joy S. and {Nowak}, Michael A. and {Plummer}, David A. and {Refsdal}, Brian L. and {Rots}, Arnold H. and {Siemiginowska}, Aneta and {Sundheim}, Beth A. and {Tibbetts}, Michael S. and {Van Stone}, David W. and {Winkelman}, Sherry L. and {Zografou}, Panagoula},
        title = "{The Chandra Source Catalog}",
      journal = {\apjs},
         year = 2010,
        month = jul,
       volume = {189},
       number = {1},
        pages = {37-82},
          doi = {10.1088/0067-0049/189/1/37},
archivePrefix = {arXiv},
       eprint = {1005.4665},
 primaryClass = {astro-ph.HE},
       adsurl = {https://ui.adsabs.harvard.edu/abs/2010ApJS..189...37E}
}

@ARTICLE{swift,
       author = {{Evans}, P.~A. and {Page}, K.~L. and {Osborne}, J.~P. and {Beardmore}, A.~P. and {Willingale}, R. and {Burrows}, D.~N. and {Kennea}, J.~A. and {Perri}, M. and {Capalbi}, M. and {Tagliaferri}, G. and {Cenko}, S.~B.},
        title = "{2SXPS: An Improved and Expanded Swift X-Ray Telescope Point-source Catalog}",
      journal = {\apjs},
         year = 2020,
        month = apr,
       volume = {247},
       number = {2},
          eid = {54},
        pages = {54},
          doi = {10.3847/1538-4365/ab7db9},
archivePrefix = {arXiv},
       eprint = {1911.11710},
 primaryClass = {astro-ph.IM},
       adsurl = {https://ui.adsabs.harvard.edu/abs/2020ApJS..247...54E}
}

@ARTICLE{erosita,
       author = {{Merloni}, A. and {Lamer}, G. and {Liu}, T. and {Ramos-Ceja}, M.~E. and {Brunner}, H. and {Bulbul}, E. and {Dennerl}, K. and {Doroshenko}, V. and {Freyberg}, M.~J. and {Friedrich}, S. and {Gatuzz}, E. and {Georgakakis}, A. and {Haberl}, F. and {Igo}, Z. and {Kreykenbohm}, I. and {Liu}, A. and {Maitra}, C. and {Malyali}, A. and {Mayer}, M.~G.~F. and {Nandra}, K. and {Predehl}, P. and {Robrade}, J. and {Salvato}, M. and {Sanders}, J.~S. and {Stewart}, I. and {Tub{\'\i}n-Arenas}, D. and {Weber}, P. and {Wilms}, J. and {Arcodia}, R. and {Artis}, E. and {Aschersleben}, J. and {Avakyan}, A. and {Aydar}, C. and {Bahar}, Y.~E. and {Balzer}, F. and {Becker}, W. and {Berger}, K. and {Boller}, T. and {Bornemann}, W. and {Br{\"u}ggen}, M. and {Brusa}, M. and {Buchner}, J. and {Burwitz}, V. and {Camilloni}, F. and {Clerc}, N. and {Comparat}, J. and {Coutinho}, D. and {Czesla}, S. and {Dannhauer}, S.~M. and {Dauner}, L. and {Dauser}, T. and {Dietl}, J. and {Dolag}, K. and {Dwelly}, T. and {Egg}, K. and {Ehl}, E. and {Freund}, S. and {Friedrich}, P. and {Gaida}, R. and {Garrel}, C. and {Ghirardini}, V. and {Gokus}, A. and {Gr{\"u}nwald}, G. and {Grandis}, S. and {Grotova}, I. and {Gruen}, D. and {Gueguen}, A. and {H{\"a}mmerich}, S. and {Hamaus}, N. and {Hasinger}, G. and {Haubner}, K. and {Homan}, D. and {Ider Chitham}, J. and {Joseph}, W.~M. and {Joyce}, A. and {K{\"o}nig}, O. and {Kaltenbrunner}, D.~M. and {Khokhriakova}, A. and {Kink}, W. and {Kirsch}, C. and {Kluge}, M. and {Knies}, J. and {Krippendorf}, S. and {Krumpe}, M. and {Kurpas}, J. and {Li}, P. and {Liu}, Z. and {Locatelli}, N. and {Lorenz}, M. and {M{\"u}ller}, S. and {Magaudda}, E. and {Mannes}, C. and {McCall}, H. and {Meidinger}, N. and {Michailidis}, M. and {Migkas}, K. and {Mu{\~n}oz-Giraldo}, D. and {Musiimenta}, B. and {Nguyen-Dang}, N.~T. and {Ni}, Q. and {Olechowska}, A. and {Ota}, N. and {Pacaud}, F. and {Pasini}, T. and {Perinati}, E. and {Pires}, A.~M. and {Pommranz}, C. and {Ponti}, G. and {Poppenhaeger}, K. and {P{\"u}hlhofer}, G. and {Rau}, A. and {Reh}, M. and {Reiprich}, T.~H. and {Roster}, W. and {Saeedi}, S. and {Santangelo}, A. and {Sasaki}, M. and {Schmitt}, J. and {Schneider}, P.~C. and {Schrabback}, T. and {Schuster}, N. and {Schwope}, A. and {Seppi}, R. and {Serim}, M.~M. and {Shreeram}, S. and {Sokolova-Lapa}, E. and {Starck}, H. and {Stelzer}, B. and {Stierhof}, J. and {Suleimanov}, V. and {Tenzer}, C. and {Traulsen}, I. and {Tr{\"u}mper}, J. and {Tsuge}, K. and {Urrutia}, T. and {Veronica}, A. and {Waddell}, S.~G.~H. and {Willer}, R. and {Wolf}, J. and {Yeung}, M.~C.~H. and {Zainab}, A. and {Zangrandi}, F. and {Zhang}, X. and {Zhang}, Y. and {Zheng}, X.},
        title = "{The SRG/eROSITA all-sky survey. First X-ray catalogues and data release of the western Galactic hemisphere}",
      journal = {\aap},
         year = 2024,
        month = feb,
       volume = {682},
          eid = {A34},
        pages = {A34},
          doi = {10.1051/0004-6361/202347165},
archivePrefix = {arXiv},
       eprint = {2401.17274},
 primaryClass = {astro-ph.HE},
       adsurl = {https://ui.adsabs.harvard.edu/abs/2024A&A...682A..34M}
}

@ARTICLE{GALEX,
       author = {{Martin}, D. Christopher and {Fanson}, James and {Schiminovich}, David and {Morrissey}, Patrick and {Friedman}, Peter G. and {Barlow}, Tom A. and {Conrow}, Tim and {Grange}, Robert and {Jelinsky}, Patrick N. and {Milliard}, Bruno and {Siegmund}, Oswald H.~W. and {Bianchi}, Luciana and {Byun}, Yong-Ik and {Donas}, Jose and {Forster}, Karl and {Heckman}, Timothy M. and {Lee}, Young-Wook and {Madore}, Barry F. and {Malina}, Roger F. and {Neff}, Susan G. and {Rich}, R. Michael and {Small}, Todd and {Surber}, Frank and {Szalay}, Alex S. and {Welsh}, Barry and {Wyder}, Ted K.},
        title = "{The Galaxy Evolution Explorer: A Space Ultraviolet Survey Mission}",
      journal = {\apjl},
         year = 2005,
        month = jan,
       volume = {619},
       number = {1},
        pages = {L1-L6},
          doi = {10.1086/426387},
archivePrefix = {arXiv},
       eprint = {astro-ph/0411302},
 primaryClass = {astro-ph},
       adsurl = {https://ui.adsabs.harvard.edu/abs/2005ApJ...619L...1M}
}

@ARTICLE{Sana,
       author = {{Sana}, H. and {de Mink}, S.~E. and {de Koter}, A. and {Langer}, N. and {Evans}, C.~J. and {Gieles}, M. and {Gosset}, E. and {Izzard}, R.~G. and {Le Bouquin}, J. -B. and {Schneider}, F.~R.~N.},
        title = "{Binary Interaction Dominates the Evolution of Massive Stars}",
      journal = {Science},
         year = 2012,
        month = jul,
       volume = {337},
       number = {6093},
        pages = {444},
          doi = {10.1126/science.1223344},
archivePrefix = {arXiv},
       eprint = {1207.6397},
 primaryClass = {astro-ph.SR},
       adsurl = {https://ui.adsabs.harvard.edu/abs/2012Sci...337..444S}
}

@ARTICLE{mink,
       author = {{de Mink}, S.~E. and {Sana}, H. and {Langer}, N. and {Izzard}, R.~G. and {Schneider}, F.~R.~N.},
        title = "{The Incidence of Stellar Mergers and Mass Gainers among Massive Stars}",
      journal = {\apj},
         year = 2014,
        month = feb,
       volume = {782},
       number = {1},
          eid = {7},
        pages = {7},
          doi = {10.1088/0004-637X/782/1/7},
archivePrefix = {arXiv},
       eprint = {1312.3650},
 primaryClass = {astro-ph.SR},
       adsurl = {https://ui.adsabs.harvard.edu/abs/2014ApJ...782....7D}
}

@ARTICLE{Moe,
       author = {{Moe}, Maxwell and {Di Stefano}, Rosanne},
        title = "{Mind Your Ps and Qs: The Interrelation between Period (P) and Mass-ratio (Q) Distributions of Binary Stars}",
      journal = {\apjs},
         year = 2017,
        month = jun,
       volume = {230},
       number = {2},
          eid = {15},
        pages = {15},
          doi = {10.3847/1538-4365/aa6fb6},
archivePrefix = {arXiv},
       eprint = {1606.05347},
 primaryClass = {astro-ph.SR},
       adsurl = {https://ui.adsabs.harvard.edu/abs/2017ApJS..230...15M}
}

@ARTICLE{crowford,
       author = {{Crawford}, J.~A.},
        title = "{On the Subgiant Components of Eclipsing Binary Systems.}",
      journal = {\apj},
         year = 1955,
        month = jan,
       volume = {121},
        pages = {71},
          doi = {10.1086/145965},
       adsurl = {https://ui.adsabs.harvard.edu/abs/1955ApJ...121...71C}
}

@ARTICLE{Xraybinaries,
       author = {{Verbunt}, Frank},
        title = "{Origin and evolution of X-ray binaries and binary radio pulsars.}",
      journal = {\araa},
         year = 1993,
        month = jan,
       volume = {31},
        pages = {93-127},
          doi = {10.1146/annurev.aa.31.090193.000521},
       adsurl = {https://ui.adsabs.harvard.edu/abs/1993ARA&A..31...93V}
}

@ARTICLE{supernovaI,
       author = {{Wang}, Bo and {Han}, Zhanwen},
        title = "{Progenitors of type Ia supernovae}",
      journal = {\nar},
         year = 2012,
        month = jun,
       volume = {56},
       number = {4},
        pages = {122-141},
          doi = {10.1016/j.newar.2012.04.001},
archivePrefix = {arXiv},
       eprint = {1204.1155},
 primaryClass = {astro-ph.SR},
       adsurl = {https://ui.adsabs.harvard.edu/abs/2012NewAR..56..122W}
}

@ARTICLE{supernova2,
       author = {{Yoon}, S. -C. and {Woosley}, S.~E. and {Langer}, N.},
        title = "{Type Ib/c Supernovae in Binary Systems. I. Evolution and Properties of the Progenitor Stars}",
      journal = {\apj},
         year = 2010,
        month = dec,
       volume = {725},
       number = {1},
        pages = {940-954},
          doi = {10.1088/0004-637X/725/1/940},
archivePrefix = {arXiv},
       eprint = {1004.0843},
 primaryClass = {astro-ph.SR},
       adsurl = {https://ui.adsabs.harvard.edu/abs/2010ApJ...725..940Y}
}

@ARTICLE{Ivanova,
       author = {{Ivanova}, Natalia and {Taam}, Ronald E.},
        title = "{Thermal Timescale Mass Transfer and the Evolution of White Dwarf Binaries}",
      journal = {\apj},
         year = 2004,
        month = feb,
       volume = {601},
       number = {2},
        pages = {1058-1066},
          doi = {10.1086/380561},
archivePrefix = {arXiv},
       eprint = {astro-ph/0310126},
 primaryClass = {astro-ph},
       adsurl = {https://ui.adsabs.harvard.edu/abs/2004ApJ...601.1058I}
}

@ARTICLE{gwsources,
       author = {{Schneider}, R. and {Ferrari}, V. and {Matarrese}, S. and {Portegies Zwart}, S.~F.},
        title = "{Low-frequency gravitational waves from cosmological compact binaries}",
      journal = {\mnras},
         year = 2001,
        month = jul,
       volume = {324},
       number = {4},
        pages = {797-810},
          doi = {10.1046/j.1365-8711.2001.04217.x},
archivePrefix = {arXiv},
       eprint = {astro-ph/0002055},
 primaryClass = {astro-ph},
       adsurl = {https://ui.adsabs.harvard.edu/abs/2001MNRAS.324..797S}
}

@ARTICLE{kraft,
       author = {{Kraft}, Robert P.},
        title = "{Binary Stars among Cataclysmic Variables. III. Ten Old Novae.}",
      journal = {\apj},
         year = 1964,
        month = feb,
       volume = {139},
        pages = {457},
          doi = {10.1086/147776},
       adsurl = {https://ui.adsabs.harvard.edu/abs/1964ApJ...139..457K}
}

@ARTICLE{kuiper,
       author = {{Kuiper}, Gerard P.},
        title = "{On the Interpretation of {\ensuremath{\beta}} Lyrae and Other Close Binaries.}",
      journal = {\apj},
         year = 1941,
        month = jan,
       volume = {93},
        pages = {133},
          doi = {10.1086/144252},
       adsurl = {https://ui.adsabs.harvard.edu/abs/1941ApJ....93..133K}
}

@INPROCEEDINGS{paczynski,
       author = {{Paczynski}, B.},
        title = "{Common Envelope Binaries}",
    booktitle = {Structure and Evolution of Close Binary Systems},
         year = 1976,
       editor = {{Eggleton}, Peter and {Mitton}, Simon and {Whelan}, John},
       volume = {73},
        month = jan,
        pages = {75},
       adsurl = {https://ui.adsabs.harvard.edu/abs/1976IAUS...73...75P}
}

@ARTICLE{Jakub,
       author = {{Cehula}, Jakub and {Pejcha}, Ond{\v{r}}ej},
        title = "{A theory of mass transfer in binary stars}",
      journal = {\mnras},
         year = 2023,
        month = sep,
       volume = {524},
       number = {1},
        pages = {471-490},
          doi = {10.1093/mnras/stad1862},
archivePrefix = {arXiv},
       eprint = {2303.05526},
 primaryClass = {astro-ph.SR},
       adsurl = {https://ui.adsabs.harvard.edu/abs/2023MNRAS.524..471C}
}

@ARTICLE{TZDra,
       author = {{Kahraman Ali{\c{c}}avu{\c{s}}}, F. and {Handler}, G. and {Ali{\c{c}}avu{\c{s}}}, F. and {De Cat}, P. and {Bedding}, T.~R. and {Lampens}, P. and {Ekinci}, {\"O}. and {G{\"u}m{\"u}{\textcommabelow s}}, D. and {Leone}, F.},
        title = "{Mass transfer and tidally tilted pulsation in the Algol-type system TZ Dra}",
      journal = {\mnras},
         year = 2022,
        month = feb,
       volume = {510},
       number = {1},
        pages = {1413-1424},
          doi = {10.1093/mnras/stab3515},
archivePrefix = {arXiv},
       eprint = {2112.03687},
 primaryClass = {astro-ph.SR},
       adsurl = {https://ui.adsabs.harvard.edu/abs/2022MNRAS.510.1413K}
}

@ARTICLE{AUMonocerotis,
       author = {{Armeni}, Antonio and {Shore}, Steven N.},
        title = "{Revisiting the high-mass transfer close binary star system AU Monocerotis}",
      journal = {\aap},
         year = 2022,
        month = aug,
       volume = {664},
          eid = {A103},
        pages = {A103},
          doi = {10.1051/0004-6361/202243610},
       adsurl = {https://ui.adsabs.harvard.edu/abs/2022A&A...664A.103A}
}

@ARTICLE{V2840Cygni,
       author = {{Pothuneni}, Ravi Raja and {Devarapalli}, Shanti Priya and {Jagirdar}, Rukmini},
        title = "{The First Photometric and Spectroscopic Study of Contact Binary V2840 Cygni}",
      journal = {Research in Astronomy and Astrophysics},
         year = 2023,
        month = feb,
       volume = {23},
       number = {2},
          eid = {025017},
        pages = {025017},
          doi = {10.1088/1674-4527/acae6e},
archivePrefix = {arXiv},
       eprint = {2303.03514},
 primaryClass = {astro-ph.SR},
       adsurl = {https://ui.adsabs.harvard.edu/abs/2023RAA....23b5017P}
}

@INPROCEEDINGS{CEPod2001,
       author = {{Podsiadlowski}, Philipp},
        title = "{Common-Envelope Evolution and Stellar Mergers}",
    booktitle = {Evolution of Binary and Multiple Star Systems},
         year = 2001,
       editor = {{Podsiadlowski}, Ph. and {Rappaport}, S. and {King}, A.~R. and {D'Antona}, F. and {Burderi}, L.},
       series = {Astronomical Society of the Pacific Conference Series},
       volume = {229},
        month = jan,
        pages = {239},
       adsurl = {https://ui.adsabs.harvard.edu/abs/2001ASPC..229..239P}
}

@INPROCEEDINGS{CEIvanova2001,
       author = {{Ivanova}, Natalia and {Podsiadlowski}, Philipp and {Spruit}, Henk},
        title = "{Common-Envelope Evolution: the Nucleosynthesis in Mergers of Massive Stars}",
    booktitle = {Evolution of Binary and Multiple Star Systems},
         year = 2001,
       editor = {{Podsiadlowski}, Ph. and {Rappaport}, S. and {King}, A.~R. and {D'Antona}, F. and {Burderi}, L.},
       series = {Astronomical Society of the Pacific Conference Series},
       volume = {229},
        month = jan,
        pages = {261},
          doi = {10.48550/arXiv.astro-ph/0102141},
archivePrefix = {arXiv},
       eprint = {astro-ph/0102141},
 primaryClass = {astro-ph},
       adsurl = {https://ui.adsabs.harvard.edu/abs/2001ASPC..229..261I}
}

@ARTICLE{CEPablo2021,
       author = {{Marchant}, Pablo and {Pappas}, Kaliro{\"e} M.~W. and {Gallegos-Garcia}, Monica and {Berry}, Christopher P.~L. and {Taam}, Ronald E. and {Kalogera}, Vicky and {Podsiadlowski}, Philipp},
        title = "{The role of mass transfer and common envelope evolution in the formation of merging binary black holes}",
      journal = {\aap},
         year = 2021,
        month = jun,
       volume = {650},
          eid = {A107},
        pages = {A107},
          doi = {10.1051/0004-6361/202039992},
archivePrefix = {arXiv},
       eprint = {2103.09243},
 primaryClass = {astro-ph.SR},
       adsurl = {https://ui.adsabs.harvard.edu/abs/2021A&A...650A.107M}
}

@ARTICLE{CEDeMarco2023,
       author = {{R{\"o}pke}, Friedrich K. and {De Marco}, Orsola},
        title = "{Simulations of common-envelope evolution in binary stellar systems: physical models and numerical techniques}",
      journal = {Living Reviews in Computational Astrophysics},
         year = 2023,
        month = dec,
       volume = {9},
       number = {1},
          eid = {2},
        pages = {2},
          doi = {10.1007/s41115-023-00017-x},
archivePrefix = {arXiv},
       eprint = {2212.07308},
 primaryClass = {astro-ph.SR},
       adsurl = {https://ui.adsabs.harvard.edu/abs/2023LRCA....9....2R}
}

@ARTICLE{Soker2003,
       author = {{Soker}, Noam and {Tylenda}, Romuald},
        title = "{Main-Sequence Stellar Eruption Model for V838 Monocerotis}",
      journal = {\apjl},
         year = 2003,
        month = jan,
       volume = {582},
       number = {2},
        pages = {L105-L108},
          doi = {10.1086/367759},
archivePrefix = {arXiv},
       eprint = {astro-ph/0210463},
 primaryClass = {astro-ph},
       adsurl = {https://ui.adsabs.harvard.edu/abs/2003ApJ...582L.105S}
}

@ARTICLE{Pejcha2014,
       author = {{Pejcha}, Ond{\v{r}}ej},
        title = "{Burying a Binary: Dynamical Mass Loss and a Continuous Optically thick Outflow Explain the Candidate Stellar Merger V1309 Scorpii}",
      journal = {\apj},
         year = 2014,
        month = jun,
       volume = {788},
       number = {1},
          eid = {22},
        pages = {22},
          doi = {10.1088/0004-637X/788/1/22},
archivePrefix = {arXiv},
       eprint = {1307.4088},
 primaryClass = {astro-ph.SR},
       adsurl = {https://ui.adsabs.harvard.edu/abs/2014ApJ...788...22P}
}

@ARTICLE{Pejcha2016,
       author = {{Pejcha}, Ond{\v{r}}ej and {Metzger}, Brian D. and {Tomida}, Kengo},
        title = "{Cool and luminous transients from mass-losing binary stars}",
      journal = {\mnras},
         year = 2016,
        month = feb,
       volume = {455},
       number = {4},
        pages = {4351-4372},
          doi = {10.1093/mnras/stv2592},
archivePrefix = {arXiv},
       eprint = {1509.02531},
 primaryClass = {astro-ph.SR},
       adsurl = {https://ui.adsabs.harvard.edu/abs/2016MNRAS.455.4351P}
}

@ARTICLE{Blagorodnova2017,
       author = {{Blagorodnova}, N. and {Kotak}, R. and {Polshaw}, J. and {Kasliwal}, M.~M. and {Cao}, Y. and {Cody}, A.~M. and {Doran}, G.~B. and {Elias-Rosa}, N. and {Fraser}, M. and {Fremling}, C. and {Gonzalez-Fernandez}, C. and {Harmanen}, J. and {Jencson}, J. and {Kankare}, E. and {Kudritzki}, R. -P. and {Kulkarni}, S.~R. and {Magnier}, E. and {Manulis}, I. and {Masci}, F.~J. and {Mattila}, S. and {Nugent}, P. and {Ochner}, P. and {Pastorello}, A. and {Reynolds}, T. and {Smith}, K. and {Sollerman}, J. and {Taddia}, F. and {Terreran}, G. and {Tomasella}, L. and {Turatto}, M. and {Vreeswijk}, P.~M. and {Wozniak}, P. and {Zaggia}, S.},
        title = "{Common Envelope Ejection for a Luminous Red Nova in M101}",
      journal = {\apj},
         year = 2017,
        month = jan,
       volume = {834},
       number = {2},
          eid = {107},
        pages = {107},
          doi = {10.3847/1538-4357/834/2/107},
archivePrefix = {arXiv},
       eprint = {1607.08248},
 primaryClass = {astro-ph.SR},
       adsurl = {https://ui.adsabs.harvard.edu/abs/2017ApJ...834..107B}
}

@ARTICLE{V838Mon1,
       author = {{Munari}, U. and {Henden}, A. and {Kiyota}, S. and {Laney}, D. and {Marang}, F. and {Zwitter}, T. and {Corradi}, R.~L.~M. and {Desidera}, S. and {Marrese}, P.~M. and {Giro}, E. and {Boschi}, F. and {Schwartz}, M.~B.},
        title = "{The mysterious eruption of V838 Mon}",
      journal = {\aap},
         year = 2002,
        month = jul,
       volume = {389},
        pages = {L51-L56},
          doi = {10.1051/0004-6361:20020715},
archivePrefix = {arXiv},
       eprint = {astro-ph/0205288},
 primaryClass = {astro-ph},
       adsurl = {https://ui.adsabs.harvard.edu/abs/2002A&A...389L..51M}
}

@ARTICLE{V838Mon2,
       author = {{Bond}, Howard E. and {Henden}, Arne and {Levay}, Zoltan G. and {Panagia}, Nino and {Sparks}, William B. and {Starrfield}, Sumner and {Wagner}, R. Mark and {Corradi}, R.~L.~M. and {Munari}, U.},
        title = "{An energetic stellar outburst accompanied by circumstellar light echoes}",
      journal = {\nat},
         year = 2003,
        month = mar,
       volume = {422},
       number = {6930},
        pages = {405-408},
          doi = {10.1038/nature01508},
archivePrefix = {arXiv},
       eprint = {astro-ph/0303513},
 primaryClass = {astro-ph},
       adsurl = {https://ui.adsabs.harvard.edu/abs/2003Natur.422..405B}
}

@ARTICLE{V4332Sag,
       author = {{Martini}, Paul and {Wagner}, R. Mark and {Tomaney}, Austin and {Rich}, R. Michael and {della Valle}, M. and {Hauschildt}, Peter H.},
        title = "{Nova Sagittarii 1994 1 (V4332 Sagittarii): The Discovery and Evolution of an Unusual Luminous Red Variable Star}",
      journal = {\aj},
         year = 1999,
        month = aug,
       volume = {118},
       number = {2},
        pages = {1034-1042},
          doi = {10.1086/300951},
archivePrefix = {arXiv},
       eprint = {astro-ph/9905016},
 primaryClass = {astro-ph},
       adsurl = {https://ui.adsabs.harvard.edu/abs/1999AJ....118.1034M}
}

@ARTICLE{Blagorodnova2021,
       author = {{Blagorodnova}, Nadejda and {Klencki}, Jakub and {Pejcha}, Ond{\v{r}}ej and {Vreeswijk}, Paul M. and {Bond}, Howard E. and {Burdge}, Kevin B. and {De}, Kishalay and {Fremling}, Christoffer and {Gehrz}, Robert D. and {Jencson}, Jacob E. and {Kasliwal}, Mansi M. and {Kupfer}, Thomas and {Lau}, Ryan M. and {Masci}, Frank J. and {Rich}, Michael R.},
        title = "{The luminous red nova AT 2018bwo in NGC 45 and its binary yellow supergiant progenitor}",
      journal = {\aap},
         year = 2021,
        month = sep,
       volume = {653},
          eid = {A134},
        pages = {A134},
          doi = {10.1051/0004-6361/202140525},
archivePrefix = {arXiv},
       eprint = {2102.05662},
 primaryClass = {astro-ph.SR},
       adsurl = {https://ui.adsabs.harvard.edu/abs/2021A&A...653A.134B}
}

@BOOK{princeton,
       author = {{Gray}, Richard O. and {Corbally}, Christopher, J.},
        title = "{Stellar Spectral Classification}",
         year = 2009,
       adsurl = {https://ui.adsabs.harvard.edu/abs/2009ssc..book.....G}
}

@ARTICLE{BeStars,
       author = {{Porter}, John M. and {Rivinius}, Thomas},
        title = "{Classical Be Stars}",
      journal = {\pasp},
         year = 2003,
        month = oct,
       volume = {115},
       number = {812},
        pages = {1153-1170},
          doi = {10.1086/378307},
       adsurl = {https://ui.adsabs.harvard.edu/abs/2003PASP..115.1153P}
}

@ARTICLE{BeBinary,
       author = {{Pols}, O.~R. and {Cote}, J. and {Waters}, L.~B.~F.~M. and {Heise}, J.},
        title = "{The formation of Be stars through close binary evolution.}",
      journal = {\aap},
         year = 1991,
        month = jan,
       volume = {241},
        pages = {419},
       adsurl = {https://ui.adsabs.harvard.edu/abs/1991A&A...241..419P}
}

@INPROCEEDINGS{CaEmissionBe,
       author = {{Polidan}, R.~S.},
        title = "{On the Detection of Binary be Stars}",
    booktitle = {Be and Shell Stars},
         year = 1976,
       editor = {{Slettebak}, Arne},
       volume = {70},
        month = jan,
        pages = {401},
       adsurl = {https://ui.adsabs.harvard.edu/abs/1976IAUS...70..401P}
}

@ARTICLE{Gehr,
       author = {{Gehrz}, R.~D. and {Hackwell}, J.~A. and {Jones}, T.~W.},
        title = "{Infrared observations of Be stars from 2.3 to 19.5 microns.}",
      journal = {\apj},
         year = 1974,
        month = aug,
       volume = {191},
        pages = {675-684},
          doi = {10.1086/153008},
       adsurl = {https://ui.adsabs.harvard.edu/abs/1974ApJ...191..675G}
}

@Article{numpy,
 title         = {Array programming with {NumPy}},
 author        = {Charles R. Harris and K. Jarrod Millman and St{\'{e}}fan J.
                 van der Walt and Ralf Gommers and Pauli Virtanen and David
                 Cournapeau and Eric Wieser and Julian Taylor and Sebastian
                 Berg and Nathaniel J. Smith and Robert Kern and Matti Picus
                 and Stephan Hoyer and Marten H. van Kerkwijk and Matthew
                 Brett and Allan Haldane and Jaime Fern{\'{a}}ndez del
                 R{\'{i}}o and Mark Wiebe and Pearu Peterson and Pierre
                 G{\'{e}}rard-Marchant and Kevin Sheppard and Tyler Reddy and
                 Warren Weckesser and Hameer Abbasi and Christoph Gohlke and
                 Travis E. Oliphant},
 year          = {2020},
 month         = sep,
 journal       = {Nature},
 volume        = {585},
 number        = {7825},
 pages         = {357--362},
 doi           = {10.1038/s41586-020-2649-2},
 publisher     = {Springer Science and Business Media {LLC}},
 url           = {https://doi.org/10.1038/s41586-020-2649-2}
}

@Article{matplotlib,
  Author    = {Hunter, J. D.},
  Title     = {Matplotlib: A 2D graphics environment},
  Journal   = {Computing in Science \& Engineering},
  Volume    = {9},
  Number    = {3},
  Pages     = {90--95},
  publisher = {IEEE COMPUTER SOC},
  doi       = {10.1109/MCSE.2007.55},
  year      = 2007
}

@ARTICLE{astroquery,
   author = {{Ginsburg}, A. and {Sip{\H o}cz}, B.~M. and {Brasseur}, C.~E. and
	{Cowperthwaite}, P.~S. and {Craig}, M.~W. and {Deil}, C. and
	{Guillochon}, J. and {Guzman}, G. and {Liedtke}, S. and {Lian Lim}, P. and
	{Lockhart}, K.~E. and {Mommert}, M. and {Morris}, B.~M. and
	{Norman}, H. and {Parikh}, M. and {Persson}, M.~V. and {Robitaille}, T.~P. and
	{Segovia}, J.-C. and {Singer}, L.~P. and {Tollerud}, E.~J. and
	{de Val-Borro}, M. and {Valtchanov}, I. and {Woillez}, J. and
	{The Astroquery collaboration} and {a subset of the astropy collaboration}
	},
    title = "{astroquery: An Astronomical Web-querying Package in Python}",
  journal = {\aj},
archivePrefix = "arXiv",
   eprint = {1901.04520},
 primaryClass = "astro-ph.IM",
     year = 2019,
    month = mar,
   volume = 157,
      eid = {98},
    pages = {98},
      doi = {10.3847/1538-3881/aafc33},
   adsurl = {https://adsabs.harvard.edu/abs/2019AJ....157...98G}
}

@ARTICLE{pastorello,
       author = {{Pastorello}, Andrea and {Fraser}, Morgan},
        title = "{Supernova impostors and other gap transients}",
      journal = {Nature Astronomy},
         year = 2019,
        month = aug,
       volume = {3},
        pages = {676-679},
          doi = {10.1038/s41550-019-0809-9},
archivePrefix = {arXiv},
       eprint = {1908.02323},
 primaryClass = {astro-ph.SR},
       adsurl = {https://ui.adsabs.harvard.edu/abs/2019NatAs...3..676P}
}

@ARTICLE{Lomb,
       author = {{Lomb}, N.~R.},
        title = "{Least-Squares Frequency Analysis of Unequally Spaced Data}",
      journal = {\apss},
         year = 1976,
        month = feb,
       volume = {39},
       number = {2},
        pages = {447-462},
          doi = {10.1007/BF00648343},
       adsurl = {https://ui.adsabs.harvard.edu/abs/1976Ap&SS..39..447L}
}

@ARTICLE{Scargle,
       author = {{Scargle}, J.~D.},
        title = "{Studies in astronomical time series analysis. II. Statistical aspects of spectral analysis of unevenly spaced data.}",
      journal = {\apj},
         year = 1982,
        month = dec,
       volume = {263},
        pages = {835-853},
          doi = {10.1086/160554},
       adsurl = {https://ui.adsabs.harvard.edu/abs/1982ApJ...263..835S}
}

@ARTICLE{SHBoost,
       author = {{Khalatyan}, A. and {Anders}, F. and {Chiappini}, C. and {Queiroz}, A.~B.~A. and {Nepal}, S. and {dal Ponte}, M. and {Jordi}, C. and {Guiglion}, G. and {Valentini}, M. and {Torralba Elipe}, G. and {Steinmetz}, M. and {Pantaleoni-Gonz{\'a}lez}, M. and {Malhotra}, S. and {Jim{\'e}nez-Arranz}, {\'O}. and {Enke}, H. and {Casamiquela}, L. and {Ard{\`e}vol}, J.},
        title = "{Transferring spectroscopic stellar labels to 217 million Gaia DR3 XP stars with SHBoost}",
      journal = {\aap},
         year = 2024,
        month = nov,
       volume = {691},
          eid = {A98},
        pages = {A98},
          doi = {10.1051/0004-6361/202451427},
archivePrefix = {arXiv},
       eprint = {2407.06963},
 primaryClass = {astro-ph.SR},
       adsurl = {https://ui.adsabs.harvard.edu/abs/2024A&A...691A..98K}
}

@ARTICLE{DR3vari,
       author = {{Ma{\'\i}z Apell{\'a}niz}, J. and {Holgado}, G. and {Pantaleoni Gonz{\'a}lez}, M. and {Caballero}, J.~A.},
        title = "{Stellar variability in Gaia DR3. I. Three-band photometric dispersions for 145 million sources}",
      journal = {\aap},
         year = 2023,
        month = sep,
       volume = {677},
          eid = {A137},
        pages = {A137},
          doi = {10.1051/0004-6361/202346759},
archivePrefix = {arXiv},
       eprint = {2304.14249},
 primaryClass = {astro-ph.SR},
       adsurl = {https://ui.adsabs.harvard.edu/abs/2023A&A...677A.137M}
}

@ARTICLE{DR2vari,
       author = {{Andrew}, Shion and {Swihart}, Samuel J. and {Strader}, Jay},
        title = "{Identifying Candidate Optical Variables Using Gaia Data Release 2}",
      journal = {\apj},
         year = 2021,
        month = feb,
       volume = {908},
       number = {2},
          eid = {180},
        pages = {180},
          doi = {10.3847/1538-4357/abd257},
archivePrefix = {arXiv},
       eprint = {2012.04006},
 primaryClass = {astro-ph.IM},
       adsurl = {https://ui.adsabs.harvard.edu/abs/2021ApJ...908..180A}
}

@ARTICLE{riello,
       author = {{Riello}, M. and {De Angeli}, F. and {Evans}, D.~W. and {Montegriffo}, P. and {Carrasco}, J.~M. and {Busso}, G. and {Palaversa}, L. and {Burgess}, P.~W. and {Diener}, C. and {Davidson}, M. and {Rowell}, N. and {Fabricius}, C. and {Jordi}, C. and {Bellazzini}, M. and {Pancino}, E. and {Harrison}, D.~L. and {Cacciari}, C. and {van Leeuwen}, F. and {Hambly}, N.~C. and {Hodgkin}, S.~T. and {Osborne}, P.~J. and {Altavilla}, G. and {Barstow}, M.~A. and {Brown}, A.~G.~A. and {Castellani}, M. and {Cowell}, S. and {De Luise}, F. and {Gilmore}, G. and {Giuffrida}, G. and {Hidalgo}, S. and {Holland}, G. and {Marinoni}, S. and {Pagani}, C. and {Piersimoni}, A.~M. and {Pulone}, L. and {Ragaini}, S. and {Rainer}, M. and {Richards}, P.~J. and {Sanna}, N. and {Walton}, N.~A. and {Weiler}, M. and {Yoldas}, A.},
        title = "{Gaia Early Data Release 3. Photometric content and validation}",
      journal = {\aap},
         year = 2021,
        month = may,
       volume = {649},
          eid = {A3},
        pages = {A3},
          doi = {10.1051/0004-6361/202039587},
archivePrefix = {arXiv},
       eprint = {2012.01916},
 primaryClass = {astro-ph.IM},
       adsurl = {https://ui.adsabs.harvard.edu/abs/2021A&A...649A...3R}
}

@ARTICLE{simbad,
       author = {{Wenger}, M. and {Ochsenbein}, F. and {Egret}, D. and {Dubois}, P. and {Bonnarel}, F. and {Borde}, S. and {Genova}, F. and {Jasniewicz}, G. and {Lalo{\"e}}, S. and {Lesteven}, S. and {Monier}, R.},
        title = "{The SIMBAD astronomical database. The CDS reference database for astronomical objects}",
      journal = {\aaps},
         year = 2000,
        month = apr,
       volume = {143},
        pages = {9-22},
          doi = {10.1051/aas:2000332},
archivePrefix = {arXiv},
       eprint = {astro-ph/0002110},
 primaryClass = {astro-ph},
       adsurl = {https://ui.adsabs.harvard.edu/abs/2000A&AS..143....9W}
}

@ARTICLE{GaiaVarXmatch,
       author = {{Gavras}, Panagiotis and {Rimoldini}, Lorenzo and {Nienartowicz}, Krzysztof and {de Fombelle}, Gr{\'e}gory Jevardat and {Holl}, Berry and {{\'A}brah{\'a}m}, P{\'e}ter and {Audard}, Marc and {Carnerero}, Maria I. and {Clementini}, Gisella and {De Ridder}, Joris and {Distefano}, Elisa and {Garcia-Lario}, Pedro and {Garofalo}, Alessia and {K{\'o}sp{\'a}l}, {\'A}gnes and {Kruszy{\'n}ska}, Katarzyna and {Kun}, M{\'a}ria and {Lecoeur-Ta{\"\i}bi}, Isabelle and {Marton}, G{\'a}bor and {Mazeh}, Tsevi and {Mowlavi}, Nami and {Raiteri}, Claudia M. and {Ripepi}, Vincenzo and {Szabados}, L{\'a}szl{\'o} and {Zucker}, Shay and {Eyer}, Laurent},
        title = "{Gaia Data Release 3. Cross-match of Gaia sources with variable objects from the literature}",
      journal = {\aap},
         year = 2023,
        month = jun,
       volume = {674},
          eid = {A22},
        pages = {A22},
          doi = {10.1051/0004-6361/202244367},
archivePrefix = {arXiv},
       eprint = {2207.01946},
 primaryClass = {astro-ph.IM},
       adsurl = {https://ui.adsabs.harvard.edu/abs/2023A&A...674A..22G}
}

@ARTICLE{QLP1,
       author = {{Huang}, Chelsea X. and {Vanderburg}, Andrew and {P{\'a}l}, Andras and {Sha}, Lizhou and {Yu}, Liang and {Fong}, Willie and {Fausnaugh}, Michael and {Shporer}, Avi and {Guerrero}, Natalia and {Vanderspek}, Roland and {Ricker}, George},
        title = "{Photometry of 10 Million Stars from the First Two Years of TESS Full Frame Images: Part I}",
      journal = {Research Notes of the American Astronomical Society},
         year = 2020,
        month = nov,
       volume = {4},
       number = {11},
          eid = {204},
        pages = {204},
          doi = {10.3847/2515-5172/abca2e},
archivePrefix = {arXiv},
       eprint = {2011.06459},
 primaryClass = {astro-ph.EP},
       adsurl = {https://ui.adsabs.harvard.edu/abs/2020RNAAS...4..204H}
}

@ARTICLE{QLP2,
       author = {{Kunimoto}, Michelle and {Huang}, Chelsea and {Tey}, Evan and {Fong}, Willie and {Hesse}, Katharine and {Shporer}, Avi and {Guerrero}, Natalia and {Fausnaugh}, Michael and {Vanderspek}, Roland and {Ricker}, George},
        title = "{Quick-look Pipeline Lightcurves for 9.1 Million Stars Observed over the First Year of the TESS Extended Mission}",
      journal = {Research Notes of the American Astronomical Society},
         year = 2021,
        month = oct,
       volume = {5},
       number = {10},
          eid = {234},
        pages = {234},
          doi = {10.3847/2515-5172/ac2ef0},
archivePrefix = {arXiv},
       eprint = {2110.05542},
 primaryClass = {astro-ph.EP},
       adsurl = {https://ui.adsabs.harvard.edu/abs/2021RNAAS...5..234K}
}

@ARTICLE{SPOC,
       author = {{Caldwell}, Douglas A. and {Tenenbaum}, Peter and {Twicken}, Joseph D. and {Jenkins}, Jon M. and {Ting}, Eric and {Smith}, Jeffrey C. and {Hedges}, Christina and {Fausnaugh}, Michael M. and {Rose}, Mark and {Burke}, Christopher},
        title = "{TESS Science Processing Operations Center FFI Target List Products}",
      journal = {Research Notes of the American Astronomical Society},
         year = 2020,
        month = nov,
       volume = {4},
       number = {11},
          eid = {201},
        pages = {201},
          doi = {10.3847/2515-5172/abc9b3},
archivePrefix = {arXiv},
       eprint = {2011.05495},
 primaryClass = {astro-ph.EP},
       adsurl = {https://ui.adsabs.harvard.edu/abs/2020RNAAS...4..201C}
}

@ARTICLE{TASOC,
       author = {{Lund}, Mikkel N. and {Handberg}, Rasmus and {Buzasi}, Derek L. and {Carboneau}, Lindsey and {Hall}, Oliver J. and {Pereira}, Filipe and {Huber}, Daniel and {Hey}, Daniel and {Van Reeth}, Timothy and {T'DA Collaboration}},
        title = "{TESS Data for Asteroseismology: Light-curve Systematics Correction}",
      journal = {\apjs},
         year = 2021,
        month = dec,
       volume = {257},
       number = {2},
          eid = {53},
        pages = {53},
          doi = {10.3847/1538-4365/ac214a},
archivePrefix = {arXiv},
       eprint = {2108.11780},
 primaryClass = {astro-ph.IM},
       adsurl = {https://ui.adsabs.harvard.edu/abs/2021ApJS..257...53L}
}

@ARTICLE{GFSC,
       author = {{Powell}, Brian P. and {Kruse}, Ethan and {Montet}, Benjamin T. and {Feinstein}, Adina D. and {Lewis}, Hannah M. and {Foreman-Mackey}, Daniel and {Barclay}, Thomas and {Quintana}, Elisa V. and {Col{\'o}n}, Knicole D. and {Kostov}, Veselin B. and {Boyd}, Patricia and {Smale}, Alan P. and {Mullally}, Susan E. and {Schlieder}, Joshua E. and {Schnittman}, Jeremy D. and {Carroll}, Mark L. and {Carriere}, Laura E. and {Salmon}, Ellen M. and {Strong}, Savannah L. and {Acks}, Nicko D. and {Pfaff}, Bruce E. and {Gerner}, Lyn E. and {Burch}, Timothy M.},
        title = "{The NASA GSFC TESS Full Frame Image Light Curve Data Set}",
      journal = {Research Notes of the American Astronomical Society},
         year = 2022,
        month = jun,
       volume = {6},
       number = {6},
          eid = {111},
        pages = {111},
          doi = {10.3847/2515-5172/ac74c4},
       adsurl = {https://ui.adsabs.harvard.edu/abs/2022RNAAS...6..111P}
}

@ARTICLE{CDIPS,
       author = {{Bouma}, L.~G. and {Hartman}, J.~D. and {Bhatti}, W. and {Winn}, J.~N. and {Bakos}, G. {\'A}.},
        title = "{Cluster Difference Imaging Photometric Survey. I. Light Curves of Stars in Open Clusters from TESS Sectors 6 and 7}",
      journal = {\apjs},
         year = 2019,
        month = nov,
       volume = {245},
       number = {1},
          eid = {13},
        pages = {13},
          doi = {10.3847/1538-4365/ab4a7e},
archivePrefix = {arXiv},
       eprint = {1910.01133},
 primaryClass = {astro-ph.SR},
       adsurl = {https://ui.adsabs.harvard.edu/abs/2019ApJS..245...13B}
}

@misc{nutmaat2024,
  author = {{El-Kholy}, R.~I. and {Hayman}, Z.~M.},
  title  = {NutMaat: A Python library for classifying stellar spectra based on the MKCLASS package},
  year   = {2024},
  doi    = {10.5281/zenodo.13945430},
  url    = {https://github.com/rehamelkholy/NutMaat},
}

@article{mcclass2014,
  author = {{Gray}, R.~O. and {Corbally}, C.~J.},
  title = "{An Expert Computer Program for Classifying Stars on the MK Spectral Classification System}",
  journal = {\aj},
  year = 2014,
  volume = {147},
  number = {4},
  pages = {80},
  doi = {10.1088/0004-6256/147/4/80},
}

@ARTICLE{dibs,
       author = {{Carvalho}, Adolfo S. and {Hillenbrand}, Lynne A.},
        title = "{Measuring Optical Extinction toward Young Stellar Objects Using Diffuse Interstellar Bands}",
      journal = {\apj},
         year = 2022,
        month = dec,
       volume = {940},
       number = {2},
          eid = {156},
        pages = {156},
          doi = {10.3847/1538-4357/ac9d8e},
archivePrefix = {arXiv},
       eprint = {2204.06061},
 primaryClass = {astro-ph.SR},
       adsurl = {https://ui.adsabs.harvard.edu/abs/2022ApJ...940..156C}
}

@ARTICLE{XMM,
       author = {{Webb}, N.~A. and {Coriat}, M. and {Traulsen}, I. and {Ballet}, J. and {Motch}, C. and {Carrera}, F.~J. and {Koliopanos}, F. and {Authier}, J. and {de la Calle}, I. and {Ceballos}, M.~T. and {Colomo}, E. and {Chuard}, D. and {Freyberg}, M. and {Garcia}, T. and {Kolehmainen}, M. and {Lamer}, G. and {Lin}, D. and {Maggi}, P. and {Michel}, L. and {Page}, C.~G. and {Page}, M.~J. and {Perea-Calderon}, J.~V. and {Pineau}, F. -X. and {Rodriguez}, P. and {Rosen}, S.~R. and {Santos Lleo}, M. and {Saxton}, R.~D. and {Schwope}, A. and {Tom{\'a}s}, L. and {Watson}, M.~G. and {Zakardjian}, A.},
        title = "{The XMM-Newton serendipitous survey. IX. The fourth XMM-Newton serendipitous source catalogue}",
      journal = {\aap},
         year = 2020,
        month = sep,
       volume = {641},
          eid = {A136},
        pages = {A136},
          doi = {10.1051/0004-6361/201937353},
archivePrefix = {arXiv},
       eprint = {2007.02899},
 primaryClass = {astro-ph.HE},
       adsurl = {https://ui.adsabs.harvard.edu/abs/2020A&A...641A.136W}
}

@INPROCEEDINGS{Algol-binaries,
       author = {{Peters}, Geraldine J.},
        title = "{The Algol-Type Binaries}",
    booktitle = {The Influence of Binaries on Stellar Population Studies},
         year = 2001,
       editor = {{Vanbeveren}, D.},
       series = {Astrophysics and Space Science Library},
       volume = {264},
        month = jan,
        pages = {79},
          doi = {10.1007/978-94-015-9723-4_6},
       adsurl = {https://ui.adsabs.harvard.edu/abs/2001ASSL..264...79P}
}

@ARTICLE{Be-puls,
       author = {{Diago}, P.~D. and {Guti{\'e}rrez-Soto}, J. and {Auvergne}, M. and {Fabregat}, J. and {Hubert}, A. -M. and {Floquet}, M. and {Fr{\'e}mat}, Y. and {Garrido}, R. and {Andrade}, L. and {de Batz}, B. and {Emilio}, M. and {Espinosa Lara}, F. and {Huat}, A. -L. and {Janot-Pacheco}, E. and {Leroy}, B. and {Martayan}, C. and {Neiner}, C. and {Semaan}, T. and {Suso}, J. and {Catala}, C. and {Poretti}, E. and {Rainer}, M. and {Uytterhoeven}, K. and {Michel}, E. and {Samadi}, R.},
        title = "{Pulsations in the late-type Be star HD 50 209 detected by CoRoT}",
      journal = {\aap},
         year = 2009,
        month = oct,
       volume = {506},
       number = {1},
        pages = {125-131},
          doi = {10.1051/0004-6361/200911901},
archivePrefix = {arXiv},
       eprint = {0909.4524},
 primaryClass = {astro-ph.SR},
       adsurl = {https://ui.adsabs.harvard.edu/abs/2009A&A...506..125D}
}

@ARTICLE{ruwe-binary,
       author = {{Castro-Ginard}, Alfred and {Penoyre}, Zephyr and {Casey}, Andrew R. and {Brown}, Anthony G.~A. and {Belokurov}, Vasily and {Cantat-Gaudin}, Tristan and {Drimmel}, Ronald and {Fouesneau}, Morgan and {Khanna}, Shourya and {Kurbatov}, Evgeny P. and {Price-Whelan}, Adrian M. and {Rix}, Hans-Walter and {Smart}, Richard L.},
        title = "{Gaia DR3 detectability of unresolved binary systems}",
      journal = {\aap},
         year = 2024,
        month = aug,
       volume = {688},
          eid = {A1},
        pages = {A1},
          doi = {10.1051/0004-6361/202450172},
archivePrefix = {arXiv},
       eprint = {2404.14127},
 primaryClass = {astro-ph.GA},
       adsurl = {https://ui.adsabs.harvard.edu/abs/2024A&A...688A...1C}
}

@ARTICLE{mesa,
       author = {{Paxton}, Bill and {Bildsten}, Lars and {Dotter}, Aaron and {Herwig}, Falk and {Lesaffre}, Pierre and {Timmes}, Frank},
        title = "{Modules for Experiments in Stellar Astrophysics (MESA)}",
      journal = {\apjs},
         year = 2011,
        month = jan,
       volume = {192},
       number = {1},
          eid = {3},
        pages = {3},
          doi = {10.1088/0067-0049/192/1/3},
archivePrefix = {arXiv},
       eprint = {1009.1622},
 primaryClass = {astro-ph.SR},
       adsurl = {https://ui.adsabs.harvard.edu/abs/2011ApJS..192....3P}
}

@ARTICLE{mist,
       author = {{Dotter}, Aaron},
        title = "{MESA Isochrones and Stellar Tracks (MIST) 0: Methods for the Construction of Stellar Isochrones}",
      journal = {\apjs},
         year = 2016,
        month = jan,
       volume = {222},
       number = {1},
          eid = {8},
        pages = {8},
          doi = {10.3847/0067-0049/222/1/8},
archivePrefix = {arXiv},
       eprint = {1601.05144},
 primaryClass = {astro-ph.SR},
       adsurl = {https://ui.adsabs.harvard.edu/abs/2016ApJS..222....8D}
}

@ARTICLE{pypeit,
       author = {{Prochaska}, J. and {Hennawi}, Joseph and {Westfall}, Kyle and {Cooke}, Ryan and {Wang}, Feige and {Hsyu}, Tiffany and {Davies}, Frederick and {Farina}, Emanuele and {Pelliccia}, Debora},
        title = "{PypeIt: The Python Spectroscopic Data Reduction Pipeline}",
      journal = {The Journal of Open Source Software},
         year = 2020,
        month = dec,
       volume = {5},
       number = {56},
          eid = {2308},
        pages = {2308},
          doi = {10.21105/joss.02308},
archivePrefix = {arXiv},
       eprint = {2005.06505},
 primaryClass = {astro-ph.IM},
       adsurl = {https://ui.adsabs.harvard.edu/abs/2020JOSS....5.2308P}
}

@ARTICLE{Negueruela,
       author = {{Negueruela}, I. and {Steele}, I.~A. and {Bernabeu}, G.},
        title = "{On the class of Oe stars}",
      journal = {Astronomische Nachrichten},
         year = 2004,
        month = dec,
       volume = {325},
       number = {9},
        pages = {749-760},
          doi = {10.1002/asna.200310258},
archivePrefix = {arXiv},
       eprint = {astro-ph/0407113},
 primaryClass = {astro-ph},
       adsurl = {https://ui.adsabs.harvard.edu/abs/2004AN....325..749N}
}

@ARTICLE{BeHRd,
       author = {{Cochetti}, Y.~R. and {Zorec}, J. and {Cidale}, L.~S. and {Arias}, M.~L. and {Aidelman}, Y. and {Torres}, A.~F. and {Fr{\'e}mat}, Y. and {Granada}, A.},
        title = "{Be and Bn stars: Balmer discontinuity and stellar-class relationship}",
      journal = {\aap},
         year = 2020,
        month = feb,
       volume = {634},
          eid = {A18},
        pages = {A18},
          doi = {10.1051/0004-6361/201936444},
archivePrefix = {arXiv},
       eprint = {1912.12994},
 primaryClass = {astro-ph.SR},
       adsurl = {https://ui.adsabs.harvard.edu/abs/2020A&A...634A..18C}
}

@ARTICLE{Be-var,
       author = {{Labadie-Bartz}, Jonathan and {Pepper}, Joshua and {McSwain}, M. Virginia and {Bjorkman}, J.~E. and {Bjorkman}, K.~S. and {Lund}, Michael B. and {Rodriguez}, Joseph E. and {Stassun}, Keivan G. and {Stevens}, Daniel J. and {James}, David J. and {Kuhn}, Rudolf B. and {Siverd}, Robert J. and {Beatty}, Thomas G.},
        title = "{Photometric Variability of the Be Star Population}",
      journal = {\aj},
         year = 2017,
        month = jun,
       volume = {153},
       number = {6},
          eid = {252},
        pages = {252},
          doi = {10.3847/1538-3881/aa6396},
archivePrefix = {arXiv},
       eprint = {1609.08449},
 primaryClass = {astro-ph.SR},
       adsurl = {https://ui.adsabs.harvard.edu/abs/2017AJ....153..252L}
}

@ARTICLE{hardXray-sb,
       author = {{Makarov}, Valeri V. and {Unwin}, Stephen C.},
        title = "{Radial velocities and binarity of southern SIM grid stars}",
      journal = {\mnras},
         year = 2015,
        month = jan,
       volume = {446},
       number = {2},
        pages = {2055-2058},
          doi = {10.1093/mnras/stu2239},
archivePrefix = {arXiv},
       eprint = {1410.7943},
 primaryClass = {astro-ph.SR},
       adsurl = {https://ui.adsabs.harvard.edu/abs/2015MNRAS.446.2055M}
}

@ARTICLE{Bebinary2014,
       author = {{Shao}, Yong and {Li}, Xiang-Dong},
        title = "{On the Formation of Be Stars through Binary Interaction}",
      journal = {\apj},
         year = 2014,
        month = nov,
       volume = {796},
       number = {1},
          eid = {37},
        pages = {37},
          doi = {10.1088/0004-637X/796/1/37},
archivePrefix = {arXiv},
       eprint = {1410.0100},
 primaryClass = {astro-ph.HE},
       adsurl = {https://ui.adsabs.harvard.edu/abs/2014ApJ...796...37S}
}

@ARTICLE{asas_vari,
       author = {{Jayasinghe}, T. and {Stanek}, K.~Z. and {Kochanek}, C.~S. and {Shappee}, B.~J. and {Holoien}, T.~W. -S. and {Thompson}, Todd A. and {Prieto}, J.~L. and {Dong}, Subo and {Pawlak}, M. and {Pejcha}, O. and {Shields}, J.~V. and {Pojmanski}, G. and {Otero}, S. and {Hurst}, N. and {Britt}, C.~A. and {Will}, D.},
        title = "{The ASAS-SN catalogue of variable stars III: variables in the southern TESS continuous viewing zone}",
      journal = {\mnras},
         year = 2019,
        month = may,
       volume = {485},
       number = {1},
        pages = {961-971},
          doi = {10.1093/mnras/stz444},
archivePrefix = {arXiv},
       eprint = {1901.00009},
 primaryClass = {astro-ph.SR},
       adsurl = {https://ui.adsabs.harvard.edu/abs/2019MNRAS.485..961J}
}

@ARTICLE{CEP-IR,
       author = {{Kervella}, P. and {M{\'e}rand}, A. and {Perrin}, G. and {Coud{\'e} du Foresto}, V.},
        title = "{Extended envelopes around Galactic Cepheids. I. {\ensuremath{\ell}} Carinae from near and mid-infrared interferometry with the VLTI}",
      journal = {\aap},
         year = 2006,
        month = mar,
       volume = {448},
       number = {2},
        pages = {623-631},
          doi = {10.1051/0004-6361:20053603},
       adsurl = {https://ui.adsabs.harvard.edu/abs/2006A&A...448..623K}
}

@ARTICLE{CEP-IR-theo,
       author = {{Hocd{\'e}}, V. and {Nardetto}, N. and {Lagadec}, E. and {Niccolini}, G. and {Domiciano de Souza}, A. and {M{\'e}rand}, A. and {Kervella}, P. and {Gallenne}, A. and {Marengo}, M. and {Trahin}, B. and {Gieren}, W. and {Pietrzy{\'n}ski}, G. and {Borgniet}, S. and {Breuval}, L. and {Javanmardi}, B.},
        title = "{A thin shell of ionized gas as the explanation for infrared excess among classical Cepheids}",
      journal = {\aap},
         year = 2020,
        month = jan,
       volume = {633},
          eid = {A47},
        pages = {A47},
          doi = {10.1051/0004-6361/201935848},
archivePrefix = {arXiv},
       eprint = {1909.12376},
 primaryClass = {astro-ph.SR},
       adsurl = {https://ui.adsabs.harvard.edu/abs/2020A&A...633A..47H}
}

@ARTICLE{cep-emission,
       author = {{Kovtyukh}, V.~V. and {Wallerstein}, G. and {Andrievsky}, S.~M. and {Gillet}, D. and {Fokin}, A.~B. and {Templeton}, M. and {Henden}, A.~A.},
        title = "{Neutral and ionized emission lines in the type II Cepheid W Virginis}",
      journal = {\aap},
         year = 2011,
        month = feb,
       volume = {526},
          eid = {A116},
        pages = {A116},
          doi = {10.1051/0004-6361/201015912},
       adsurl = {https://ui.adsabs.harvard.edu/abs/2011A&A...526A.116K}
}

@ARTICLE{deltascuPMS,
       author = {{D{\'\i}az-Fraile}, D. and {Rodr{\'\i}guez}, E. and {Amado}, P.~J.},
        title = "{Searching for {\ensuremath{\delta}} Scuti-type pulsation and characterising northern pre-main-sequence field stars}",
      journal = {\aap},
         year = 2014,
        month = aug,
       volume = {568},
          eid = {A32},
        pages = {A32},
          doi = {10.1051/0004-6361/201423478},
archivePrefix = {arXiv},
       eprint = {1407.0391},
 primaryClass = {astro-ph.SR},
       adsurl = {https://ui.adsabs.harvard.edu/abs/2014A&A...568A..32D}
}

@ARTICLE{Banerjee2021,
       author = {{Banerjee}, Gourav and {Mathew}, Blesson and {Paul}, K.~T. and {Subramaniam}, Annapurni and {Bhattacharyya}, Suman and {Anusha}, R.},
        title = "{Optical spectroscopy of Galactic field classical Be stars}",
      journal = {\mnras},
         year = 2021,
        month = jan,
       volume = {500},
       number = {3},
        pages = {3926-3943},
          doi = {10.1093/mnras/staa3469},
archivePrefix = {arXiv},
       eprint = {2011.08622},
 primaryClass = {astro-ph.SR},
       adsurl = {https://ui.adsabs.harvard.edu/abs/2021MNRAS.500.3926B}
}

@ARTICLE{PMBeStars,
       author = {{Dodd}, Jonathan M. and {Oudmaijer}, Ren{\'e} D. and {Radley}, Isaac C. and {Vioque}, Miguel and {Frost}, Abigail J.},
        title = "{Gaia uncovers difference in B and Be star binarity at small scales: evidence for mass transfer causing the Be phenomenon}",
      journal = {\mnras},
         year = 2024,
        month = jan,
       volume = {527},
       number = {2},
        pages = {3076-3086},
          doi = {10.1093/mnras/stad3105},
archivePrefix = {arXiv},
       eprint = {2310.05653},
 primaryClass = {astro-ph.SR},
       adsurl = {https://ui.adsabs.harvard.edu/abs/2024MNRAS.527.3076D}
}

@inproceedings{pandas,
  title={Data structures for statistical computing in python},
  author={McKinney, Wes and others},
  booktitle={Proceedings of the 9th Python in Science Conference},
  volume={445},
  pages={51--56},
  year={2010},
  organization={Austin, TX}
}

@ARTICLE{theGaiaMission,
       author = {{Gaia Collaboration} and {Prusti}, T. and {de Bruijne}, J.~H.~J. and {Brown}, A.~G.~A. and {Vallenari}, A. and {Babusiaux}, C. and {Bailer-Jones}, C.~A.~L. and {Bastian}, U. and {Biermann}, M. and {Evans}, D.~W. and {Eyer}, L. and {Jansen}, F. and {Jordi}, C. and {Klioner}, S.~A. and {Lammers}, U. and {Lindegren}, L. and {Luri}, X. and {Mignard}, F. and {Milligan}, D.~J. and {Panem}, C. and {Poinsignon}, V. and {Pourbaix}, D. and {Randich}, S. and {Sarri}, G. and {Sartoretti}, P. and {Siddiqui}, H.~I. and {Soubiran}, C. and {Valette}, V. and {van Leeuwen}, F. and {Walton}, N.~A. and {Aerts}, C. and {Arenou}, F. and {Cropper}, M. and {Drimmel}, R. and {H{\o}g}, E. and {Katz}, D. and {Lattanzi}, M.~G. and {O'Mullane}, W. and {Grebel}, E.~K. and {Holland}, A.~D. and {Huc}, C. and {Passot}, X. and {Bramante}, L. and {Cacciari}, C. and {Casta{\~n}eda}, J. and {Chaoul}, L. and {Cheek}, N. and {De Angeli}, F. and {Fabricius}, C. and {Guerra}, R. and {Hern{\'a}ndez}, J. and {Jean-Antoine-Piccolo}, A. and {Masana}, E. and {Messineo}, R. and {Mowlavi}, N. and {Nienartowicz}, K. and {Ord{\'o}{\~n}ez-Blanco}, D. and {Panuzzo}, P. and {Portell}, J. and {Richards}, P.~J. and {Riello}, M. and {Seabroke}, G.~M. and {Tanga}, P. and {Th{\'e}venin}, F. and {Torra}, J. and {Els}, S.~G. and {Gracia-Abril}, G. and {Comoretto}, G. and {Garcia-Reinaldos}, M. and {Lock}, T. and {Mercier}, E. and {Altmann}, M. and {Andrae}, R. and {Astraatmadja}, T.~L. and {Bellas-Velidis}, I. and {Benson}, K. and {Berthier}, J. and {Blomme}, R. and {Busso}, G. and {Carry}, B. and {Cellino}, A. and {Clementini}, G. and {Cowell}, S. and {Creevey}, O. and {Cuypers}, J. and {Davidson}, M. and {De Ridder}, J. and {de Torres}, A. and {Delchambre}, L. and {Dell'Oro}, A. and {Ducourant}, C. and {Fr{\'e}mat}, Y. and {Garc{\'\i}a-Torres}, M. and {Gosset}, E. and {Halbwachs}, J. -L. and {Hambly}, N.~C. and {Harrison}, D.~L. and {Hauser}, M. and {Hestroffer}, D. and {Hodgkin}, S.~T. and {Huckle}, H.~E. and {Hutton}, A. and {Jasniewicz}, G. and {Jordan}, S. and {Kontizas}, M. and {Korn}, A.~J. and {Lanzafame}, A.~C. and {Manteiga}, M. and {Moitinho}, A. and {Muinonen}, K. and {Osinde}, J. and {Pancino}, E. and {Pauwels}, T. and {Petit}, J. -M. and {Recio-Blanco}, A. and {Robin}, A.~C. and {Sarro}, L.~M. and {Siopis}, C. and {Smith}, M. and {Smith}, K.~W. and {Sozzetti}, A. and {Thuillot}, W. and {van Reeven}, W. and {Viala}, Y. and {Abbas}, U. and {Abreu Aramburu}, A. and {Accart}, S. and {Aguado}, J.~J. and {Allan}, P.~M. and {Allasia}, W. and {Altavilla}, G. and {{\'A}lvarez}, M.~A. and {Alves}, J. and {Anderson}, R.~I. and {Andrei}, A.~H. and {Anglada Varela}, E. and {Antiche}, E. and {Antoja}, T. and {Ant{\'o}n}, S. and {Arcay}, B. and {Atzei}, A. and {Ayache}, L. and {Bach}, N. and {Baker}, S.~G. and {Balaguer-N{\'u}{\~n}ez}, L. and {Barache}, C. and {Barata}, C. and {Barbier}, A. and {Barblan}, F. and {Baroni}, M. and {Barrado y Navascu{\'e}s}, D. and {Barros}, M. and {Barstow}, M.~A. and {Becciani}, U. and {Bellazzini}, M. and {Bellei}, G. and {Bello Garc{\'\i}a}, A. and {Belokurov}, V. and {Bendjoya}, P. and {Berihuete}, A. and {Bianchi}, L. and {Bienaym{\'e}}, O. and {Billebaud}, F. and {Blagorodnova}, N. and {Blanco-Cuaresma}, S. and {Boch}, T. and {Bombrun}, A. and {Borrachero}, R. and {Bouquillon}, S. and {Bourda}, G. and {Bouy}, H. and {Bragaglia}, A. and {Breddels}, M.~A. and {Brouillet}, N. and {Br{\"u}semeister}, T. and {Bucciarelli}, B. and {Budnik}, F. and {Burgess}, P. and {Burgon}, R. and {Burlacu}, A. and {Busonero}, D. and {Buzzi}, R. and {Caffau}, E. and {Cambras}, J. and {Campbell}, H. and {Cancelliere}, R. and {Cantat-Gaudin}, T. and {Carlucci}, T. and {Carrasco}, J.~M. and {Castellani}, M. and {Charlot}, P. and {Charnas}, J. and {Charvet}, P. and {Chassat}, F. and {Chiavassa}, A. and {Clotet}, M. and {Cocozza}, G. and {Collins}, R.~S. and {Collins}, P. and {Costigan}, G.},
        title = "{The Gaia mission}",
      journal = {\aap},
         year = 2016,
        month = nov,
       volume = {595},
          eid = {A1},
        pages = {A1},
          doi = {10.1051/0004-6361/201629272},
archivePrefix = {arXiv},
       eprint = {1609.04153},
 primaryClass = {astro-ph.IM},
       adsurl = {https://ui.adsabs.harvard.edu/abs/2016A&A...595A...1G}
}

@ARTICLE{Be-review,
       author = {{Rivinius}, Thomas and {Carciofi}, Alex C. and {Martayan}, Christophe},
        title = "{Classical Be stars. Rapidly rotating B stars with viscous Keplerian decretion disks}",
      journal = {\aapr},
         year = 2013,
        month = oct,
       volume = {21},
          eid = {69},
        pages = {69},
          doi = {10.1007/s00159-013-0069-0},
archivePrefix = {arXiv},
       eprint = {1310.3962},
 primaryClass = {astro-ph.SR},
       adsurl = {https://ui.adsabs.harvard.edu/abs/2013A&ARv..21...69R}
}

@ARTICLE{Cai2022,
       author = {{Cai}, Y. -Z. and {Pastorello}, A. and {Fraser}, M. and {Wang}, X. -F. and {Filippenko}, A.~V. and {Reguitti}, A. and {Patra}, K.~C. and {Goranskij}, V.~P. and {Barsukova}, E.~A. and {Brink}, T.~G. and {Elias-Rosa}, N. and {Stevance}, H.~F. and {Zheng}, W. and {Yang}, Y. and {Atapin}, K.~E. and {Benetti}, S. and {de Boer}, T.~J.~L. and {Bose}, S. and {Burke}, J. and {Byrne}, R. and {Cappellaro}, E. and {Chambers}, K.~C. and {Chen}, W. -L. and {Emami}, N. and {Gao}, H. and {Hiramatsu}, D. and {Howell}, D.~A. and {Huber}, M.~E. and {Kankare}, E. and {Kelly}, P.~L. and {Kotak}, R. and {Kravtsov}, T. and {Lander}, V. Yu. and {Li}, Z. -T. and {Lin}, C. -C. and {Lundqvist}, P. and {Magnier}, E.~A. and {Malygin}, E.~A. and {Maslennikova}, N.~A. and {Matilainen}, K. and {Mazzali}, P.~A. and {McCully}, C. and {Mo}, J. and {Moran}, S. and {Newsome}, M. and {Oparin}, D.~V. and {Padilla Gonzalez}, E. and {Reynolds}, T.~M. and {Shatsky}, N.~I. and {Smartt}, S.~J. and {Smith}, K.~W. and {Stritzinger}, M.~D. and {Tatarnikov}, A.~M. and {Terreran}, G. and {Uklein}, R.~I. and {Valerin}, G. and {Vallely}, P.~J. and {Vozyakova}, O.~V. and {Wainscoat}, R. and {Yan}, S. -Y. and {Zhang}, J. -J. and {Zhang}, T. -M. and {Zheltoukhov}, S.~G. and {Dastidar}, R. and {Fulton}, M. and {Galbany}, L. and {Gangopadhyay}, A. and {Ge}, H. -W. and {Guti{\'e}rrez}, C.~P. and {Lin}, H. and {Misra}, K. and {Ou}, Z. -W. and {Salmaso}, I. and {Tartaglia}, L. and {Xiao}, L. and {Zhang}, X. -H.},
        title = "{Forbidden hugs in pandemic times. III. Observations of the luminous red nova AT 2021biy in the nearby galaxy NGC 4631}",
      journal = {\aap},
         year = 2022,
        month = nov,
       volume = {667},
          eid = {A4},
        pages = {A4},
          doi = {10.1051/0004-6361/202244393},
archivePrefix = {arXiv},
       eprint = {2207.00734},
 primaryClass = {astro-ph.SR},
       adsurl = {https://ui.adsabs.harvard.edu/abs/2022A&A...667A...4C}
}

@article{MacLeod2017,
doi = {10.3847/1538-4357/835/2/282},
url = {https://dx.doi.org/10.3847/1538-4357/835/2/282},
year = {2017},
month = {feb},
publisher = {The American Astronomical Society},
volume = {835},
number = {2},
pages = {282},
author = {MacLeod, Morgan and Macias, Phillip and Ramirez-Ruiz, Enrico and Grindlay, Jonathan and Batta, Aldo and Montes, Gabriela},
title = {Lessons from the Onset of a Common Envelope Episode: the Remarkable M31 2015 Luminous Red Nova Outburst},
journal = {The Astrophysical Journal}
}

@ARTICLE{Pastorello2023,
       author = {{Pastorello}, A. and {Valerin}, G. and {Fraser}, M. and {Reguitti}, A. and {Elias-Rosa}, N. and {Filippenko}, A.~V. and {Rojas-Bravo}, C. and {Tartaglia}, L. and {Reynolds}, T.~M. and {Valenti}, S. and {Andrews}, J.~E. and {Ashall}, C. and {Bostroem}, K.~A. and {Brink}, T.~G. and {Burke}, J. and {Cai}, Y. -Z. and {Cappellaro}, E. and {Coulter}, D.~A. and {Dastidar}, R. and {Davis}, K.~W. and {Dimitriadis}, G. and {Fiore}, A. and {Foley}, R.~J. and {Fugazza}, D. and {Galbany}, L. and {Gangopadhyay}, A. and {Geier}, S. and {Guti{\'e}rrez}, C.~P. and {Haislip}, J. and {Hiramatsu}, D. and {Holmbo}, S. and {Howell}, D.~A. and {Hsiao}, E.~Y. and {Hung}, T. and {Jha}, S.~W. and {Kankare}, E. and {Karamehmetoglu}, E. and {Kilpatrick}, C.~D. and {Kotak}, R. and {Kouprianov}, V. and {Kravtsov}, T. and {Kumar}, S. and {Li}, Z. -T. and {Lundquist}, M.~J. and {Lundqvist}, P. and {Matilainen}, K. and {Mazzali}, P.~A. and {McCully}, C. and {Misra}, K. and {Morales-Garoffolo}, A. and {Moran}, S. and {Morrell}, N. and {Newsome}, M. and {Padilla Gonzalez}, E. and {Pan}, Y. -C. and {Pellegrino}, C. and {Phillips}, M.~M. and {Pignata}, G. and {Piro}, A.~L. and {Reichart}, D.~E. and {Rest}, A. and {Salmaso}, I. and {Sand}, D.~J. and {Siebert}, M.~R. and {Smartt}, S.~J. and {Smith}, K.~W. and {Srivastav}, S. and {Stritzinger}, M.~D. and {Taggart}, K. and {Tinyanont}, S. and {Yan}, S. -Y. and {Wang}, L. and {Wang}, X. -F. and {Williams}, S.~C. and {Wyatt}, S. and {Zhang}, T. -M. and {de Boer}, T. and {Chambers}, K. and {Gao}, H. and {Magnier}, E.},
        title = "{Forbidden hugs in pandemic times. IV. Panchromatic evolution of three luminous red novae}",
      journal = {\aap},
         year = 2023,
        month = mar,
       volume = {671},
          eid = {A158},
        pages = {A158},
          doi = {10.1051/0004-6361/202244684},
archivePrefix = {arXiv},
       eprint = {2208.02782},
 primaryClass = {astro-ph.SR},
       adsurl = {https://ui.adsabs.harvard.edu/abs/2023A&A...671A.158P}
}

@ARTICLE{Fitzpatrick99,
       author = {{Fitzpatrick}, Edward L.},
        title = "{Correcting for the Effects of Interstellar Extinction}",
      journal = {\pasp},
         year = 1999,
        month = jan,
       volume = {111},
       number = {755},
        pages = {63-75},
          doi = {10.1086/316293},
archivePrefix = {arXiv},
       eprint = {astro-ph/9809387},
 primaryClass = {astro-ph},
       adsurl = {https://ui.adsabs.harvard.edu/abs/1999PASP..111...63F}
}

@software{pysynphot,
       author = {{STScI Development Team}},
        title = "{pysynphot: Synthetic photometry software package}",
 howpublished = {Astrophysics Source Code Library, record ascl:1303.023},
         year = 2013,
        month = mar,
          eid = {ascl:1303.023},
archivePrefix = {ascl},
       eprint = {1303.023},
       adsurl = {https://ui.adsabs.harvard.edu/abs/2013ascl.soft03023S}
}

@INPROCEEDINGS{CastelliANDKurucz,
       author = {{Castelli}, F. and {Kurucz}, R.~L.},
        title = "{New Grids of ATLAS9 Model Atmospheres}",
    booktitle = {Modelling of Stellar Atmospheres},
         year = 2003,
       editor = {{Piskunov}, N. and {Weiss}, W.~W. and {Gray}, D.~F.},
       series = {IAU Symposium},
       volume = {210},
        month = jan,
        pages = {A20},
          doi = {10.48550/arXiv.astro-ph/0405087},
archivePrefix = {arXiv},
       eprint = {astro-ph/0405087},
 primaryClass = {astro-ph},
       adsurl = {https://ui.adsabs.harvard.edu/abs/2003IAUS..210P.A20C}
}

@ARTICLE{V1309ScoDiscovery,
       author = {{Mason}, E. and {Diaz}, M. and {Williams}, R.~E. and {Preston}, G. and {Bensby}, T.},
        title = "{The peculiar nova V1309 Scorpii/nova Scorpii 2008. A candidate twin of V838 Monocerotis}",
      journal = {\aap},
         year = 2010,
        month = jun,
       volume = {516},
          eid = {A108},
        pages = {A108},
          doi = {10.1051/0004-6361/200913610},
archivePrefix = {arXiv},
       eprint = {1004.3600},
 primaryClass = {astro-ph.SR},
       adsurl = {https://ui.adsabs.harvard.edu/abs/2010A&A...516A.108M}
}

@ARTICLE{Reguitti2025,
       author = {{Reguitti}, A. and {Pastorello}, A. and {Valerin}, G.},
        title = "{The fate of the progenitors of Luminous Red Novae: Infrared detection of LRNe years after the outburst}",
      journal = {arXiv e-prints},
         year = 2025,
        month = apr,
          eid = {arXiv:2504.14592},
        pages = {arXiv:2504.14592},
          doi = {10.48550/arXiv.2504.14592},
archivePrefix = {arXiv},
       eprint = {2504.14592},
 primaryClass = {astro-ph.SR},
       adsurl = {https://ui.adsabs.harvard.edu/abs/2025arXiv250414592R}
}

@ARTICLE{Be_xray,
       author = {{Gies}, Douglas R. and {Wang}, Luqian and {Klement}, Robert},
        title = "{Gamma Cas Stars as Be+White Dwarf Binary Systems}",
      journal = {\apjl},
         year = 2023,
        month = jan,
       volume = {942},
       number = {1},
          eid = {L6},
        pages = {L6},
          doi = {10.3847/2041-8213/acaaa1},
archivePrefix = {arXiv},
       eprint = {2212.06916},
 primaryClass = {astro-ph.SR},
       adsurl = {https://ui.adsabs.harvard.edu/abs/2023ApJ...942L...6G}
}

@ARTICLE{Erasmus2024,
       author = {{Erasmus}, Nicolas and {Steele}, Iain A. and {Piascik}, Andrzej S. and {Bates}, Stuart D. and {Mottram}, Chris. J. and {Rosie}, Kathryn A. and {van Gend}, Carel H.~D.~R. and {Geen}, Ulrich and {Pretorius}, Magaretha L. and {Potter}, Stephen B. and {Loubser}, Egan and {Koorts}, Willie and {Gajjar}, Hitesh and {Titus}, Keegan and {Worters}, Hannah L. and {Sickafoose}, Amanda A. and {Chandra}, Sunil and {O'Connor}, James E. and {Matlala}, Kgothatso and {Crook-Mansour}, Justine and {Ranjbar}, Ali and {Smith}, Robert J. and {Jermak}, Helen and {Abiodun}, Shalom and {Egbo}, Okwudili D.},
        title = "{Mookodi: multi-purpose low-resolution spectrograph and multi-filter photometric imager for rapid follow-up observations of astronomical transient events}",
      journal = {Journal of Astronomical Telescopes, Instruments, and Systems},
         year = 2024,
        month = apr,
       volume = {10},
          eid = {025005},
        pages = {025005},
          doi = {10.1117/1.JATIS.10.2.025005},
       adsurl = {https://ui.adsabs.harvard.edu/abs/2024JATIS..10b5005E}
}

@INPROCEEDINGS{Worters2016,
       author = {{Worters}, Hannah L. and {O'Connor}, James E. and {Carter}, David B. and {Loubser}, Egan and {Fourie}, Pieter A. and {Sickafoose}, Amanda and {Swanevelder}, Pieter},
        title = "{SAAO's new robotic telescope and WiNCam (Wide-field Nasmyth Camera)}",
    booktitle = {Ground-based and Airborne Instrumentation for Astronomy VI},
         year = 2016,
       editor = {{Evans}, Christopher J. and {Simard}, Luc and {Takami}, Hideki},
       series = {Society of Photo-Optical Instrumentation Engineers (SPIE) Conference Series},
       volume = {9908},
        month = aug,
          eid = {99083Y},
        pages = {99083Y},
          doi = {10.1117/12.2231636},
       adsurl = {https://ui.adsabs.harvard.edu/abs/2016SPIE.9908E..3YW}
}


\begin{appendix} 

\onecolumn
\begin{landscape}
\section{Additional tables} \label{app:tables}
\begin{longtable}{lllllllllll}
\caption{Information of the candidates in our sample: sky position, distance, apparent magnitude, colour, line-of-sight extinction, mass, surface gravity and effective temperature.}\\
\hline \hline 
\label{longtable1:data}

Gaia source ID$^1$ & RA$^1$ & DEC$^1$ & Distance$^3$ & G$^1$ & BP-RP$^1$ & E(B-V)$^2$ & E(B-V) (2024)$^3$ & M$^3$ & $\log\,g$$^3$ & T$_{\mathrm{eff}}$$^3$ \\
& (deg) & (deg) & (parsec) & (mag) & (mag) & (mag) & (mag) & (M$_\odot$) & (cgs) & (K) \\
\hline
\endfirsthead
\caption{continued.}\\
\hline \hline 
Gaia source ID$^1$ & RA$^1$ & DEC$^1$ & Distance$^3$ & G$^1$ & BP-RP$^1$ & E(B-V)$^2$ & E(B-V) (2024)$^3$ & M$^3$ & $\log\,g$$^3$ & T$_{\mathrm{eff}}$$^3$ \\
& (deg) & (deg) & (parsec) & (mag) & (mag) & (mag) & (mag) & (M$_\odot$) & (cgs) & (K) \\
\hline 
\endhead
\hline
\endfoot
187219239343050880 & 78.5501 & 37.0042 & $3091^{+106}_{-191}$ & 13.95 & 1.51 & $0.56^{+0.00}_{-0.12}$ & $1.14 \pm 0.17$ &  $10.01 \pm 4.57$ & $3.90 \pm 0.74$ & $14127 \pm 4248$\\
  428103652673049728 & 7.3050 & 59.3159 & $3380^{+208}_{-157}$ & 14.49 & 1.28 & $0.57^{+0.00}_{-0.05}$  & $0.66 \pm 0.14$ & $6.00 \pm 13.69$ & $3.84 \pm 0.31$ & $9184 \pm 1767$\\
  460686648965862528 & 44.2189 & 57.2332 & $2338^{+45}_{-57}$ & 15.72 & 2.19 & $1.30^{+0.01}_{-0.04}$ & $1.32 \pm 0.10$ & $1.81 \pm 0.50$ & $3.63 \pm 0.31$ & $7258 \pm 523$\\
  461193695624775424 & 44.8338 & 58.7449 & $3214^{+928}_{-146}$ & 10.97 & 1.29 & $0.90^{+0.01}_{-0.22}$ & $1.26 \pm 0.28$ & $9.75 \pm 37.69$ & $3.12 \pm 0.64$ & $34393 \pm 13005$\\
  473575777103322496 & 60.9224 & 60.2468 & $3878^{+109}_{-106}$ & 12.71 & 1.14 & $0.29^{+0.04}_{-0.12}$ & $0.72 \pm 0.14$ & $2.91 \pm 1.13$ & $2.60 \pm 0.60$ & $10371 \pm 713$\\
  508419369310190976 & 30.6300 & 61.0035 & $4365^{+286}_{-952}$ & 11.25 & 0.99 & $0.65^{+0.00}_{-0.01}$ & $0.89 \pm 0.21$ & $7.62 \pm 6.57$ & $3.36 \pm 0.47$ & $19215 \pm 7082$\\
  512721444765993472 & 22.2065 & 63.8494 & $5380^{+396}_{-1055}$ & 11.75 & 1.21 & $0.85^{+0.02}_{-0.05}$ & $1.15 \pm 0.16$ & $12.01 \pm 69.03$ & $3.54 \pm 0.60$ & $18709 \pm 6265$\\
  526939882463713152 & 9.2763 & 63.7070 & $4364^{+947}_{-364}$ & 13.63 & 1.64 & $0.69^{+0.45}_{-0.11}$ & $1.33 \pm 0.18$ & $8.75 \pm 2.86$ & $3.16 \pm 0.49$ & $16675 \pm 2881$\\
  527155253604491392 & 8.1916 & 64.9694 & $3067^{+194}_{-76}$ & 13.24 & 1.86 & $0.76^{+0.00}_{-0.05}$ & $1.03 \pm 0.14$ & $4.38 \pm 1.57$ & $2.38 \pm 0.40$ & $7916 \pm 770$\\
  1870955515858422656 & 312.9137 & 37.5701 & $2136^{+11}_{-79}$ & 14.16 & 1.90 & $0.75^{+0.01}_{-0.09}$ & $0.54 \pm 0.07$ & $1.25 \pm 0.23$ & $2.66 \pm 0.20$ & $4765 \pm 73$\\
  2002117151282819840 & 341.3075 & 53.2094 & $2461^{+20}_{-88}$ & 9.79 & 0.21 & $0.14^{+0.00}_{-0.02}$ & $0.50 \pm 0.11$ & $7.74 \pm 2.06$ & $3.59 \pm 0.44$ & $15929 \pm 1900$\\
  2006088484204609408 & 333.7223 & 55.6014 & $4913^{+318}_{-976}$ & 14.23 & 1.39 & $0.72^{+0.06}_{-0.22}$ & $0.96 \pm 0.14$ & $5.85 \pm 5.01$ & $3.46 \pm 0.37$ & $13909 \pm 3858$\\
  2006912396372680960 & 339.2652 & 57.1667 & $2455^{+35}_{-33}$ & 12.81 & 1.31 & $0.43^{+0.00}_{-0.03}$ & $0.80 \pm 0.16$ & $5.97 \pm 0.74$ & $3.75 \pm 0.27$ & $9145 \pm 1079$\\
  2007318661608788096 & 341.0511 & 57.8325 & $3240^{+612}_{-137}$ & 15.61 & 1.82 & $0.72^{+0.01}_{-0.06}$ & $0.65 \pm 0.10$ & $1.36 \pm 0.29$ & $2.85 \pm 0.25$ & $5072 \pm 134$\\
  2013187240507011456 & 343.6522 & 58.0566 & $3153^{+178}_{-431}$ & 14.35 & 1.87 & $1.26^{+0.04}_{-0.26}$ & $0.98 \pm 0.19$ & $5.53 \pm 27.20$ & $3.39 \pm 0.35$ & $8676 \pm 1173$\\
  2027563492489195520 & 298.5404 & 27.3946 & $7715^{+1371}_{-2552}$ & 14.89 & 2.23 & $1.34^{+0.03}_{-0.01}$ & $1.52 \pm 0.15$ & $6.20 \pm 1.51$ & $2.56 \pm 0.63$ & $10210 \pm 1630$\\
  2030965725082217088 & 300.0884 & 31.6290 & $2677^{+2299}_{-42}$ & 13.21 & 1.75 & $1.15^{+0.11}_{-0.05}$ & $1.12 \pm 0.16$ & $5.45 \pm 1.43$ & $3.48 \pm 0.54$ & $10482 \pm 799$\\
  2054338249889402880 & 305.1283 & 32.5679 & $3091^{+328}_{-170}$ & 13.92 & 2.30 & $1.65^{+0.07}_{-0.08}$ & $1.78 \pm 0.50$ & $7.22 \pm 2.78$ & $3.82 \pm 0.55$ & $19691 \pm 2232$\\
  2060841448854265216 & 303.4874 & 37.5023 & $3668^{+30}_{-90}$ & 14.57 & 2.30 & $1.44^{+0.00}_{-0.01}$ & $1.40 \pm 0.15$ & $3.51 \pm 0.56$ & $2.17 \pm 0.32$ & $5089 \pm 173$\\
  2061252975440642816 & 304.7922 & 38.2985 & $3869^{+1742}_{-948}$ & 16.33 & 2.13 & $0.97^{+0.07}_{-0.03}$ & $1.62 \pm 0.15$ & $2.07 \pm 1.14$ & $3.32 \pm 0.39$ & $6629 \pm 614$\\
  2074693061975359232 & 302.9374 & 41.7822 & $2242^{+8}_{-49}$ & 13.23 & 1.10 & $0.44^{+0.02}_{-0.01}$ & $0.62 \pm 0.17$ & $3.69 \pm 0.60$ & $3.77 \pm 0.25$ & $8711 \pm 628$\\
  2083649030845658624 & 307.8765 & 47.9782 & $2554^{+63}_{-3}$ & 15.64 & 1.98 & $0.89^{+0.01}_{-0.00}$ & $0.82 \pm 0.18$ & $1.39 \pm 0.37$ & $3.15 \pm 0.33$ & $4912 \pm 216$\\
  2164630463117114496 & 319.5385 & 47.5318 & $3023^{+233}_{-49}$ & 13.41 & 1.77 & $0.79^{+0.11}_{-0.04}$ & $1.39 \pm 0.16$ & $5.47 \pm 1.04$ & $3.86 \pm 0.65$ & $11948 \pm 1467$\\
  2166378312964576256 & 312.6099 & 46.0180 & $3524^{+70}_{-128}$ & 14.63 & 2.52 & $1.20^{+0.00}_{-0.00}$ & $1.88 \pm 0.15$ & $5.97 \pm 0.81$ & $2.43 \pm 0.38$ & $11232 \pm 756$\\
  2169083008475385856 & 317.1013 & 50.1992 & $3797^{+2776}_{-528}$ & 13.83 & 3.22 & $2.21^{+0.31}_{-0.07}$ & $2.24 \pm 0.31$ & $5.63 \pm 1.24$ & $1.71 \pm 0.49$ & $6603 \pm 717$\\
  2173852964799716480 & 326.9270 & 53.9998 & $6441^{+336}_{-305}$ & 13.65 & 1.80 & $1.26^{+0.01}_{-0.08}$ & $1.46 \pm 0.16$ & $10.61 \pm 4.60$ & $3.26 \pm 0.91$ & $13979 \pm 2477$\\
  2175699216614191360 & 319.3053 & 53.4088 & $4169^{+82}_{-80}$ & 14.98 & 2.52 & $1.73^{+0.09}_{-0.03}$ & $1.67 \pm 0.12$ & $4.64 \pm 1.03$ & $2.38 \pm 0.47$ & $9464 \pm 557$\\
  2200433413577635840 & 336.7675 & 59.0930 & $3097^{+252}_{-170}$ & 12.23 & 1.67 & $1.19^{+0.01}_{-0.02}$ & $1.32 \pm 0.19$ & $10.85 \pm 3.47\times10^7$ & $3.26 \pm 0.63$ & $19734 \pm 3965$\\ 
  2934216142176785920 & 107.0585 & -18.5093 & $3113^{+37}_{-14}$ & 9.60 & 1.66 & $0.81^{+0.00}_{-0.02}$ & $0.41 \pm 0.07$ & $5.14 \pm 1.60$ & $1.37 \pm 0.41$ & $4588 \pm 357$\\
  3355776901779440384 & 99.6750 & 13.7071 & $6793^{+126}_{-1243}$ & 9.71 & 0.41 & $0.22^{+0.05}_{-0.03}$ & $0.30 \pm 0.13$ & $10.42 \pm 1.33$ & $3.47 \pm 0.53$ & $20345 \pm 3824$\\
  3369399099232812160 & 97.8989 & 16.4806 & $3311^{+495}_{-48}$ & 11.65 & 0.64 & $0.42^{+0.03}_{-0.00}$ & $0.63 \pm 0.13$ & $6.89 \pm 0.71$ & $3.42 \pm 0.59$ & $10909 \pm 1593$\\
  3444168325163139840 & 86.0562 & 29.1531 & $2014^{+109}_{-172}$ & 10.53 & 1.03 & $0.48^{+0.07}_{-0.13}$ & $1.03 \pm 0.28$ & $11.96 \pm 50.69$ & $3.81 \pm 0.53$ & $23089 \pm 11545$\\
  4054010697162430592 & 266.6885 & -32.6311 & $1131^{+33}_{-10}$ & 9.71 & 1.43 & $0.44^{+0.01}_{-0.04}$ & $0.96 \pm 4.55$ & $4.02 \pm 0.78$ & $1.84 \pm 0.33$ & $7136 \pm 718$\\
  4076568861833452160 & 280.5377 & -24.5018 & $5576^{+25}_{-86}$ & 11.84 & 1.17 & $0.36^{+0.01}_{-0.01}$ & $0.38 \pm 0.13$ & $1.85 \pm 1.23$ & $1.68 \pm 1.24$ & $5623 \pm 528$\\
  4094491141885400576 & 275.0983 & -19.7796 & $5890^{+180}_{-105}$ & 13.27 & 2.59 & $1.53^{+0.02}_{-0.01}$ & $1.65 \pm 0.18$ & $7.40 \pm 1.50$ & $1.72 \pm 0.53$ & $9585 \pm 1492$\\
  4096527235637366912 & 275.4238 & -17.6223 & $1970^{+34}_{-578}$ & 10.36 & 0.74 & $0.40^{+0.15}_{-0.02}$ & $0.58 \pm 0.13$ & $4.45 \pm 1.47$ & $3.63 \pm 0.39$ & $14307 \pm 2117$\\
  4260141158544875008 & 280.6910 & -1.3850 & $3172^{+161}_{-540}$ & 15.61 & 3.11 & $1.79^{+0.07}_{-0.05}$ & $1.43 \pm 0.07$ & $1.51 \pm 0.29$ & $2.19 \pm 0.27$ & $4485 \pm 89$\\
  4263591911398361472 & 289.9265 & -0.0499 & $1031^{+8}_{-4}$ & 8.57 & 0.44 & $0.24^{+0.01}_{-0.00}$ & $0.36 \pm 0.19$ & $3.95 \pm 1.38$ & $3.42 \pm 0.39$ & $10853 \pm 2581$\\
  4271992661242707200 & 278.4116 & -1.1951 & $1260^{+96}_{-7}$ & 15.59 & 3.89 & $2.67^{+0.07}_{-0.01}$ & $2.22 \pm 0.11$ & $1.40 \pm 0.29$ & $2.55 \pm 0.18$ & $4744 \pm 124$\\
  4272588356022299520 & 278.9687 & -0.0870 & $1904^{+118}_{-138}$ & 15.40 & 3.74 & $2.34^{+0.08}_{-0.03}$ & $1.90 \pm 0.10$ & $1.66 \pm 0.39$ & $1.84 \pm 0.26$ & $4510 \pm 124$\\
  4281886474885416064 & 282.0665 & 4.0900 & $1036^{+162}_{-46}$ & 15.34 & 3.05 & $2.01^{+0.02}_{-0.25}$ & $1.39 \pm 0.12$ & $1.07 \pm 0.18$ & $2.99 \pm 0.16$ & $4617 \pm 130$\\
  4299904519833646080 & 298.9082 & 8.9623 & $7318^{+770}_{-1579}$ & 9.58 & 0.55 & $0.23^{+0.01}_{-0.06}$ & $0.13 \pm 0.07$ & $2.86 \pm 0.56$ & $3.00 \pm 0.71$ & $6836 \pm 226$\\
  4321276689423536384 & 292.2191 & 14.8869 & $2927^{+611}_{-155}$ & 15.12 & 2.78 & $1.49^{+0.02}_{-0.09}$ & $1.26 \pm 0.09$ & $1.71 \pm 0.40$ & $2.14 \pm 0.27$ & $4598 \pm 128$\\
  4515124540765147776 & 290.2013 & 18.4955 & $2496^{+56}_{-8}$ & 16.48 & 4.09 & $2.62^{+0.01}_{-0.00}$ & $1.96 \pm 0.17$ & $1.39 \pm 0.35$ & $1.97 \pm 0.43$ & $3879 \pm 107$\\
  4519475166529738112 & 286.9374 & 20.3501 & $4917^{+80}_{-1218}$ & 13.29 & 2.71 & $1.45^{+0.04}_{-0.27}$ & $0.81 \pm 0.26$ & $1.12 \pm 0.32$ & $0.74 \pm 0.31$ & $3764 \pm 92$\\
  5311969857556479616 & 138.7208 & -51.7624 & $3876^{+156}_{-173}$ & 14.33 & 1.92 & $0.76^{+0.01}_{-0.01}$ & $1.49 \pm 0.17$ & $6.73 \pm 2.96$ & $3.10 \pm 0.41$ & $15212 \pm 3692$\\
  5323384162646755712 & 134.6810 & -53.1388 & $3757^{+1528}_{-152}$ & 16.00 & 1.75 & $0.85^{+0.15}_{-0.28}$ & $0.69 \pm 0.10$ & $1.42 \pm 0.42$ & $2.70 \pm 0.22$ & $5446 \pm 131$\\
  5326288831829996416 & 140.1741 & -48.1982 & $2595^{+125}_{-60}$ & 17.25 & 3.37 & $2.06^{+0.00}_{-0.00}$ & $1.85 \pm 0.14$ & $1.75 \pm 0.42$ & $2.73 \pm 0.19$ & $4818 \pm 123$\\
  5328449200388495616 & 133.0337 & -48.5787 & $4153^{+48}_{-6}$ & 16.19 & 1.98 & $0.90^{+0.01}_{-0.01}$ & $0.75 \pm 0.10$ & $1.26 \pm 0.36$ & $2.62 \pm 0.24$ & $5025 \pm 109$\\
  5338183383022960512 & 163.8516 & -60.8902 & $5858^{+275}_{-351}$ & 14.55 & 1.30 & $0.82^{+0.14}_{-0.13}$ & $0.92 \pm 0.18$ & $3.82 \pm 2.86$ & $2.66 \pm 0.37$ & $10494 \pm 2176$\\
  5350869719969619840 & 162.5043 & -58.2992 & $2457^{+78}_{-181}$ & 10.54 & 1.00 & $0.69^{+0.05}_{-0.17}$ & $0.61 \pm 0.12$ & $3.71 \pm 7.33$ & $2.13 \pm 0.69$ & $7552 \pm 626$\\
  5524022735225482624 & 130.8649 & -43.5837 & $2594^{+56}_{-25}$ & 13.70 & 2.33 & $1.52^{+0.01}_{-0.02}$ & $1.88 \pm 0.16$ & $5.43 \pm 1.31$ & $3.55 \pm 0.79$ & $13174 \pm 1351$\\
  5593826360487373696 & 117.0976 & -34.3848 & $4244^{+219}_{-114}$ & 12.01 & 2.36 & $1.64^{+0.04}_{-0.02}$ & $1.56 \pm 0.55$ & $8.33 \pm 125.53$ & $2.02 \pm 0.50$ & $11664 \pm 1513$\\
  5599309216965305728 & 114.9433 & -29.8086 & $3229^{+416}_{-455}$ & 10.41 & 0.25 & $0.17^{+0.04}_{-0.05}$ & $0.42 \pm 0.07$ & $5.83 \pm 1.17$ & $2.72 \pm 0.63$ & $12988 \pm 1080$\\
  5617186348318629248 & 107.9070 & -25.5237 & $2845^{+184}_{-404}$ & 14.06 & 1.27 & $0.30^{+0.00}_{-0.04}$ & $0.33 \pm 0.08$ & $1.12 \pm 0.20$ & $2.82 \pm 0.22$ & $5309 \pm 167$\\
  5866345647515558400 & 209.6089 & -61.6174 & $2501^{+20}_{-4}$ & 14.97 & 3.10 & $1.85^{+0.01}_{-0.05}$ & $1.46 \pm 0.12$ & $1.76 \pm 0.40$ & $1.96 \pm 0.21$ & $4391 \pm 78$\\
  5866474526572151936 & 211.2822 & -60.8993 & $2710^{+20}_{-197}$ & 12.39 & 1.41 & $0.97^{+0.06}_{-0.44}$ & $1.16 \pm 0.27$ & $11.88 \pm 26.53$ & $3.13 \pm 0.41$ & $30216 \pm 7958$\\
  5868425648663616768 & 201.6601 & -61.9116 & $3706^{+595}_{-8}$ & 15.68 & 3.25 & $1.93^{+0.03}_{-0.13}$ & $1.67 \pm 0.13$ & $1.87 \pm 0.67$ & $1.89 \pm 0.25$ & $4470 \pm 96$\\
  5880159842877908352 & 228.1479 & -57.9014 & $2948^{+64}_{-70}$ & 14.10 & 2.87 & $2.24^{+0.02}_{-0.18}$ & $1.65 \pm 0.10$ & $2.81 \pm 0.52$ & $1.95 \pm 0.36$ & $7429 \pm 293$\\
  5882737819707242240 & 235.7697 & -56.5290 & $2067^{+213}_{-58}$ & 11.56 & 1.31 & $0.51^{+0.06}_{-0.02}$ & $1.16 \pm 0.22$ & $6.38 \pm 1.45$ & $3.29 \pm 0.52$ & $17763 \pm 5569$\\
  5962956195185292288 & 257.2450 & -46.5970 & $3989^{+129}_{-77}$ & 13.62 & 3.80 & $2.48^{+0.02}_{-0.01}$ & $1.76 \pm 0.20$ & $1.60 \pm 0.48$ & $0.86 \pm 0.32$ & $4340 \pm 162$\\
  5965503866703572480 & 256.8225 & -42.4636 & $3090^{+30}_{-72}$ & 14.13 & 2.36 & $1.71^{+0.00}_{-0.01}$ & $1.73 \pm 0.15$ & $7.65 \pm 7.00$ & $3.29 \pm 0.44$ & $18218 \pm 5978$\\
  5966574133883852416 & 254.5209 & -41.4761 & $2496^{+161}_{-736}$ & 17.00 & 3.04 & $2.02^{+0.32}_{-0.09}$ & $1.77 \pm 0.15$ & $1.96 \pm 0.70$ & $3.14 \pm 0.29$ & $5742 \pm 268$\\
  6052463412400724992 & 248.7806 & -20.7259 & $5158^{+63}_{-1054}$ & 16.50 & 1.75 & $0.66^{+0.02}_{-0.04}$ & $0.47 \pm 0.08$ & $0.91 \pm 0.11$ & $3.56 \pm 0.34$ & $5029 \pm 157$\\
  6053890788968694656 & 189.2723 & -62.4698 & $3559^{+62}_{-223}$ & 14.07 & 2.37 & $1.58^{+0.00}_{-0.00}$ & $1.80 \pm 0.15$ & $8.59 \pm 4.78$ & $3.28 \pm 0.55$ & $17548 \pm 2141$\\
  6123873398383875456 & 212.0047 & -33.2620 & $1316^{+12}_{-139}$ & 11.56 & 0.64 & $0.07^{+0.09}_{-0.01}$ & $0.04 \pm 0.02$ & $1.58 \pm 0.26$ & $3.75 \pm 0.19$ & $6469 \pm 106$\\
  6228685649971375616 & 223.1954 & -25.7899 & $7094^{+718}_{-3144}$ & 15.15 & 0.83 & $0.13^{+0.07}_{-0.05}$ & $0.08 \pm 0.03$ & $1.01 \pm 0.14$ & $3.75 \pm 0.17$ & $6112 \pm 113$\\
\end{longtable}
\tablebib{(1)~\citet{GaiaDR3};
(2) \citet{Starhorse}; (3) \citet{SHBoost}}
\end{landscape}
\twocolumn

\begin{table*}[h]
\caption{Observation date, instrument and instrument details for the follow-up spectra obtained for a subsample of our candidates. }             
\label{table:spec}      
\begin{tabular}{cllllll}
\hline \hline 
Index & Gaia DR3 ID & Date & Telescope/Instr. & Setup & Range & R \\
& & & & & (\AA{}) & \\ 
\hline
  1 & 1870955515858422656 & 2024-04-14 03:38:36 & CAHA/CAFOS & G-200/1.5$^{\prime\prime}$ Slit & 4000--8500 & ${\approx}$300\\
  2 & 187219239343050880 & 2024-10-30 03:59:50 & NOT/ALFOSC & G-19/0.5$^{\prime\prime}$ Slit & 4400--6950 & 1940\\
    & 187219239343050880 & 2024-10-30 05:17:55 & NOT/ALFOSC & G-7/1$^{\prime\prime}$ Slit & 3650-7110 & 650\\
    & 187219239343050880 & 2012-02-19 12:01:14 & LAMOST & Low res & 3700--9000 & 1800\\
    & 187219239343050880 & 2014-01-29 12:16:59 & LAMOST & Low res & 3700--9000 & 1800\\
    & 187219239343050880 & 2014-10-17 19:17:55 & LAMOST & Low res & 3700--9000 & 1800\\
    & 187219239343050880 & 2024-10-29 04:05:44 & NOT/FIES-L & Low res & 3700--8300 & 25000\\
    & 187219239343050880 & 2024-10-29 04:36:32 & NOT/FIES-L & Low res & 3700--8300 & 25000\\
    & 187219239343050880 & 2024-10-30 02:52:46 & NOT/FIES-L & Low res & 3700--8300 & 25000\\
    & 187219239343050880 & 2024-10-30 03:23:33 & NOT/FIES-L & Low res & 3700--8300 & 25000\\
  3 & 2006088484204609408 & 2024-10-30 00:11:18 & NOT/ALFOSC  & G-7/1$^{\prime\prime}$ Slit & 3650-7110 & 650\\
    & 2006088484204609408 & 2024-10-28 20:57:13 & NOT/FIES-L & Low res & 3700--8300 & 25000\\
    & 2006088484204609408 & 2024-10-28 23:05:09 & NOT/FIES-L & Low res & 3700--8300 & 25000\\
    & 2006088484204609408 & 2024-10-29 00:33:53 & NOT/FIES-L & Low res & 3700--8300 & 25000\\
    & 2006088484204609408 & 2024-10-29 21:24:41 & NOT/FIES-L & Low res & 3700--8300 & 25000\\
    & 2006088484204609408 & 2024-10-29 21:55:28 & NOT/FIES-L & Low res & 3700--8300 & 25000\\
  4 & 2013187240507011456 & 2024-04-14 03:45:35 & CAHA/CAFOS & G-200/1.5$^{\prime\prime}$ Slit & 4000--8500 & ${\approx}$300\\
  5 & 2030965725082217088 & 2025-06-02 03:17:24 & NOT/ALFOSC & G-4/1$^{\prime\prime}$ Slit & 3200--9600 & 360\\
  6 & 2060841448854265216 & 2024-04-14 03:55:16 & CAHA/CAFOS & G-200/1.5$^{\prime\prime}$ Slit & 4000--8500 & ${\approx}$300\\
  7 & 2061252975440642816 & 2024-10-29 19:52:13 & NOT/ALFOSC & G-19/0.5$^{\prime\prime}$ Slit & 4400--6950 & 1940\\
    & 2061252975440642816 & 2024-10-29 20:02:23 & NOT/ALFOSC & G-19/0.5$^{\prime\prime}$ Slit & 4400--6950 & 1940\\
  8 & 2074693061975359232 & 2024-04-13 04:06:39 & CAHA/CAFOS & G-200/1.5$^{\prime\prime}$ Slit & 4000--8500 & ${\approx}$300\\
  9 & 2164630463117114496 & 2024-04-14 03:25:02 & CAHA/CAFOS & G-200/1.5$^{\prime\prime}$ Slit & 4000--8500 & ${\approx}$300\\
  10& 2166378312964576256 & 2025-06-02 02:56:27 & NOT/ALFOSC & G-4/1$^{\prime\prime}$ Slit & 3200--9600 & 360\\
  11& 2175699216614191360 & 2024-10-30 00:02:17 & NOT/ALFOSC & G-7/1$^{\prime\prime}$ Slit & 3650-7110 & 650\\
    & 2175699216614191360 & 2025-06-02 03:03:37 & NOT/ALFOSC & G-4/1$^{\prime\prime}$ Slit & 3200--9600 & 360\\
  12& 2200433413577635840 & 2024-04-14 04:10:48 & CAHA/CAFOS & G-200/1.5$^{\prime\prime}$ Slit & 4000--8500 & ${\approx}$300\\
    & 2200433413577635840 & 2024-05-27 01:16:01 & NOT/FIES-M & Med res & 3700--8300 & 46000\\
  13& 3355776901779440384 & 2024-10-29 05:46:33 & NOT/FIES-L & Low res & 3700--8300 & 25000\\
  14& 3369399099232812160 & 2024-10-30 06:03:12 & NOT/ALFOSC & G-4/0.5$^{\prime\prime}$ Slit & 3200--9600 & 710\\
    & 3369399099232812160 & 2024-10-29 05:11:32 & NOT/FIES-L & Low res & 3700--8300 & 25000\\
    & 3369399099232812160 & 2024-10-29 05:27:19 & NOT/FIES-L & Low res & 3700--8300 & 25000\\
    & 3369399099232812160 & 2024-10-30 04:37:57 & NOT/FIES-L & Low res & 3700--8300 & 25000\\
    & 3369399099232812160 & 2024-10-30 04:53:44 & NOT/FIES-L & Low res & 3700--8300 & 25000\\
  15& 3444168325163139840 & 2024-10-29 01:32:19 & NOT/FIES-L & Low res & 3700--8300 & 25000\\
  16& 4076568861833452160 & 2024-05-27 00:37:58 & NOT/FIES-M & Med res & 3700--8300 & 46000\\
  17& 428103652673049728 & 2024-10-30 00:27:35 & NOT/ALFOSC & G-7/1$^{\prime\prime}$ Slit & 3650-7110 & 650\\
  18& 4299904519833646080 & 2024-05-26 02:21:19 & NOT/FIES-M & Med res & 3700--8300 & 46000\\
  19& 461193695624775424 & 2024-10-29 03:50:56 & NOT/FIES-L & Low res & 3700--8300 & 25000\\
  20& 473575777103322496 & 2024-10-30 00:38:10 & NOT/ALFOSC & G-7/1$^{\prime\prime}$ Slit & 3650-7110 & 650\\
  21& 508419369310190976 & 2024-10-29 01:17:57 & NOT/FIES-L & Low res & 3700--8300 & 25000\\ 
  22& 512721444765993472 & 2024-10-30 04:08:31 & NOT/ALFOSC & G-4/0.5$^{\prime\prime}$ Slit & 3200--9600 & 710\\
    & 512721444765993472 & 2024-10-29 22:29:36 & NOT/FIES-L & Low res & 3700--8300 & 25000\\
    & 512721444765993472 & 2024-10-29 22:40:23 & NOT/FIES-L & Low res & 3700--8300 & 25000\\
  23& 526939882463713152 & 2024-10-30 01:39:21 & NOT/ALFOSC & G-4/0.5$^{\prime\prime}$ Slit & 3200--9600 & 710\\
    & 526939882463713152 & 2024-10-29 01:49:16 & NOT/FIES-L & Low res & 3700--8300 & 25000\\
    & 526939882463713152 & 2024-10-29 02:20:04 & NOT/FIES-L & Low res & 3700--8300 & 25000\\
    & 526939882463713152 & 2024-10-30 01:46:22 & NOT/FIES-L & Low res & 3700--8300 & 25000\\
    &  526939882463713152 & 2024-10-30 02:17:09 & NOT/FIES-L & Low res & 3700--8300 & 25000\\
  24& 527155253604491392 & 2024-10-30 00:21:14 & NOT/ALFOSC & G-7/1$^{\prime\prime}$ Slit & 3650-7110 & 650\\
  25& 6123873398383875456 & 2024-05-11 23:09:48 & Mercator/HERMES &  & 3770--9000 & 85000\\
    & 6123873398383875456 & 2024-05-13 23:03:37 & Mercator/HERMES &  & 3770--9000 & 85000\\
  26& 5962956195185292288 & 2025-10-29 18:00:00 & Lesedi/Mookodi & $4^{\prime\prime}$ Slit & 4000--8000 & ${\approx}$175\\

\hline
\end{tabular}
\tablefoot{The date of observations is in format YYYY-MM-DD HH:MM:SS UT. We aimed to achieve a signal-to-noise ratio (SNR) greater than 80 for the low-resolution spectra (CAFOS, ALFOSC, and Mookodi), and an SNR above 40 for the higher-resolution spectra (FIES and HERMES).}
\end{table*}

\onecolumn
\begin{landscape}
\begin{longtable}{lllllll}
\caption{Summary of our results, where the spectral classification and properties observed in the light curve and spectra of our candidates, together with the periods found, are reported.}\\
\hline \hline 
\label{longtable1:results}

Gaia DR3 ID & Spec. class & Properties & Gaia-Xmatch Class & Simbad Class & Period & Period  \\
& (Gaia) & & (Subclass) & (Subclass) &  (this work) & (Gaia-Xmatch) \\
& & & & & (days) & (days) \\
\hline
\endfirsthead
\caption{continued.}\\
\hline \hline
Gaia DR3 ID & Spec. class & Properties & Gaia-Xmatch Class & Simbad Class & Period & Period  \\
& (Gaia) & & (Subclass) & (Subclass) &  (this work) & (Gaia-Xmatch)\\
& & & & & (days) & (days) \\
\hline
\endhead
\hline
\endfoot
  187219239343050880 & B (B) & Diff,SR,H$\alpha$ PCyg & RAD\_VEL\_VAR &  &  & 6.02047\\
  428103652673049728 & Be (G)  & Diff,H$\alpha$ emi &  & EmLine* (*|Em*) &  &\\
  460686648965862528 &  &  Sin,Diff & DSCT & delSctV* (dS*) & 0.15669 & 0.15669 \\
  461193695624775424 & Be (B) &  Sin,2p H$\alpha$ emi &  & Star (*|NIR) & 0.39407 & \\
  473575777103322496 & Be &  Sin,Outb,Far UV,2p H$\alpha$ emi &  &  & 0.45878 &  \\
  508419369310190976 & B &  Sin,Outb,H$\alpha$ abs &  & EmLine* (*|Em*|NIR) & 0.4011 , 1.4682 &  \\
  512721444765993472 & Oe (O) &  Sin,2p H$\alpha$ emi &  & Star (*|NIR) & 0.57251 & \\
  526939882463713152 & Be (B) &  Outb,2p H$\alpha$ emi &  &  &  & \\
  527155253604491392 & Be (B) &  Puls,Diff,2p H$\alpha$ emi & L & SB* (*|Em*|NIR|SB*) & 1.00037 & 149.15505\\
  1870955515858422656 & K (K) &  Sin,Diff &  &  & 23.10592 &  \\
  2002117151282819840 & (B) &  ECL,SR &  & EclBin (*|EB*|Em*|NIR) & 2.58051 &  \\
  2006088484204609408 & Be (B) &  ECL,Diff,SR,2p H$\alpha$ emi & ECL (EA) & EclBin (*|EB*) & 2.59606 & 5.19160 \\
  2006912396372680960 & (G) &  Outb & OMIT & Variable* (*|MIR|NIR|Opt|V*) &  & 1488.28626* \\
  2007318661608788096 & (K) &  Diff & SOLAR\_LIKE (RS) & RSCVnV* (*|MIR|NIR|Opt|RS*|V*) &  & 12.32042 \\
  2013187240507011456 & Oe (O) & Sin,Diff,SR,H$\alpha$ emi &  & EmLine* (*|Em*) & 0.35344 , 0.6081 &  \\
  2027563492489195520 & (B) &  Outb & OMIT & Variable* (*|MIR|NIR|Opt|V*) &  & 224.60175*  \\
  2030965725082217088 & Be (B) &  Sin,SR,H$\alpha$ emi &  &  & 0.43859 &   \\
  2054338249889402880 &  &  & YSO & EmLine* (*|Em*) &  &   \\
  2060841448854265216 & G (G) &  ECL,H$\alpha$ emi & ECL (EW) & EclBin (*|EB*|MIR|NIR|Opt|V) & 60.54500 & 60.70308  \\
  2061252975440642816 & ? (B) &  ECL,H$\alpha$ abs & ECL (EA) & ** (*|**|MIR|NIR|Opt|V*) & 2.13918 & 2.13901 \\
  2074693061975359232 & Be (G) &  Sin,Diff,SR,H$\alpha$ emi &  & EmLine* (*|Em*) & 0.11275,  0.1328 &   \\
  2083649030845658624 & (G) &  Hard X-ray &  &  &  &   \\
  2164630463117114496 & ? (B) &  Sin,Diff,SR,H$\alpha$ emi &  & EmLine* (*|Em*) & 0.59151 &   \\
  2166378312964576256 & Be (B) &  ECL,H$\alpha$ emi & ECL (EA) & EclBin (*|EB*|Em*|MIR|NIR|Opt|V*) & 72.42781 & 72.56402  \\
  2169083008475385856 & (O) &   &  &  &  &   \\
  2173852964799716480 & (O) & Outb &  &  &  &   \\
  2175699216614191360 & ? (B) &  Sin,Diff,H$\alpha$ emi & OMIT & EmLine* (*|Em*) & 4.76998 & 61.82265*  \\
  2200433413577635840 & Be (O) &  Sin,Outb,2p H$\alpha$ emi &  & LPV*\_Candidate (*|Em*|LP?|NIR) & 0.55163 &   \\
  2934216142176785920 & (K) &  Diff & L & LPV* (*|LP*|NIR|V*|V*?) &  &   \\
  3355776901779440384 & Be (B) &  Sin,SR,2p H$\alpha$ emi &  & Star (*|NIR|UV) & 0.30354 &   \\
  3369399099232812160 & Be (B) &  Sin,2p H$\alpha$ emi & OMIT & Star (*|NIR) & 0.47404 & 554.63452*  \\
  3444168325163139840 & Oe (O) &  Sin,Outb,2p H$\alpha$ emi &  & EmLine* (*|Em*|NIR|Opt) & 0.52602 &   \\
  4054010697162430592 & (B) & Sin,SR,Hard X-ray &  & EmLine* (*|Em*|NIR|SB*) & 0.56017 &   \\
  4076568861833452160 & F (F) & Sin,H$\alpha$ abs & CEP (CW) &  & 28.45017 & 28.45893  \\
  4094491141885400576 & (O) &  &  & LPV*\_Candidate (*|IR|LP?|MIR|NIR) &  &   \\
  4096527235637366912 & (B) &  & GCAS & Star (*|NIR) &  &   \\
  4260141158544875008 &  &  &  &  &  &   \\
  4263591911398361472 & (B) & Sin,SR & CST & Be* (*|Be*|Em*|NIR|UV) & 1.43548 &   \\
  4271992661242707200 & (K) &  & LPV (SR) & LPV* (LP*) &  & 60.73273*  \\
  4272588356022299520 & (O) &  &  &  &  &   \\
  4281886474885416064 & (K) & Diff & LPV (SR) & LPV* (LP*) &  & 46.35025  \\
  4299904519833646080 & A-F, shell? (F) & H$\alpha$ abs & OMIT & Star (*|NIR) &  &   \\
  4321276689423536384 &  & Diff,SR &  & LPV*\_Candidate (*|LP?|NIR) &  &   \\
  4515124540765147776 &  & Diff &  & LPV*\_Candidate (*|C*?|LP?|NIR) &  &   \\
  4519475166529738112 & (K) &  & OMIT & LPV*\_Candidate (*|LP?|NIR) &  &   \\
  5311969857556479616 & (B) & ECL & ECL (EB) & EclBin (*|EB*) & 13.74443 & 13.7453  \\
  5323384162646755712 &  & ECL &  &  & 2.79201 &   \\
  5326288831829996416 & (K) &  &  &  &  &   \\
  5328449200388495616 & (G) &  &  &  &  &   \\
  5338183383022960512 & (G) &  &  &  &  &   \\
  5350869719969619840 & (B) & Sin &  & EmLine* (*|Em*|NIR) & 0.56011 &   \\
  5524022735225482624 & (O) & ECL & OMIT & EclBin (*|EB*) & 7.98856 &   \\
  5593826360487373696 & (F) &  & OMIT & Variable* (*|MIR|NIR|Opt|V*) &  &   \\
  5599309216965305728 & (B) & Sin & OMIT & Star (*|NIR) & 0.47592 &   \\
  5617186348318629248 & (G) & Sin &  &  & 3.87180 &   \\
  5866345647515558400 & (K) & Diff,Hard X-ray &  &  &  &   \\
  5866474526572151936 & (O) & Outb & OMIT &  &  &   \\
  5868425648663616768 & (K) &  &  &  &  &   \\
  5880159842877908352 & (B) & Sin &  &  & 31.77066 &   \\
  5882737819707242240 & (B) & Outb & OMIT & Star (*|NIR) &  &   \\
  5962956195185292288 & M (M) & Sin,Diff & OMIT & LPV* (*|C*?|LP*|NIR) & 106.75990 &  \\ 
  5965503866703572480 & (O) & Sin &  &  & 0.15252 &  \\
  5966574133883852416 & (K) & Diff &  &  &  &   \\
  6052463412400724992 & (K) & Diff &  &  &  &   \\
  6053890788968694656 & (O) & Sin &  &  & 0.22931 &   \\
  6123873398383875456 & F (F) & Sin,H$\alpha$ abs & RR (RRC) & Star (*|NIR) & 0.39674 & 0.39674  \\
  6228685649971375616 & (F) & ECL & ECL & EclBin (*|EB*|MIR|NIR|Opt|V*) & 1.27398 & 1.27391  \\
\end{longtable}
\tablefoot{Apart from our visual spectral classification, we also report the spectral classes in the \textit{Gaia} DR3 (in parenthesis), and the classes given in the \textit{Gaia} cross-match table and \texttt{SIMBAD}. The \texttt{SIMBAD} class definitions can be found in the \textit{\texttt{SIMBAD}: Object types} documentation ({\url{https://simbad.cds.unistra.fr/guide/otypes.htx}}). An asterisk (*) next to the periods from the \textit{Gaia} cross-match indicates those that could not be reproduced with our data.}
\end{landscape}
\twocolumn

\begin{figure*} [h!]
\section{Additional figures} \label{lineslist}
   \centering
   \includegraphics[scale=0.585]{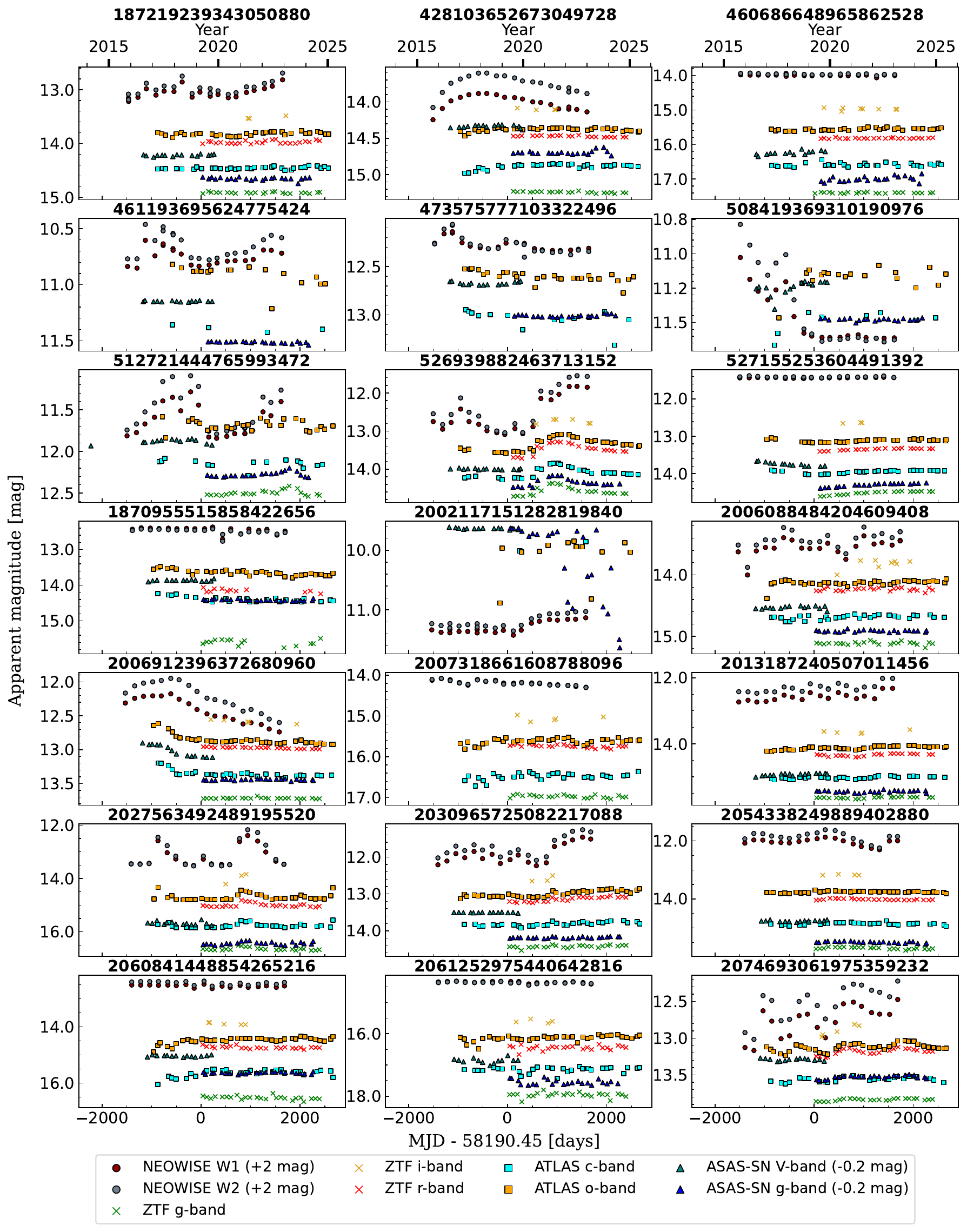} 
   \caption{Optical and infrared light curves for the sources in our sample, binned in 50-day intervals. The reference Modified Julian Date (MJD) corresponds to the start of observations by ZTF. The number shown above each panel corresponds to the \textit{Gaia} DR3 source ID.}
              \label{all_lcs_1}
\end{figure*}

\begin{figure*} [h!]
    \ContinuedFloat
   \centering
   \includegraphics[width=\textwidth]{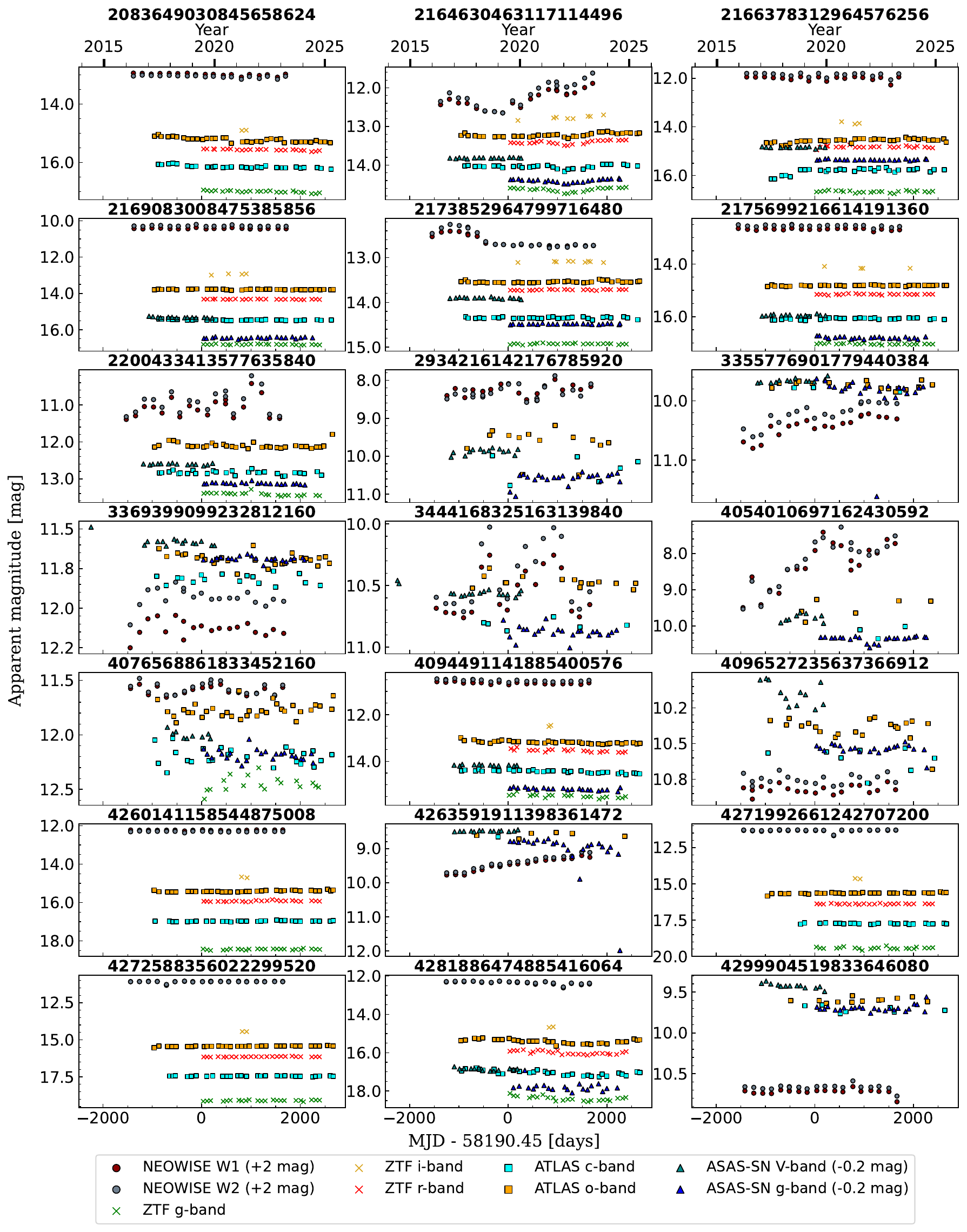}
   \caption{continued.}
              \label{all_lcs_2}
\end{figure*}

\begin{figure*} [h!]
    \ContinuedFloat
   \centering
   \includegraphics[width=\textwidth]{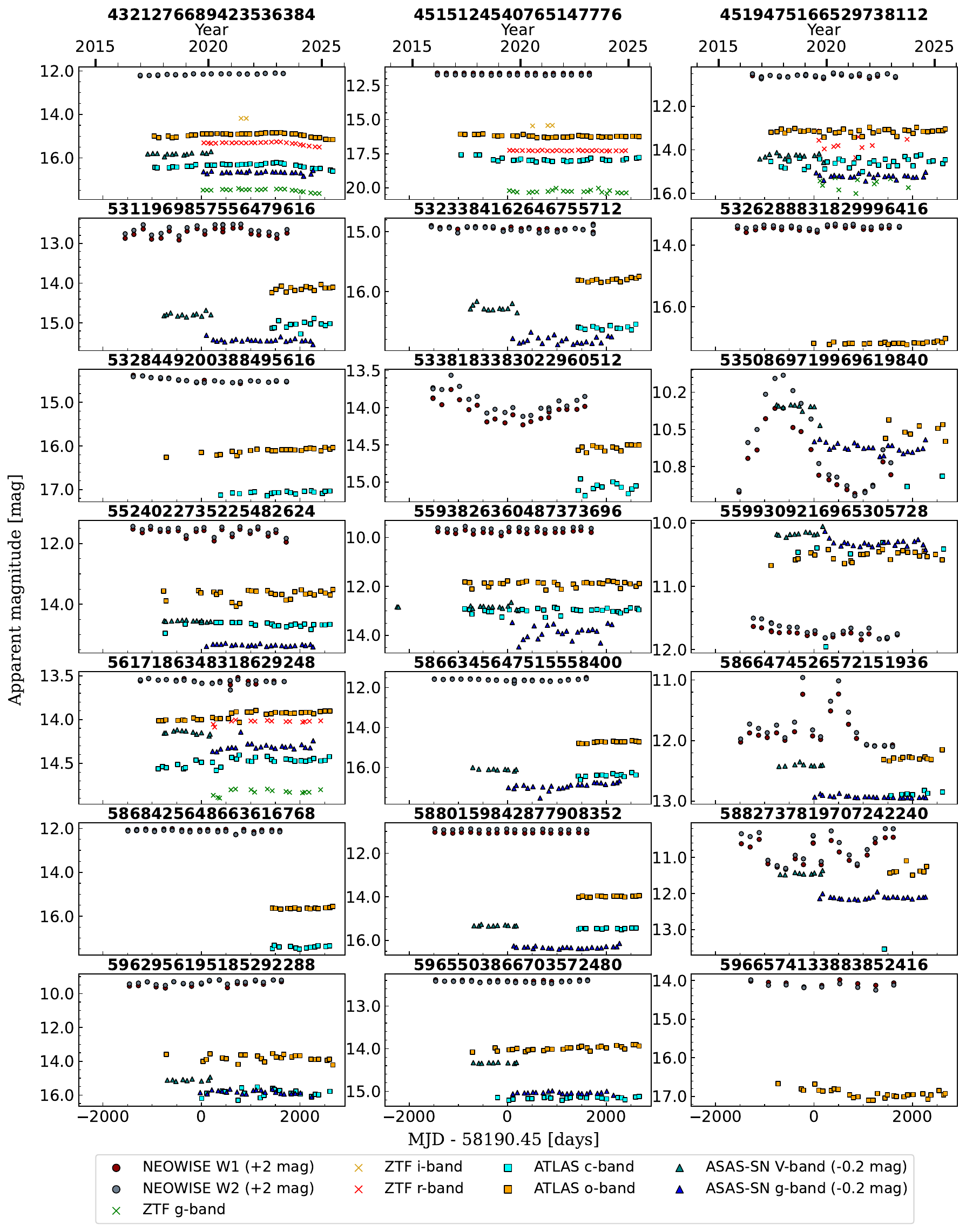}
   \caption{continued.}
              \label{all_lcs_3}
\end{figure*}

\begin{figure*} [h!]
    \ContinuedFloat
   \centering
   \includegraphics[width=\textwidth]{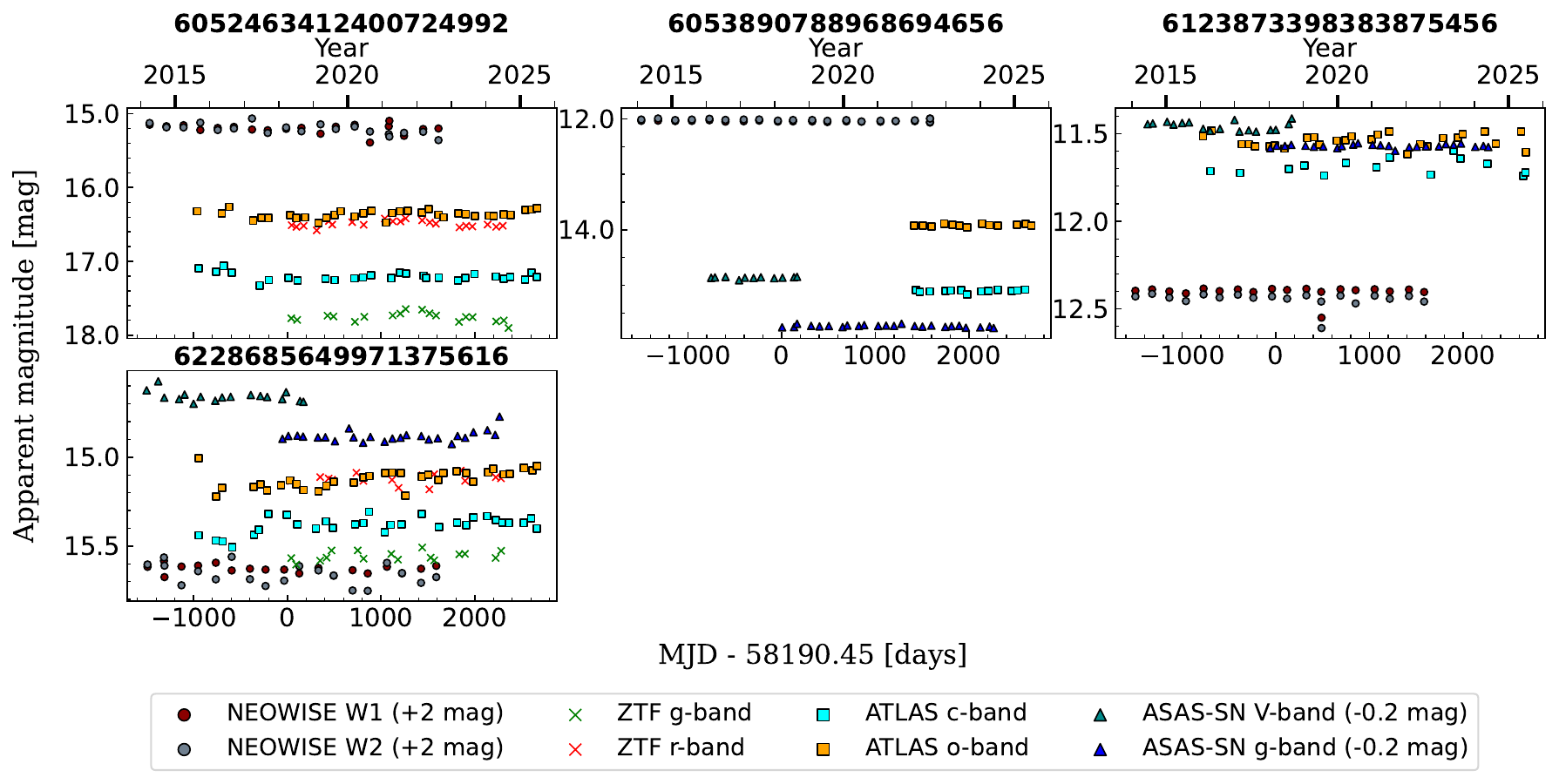}
   \caption{continued.}
              \label{all_lcs_4}
\end{figure*}


\begin{figure*} [h!]
   \centering
   \includegraphics[width=\textwidth]{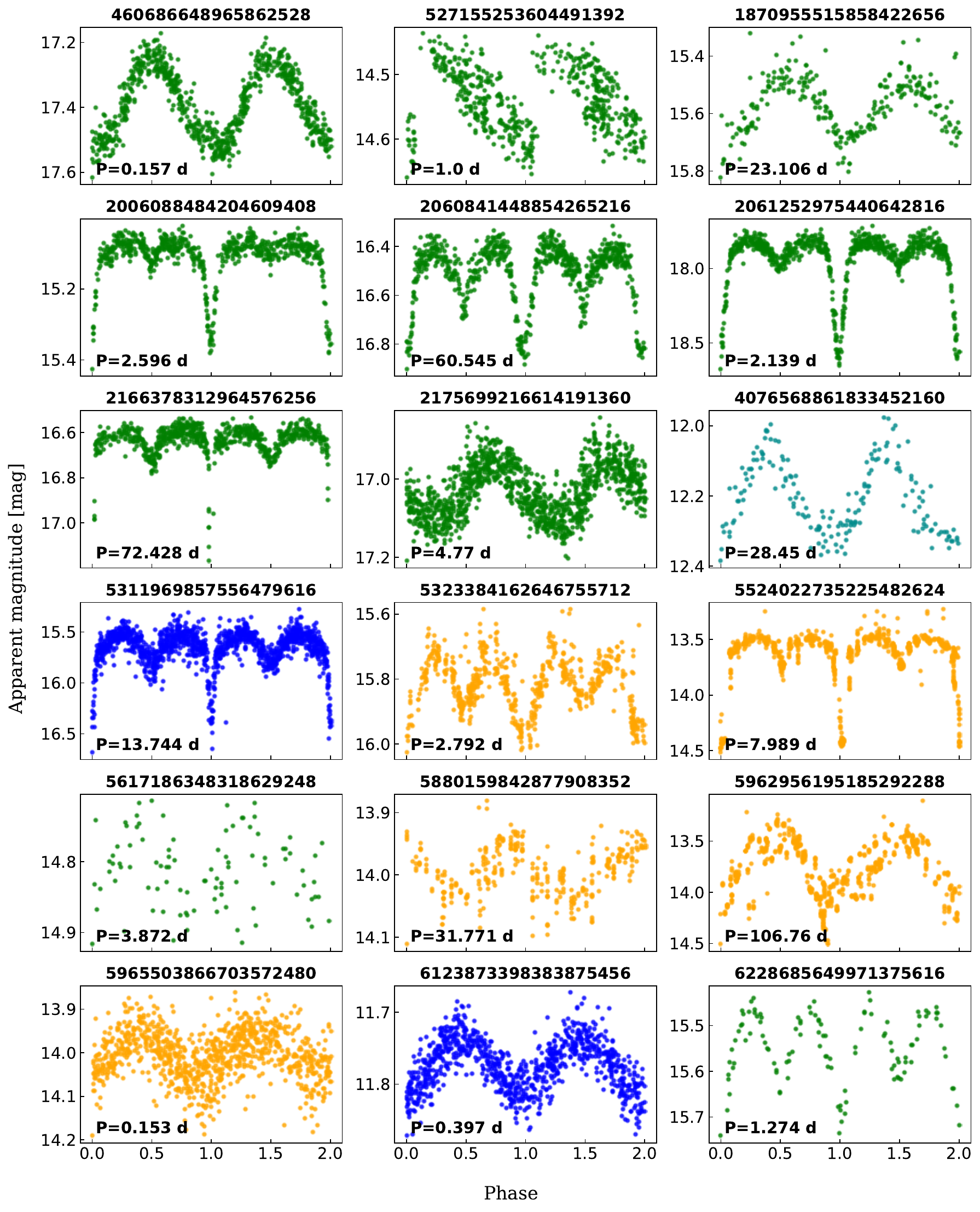}
   \caption{Folded light curves of the periodic variables in our sample. Colours represent different bands: ZTF g-band (green), ASAS-SN V-band (cyan) and g-band (blue), and ATLAS o-band (orange). The periods used to fold the light curves are shown in the bottom-left corner of each subplot.}
              \label{grid_lcs1}
\end{figure*}

\begin{figure*} [h!]
   \centering
   \includegraphics[width=\textwidth]{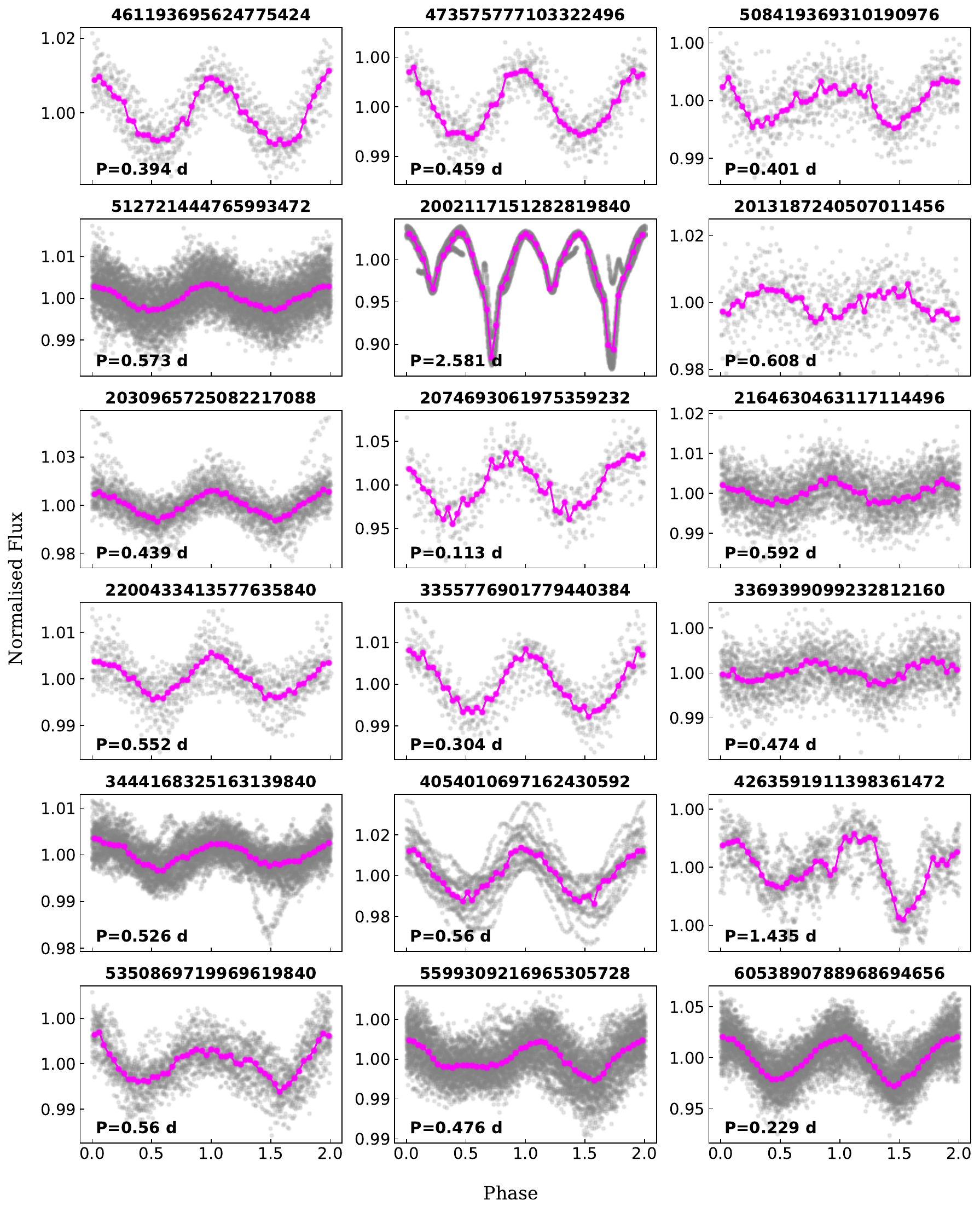}
   \caption{Same figure as Fig. \ref{grid_lcs1}, but with the light curve data from TESS. Grey points represent the original unbinned data, while pink points show the data binned into 50 bins, each with a width of 0.04 in phase space.}
              \label{grid_lcs2}
\end{figure*}

\begin{figure*} [h!]
   \centering
   \includegraphics[width=\textwidth]{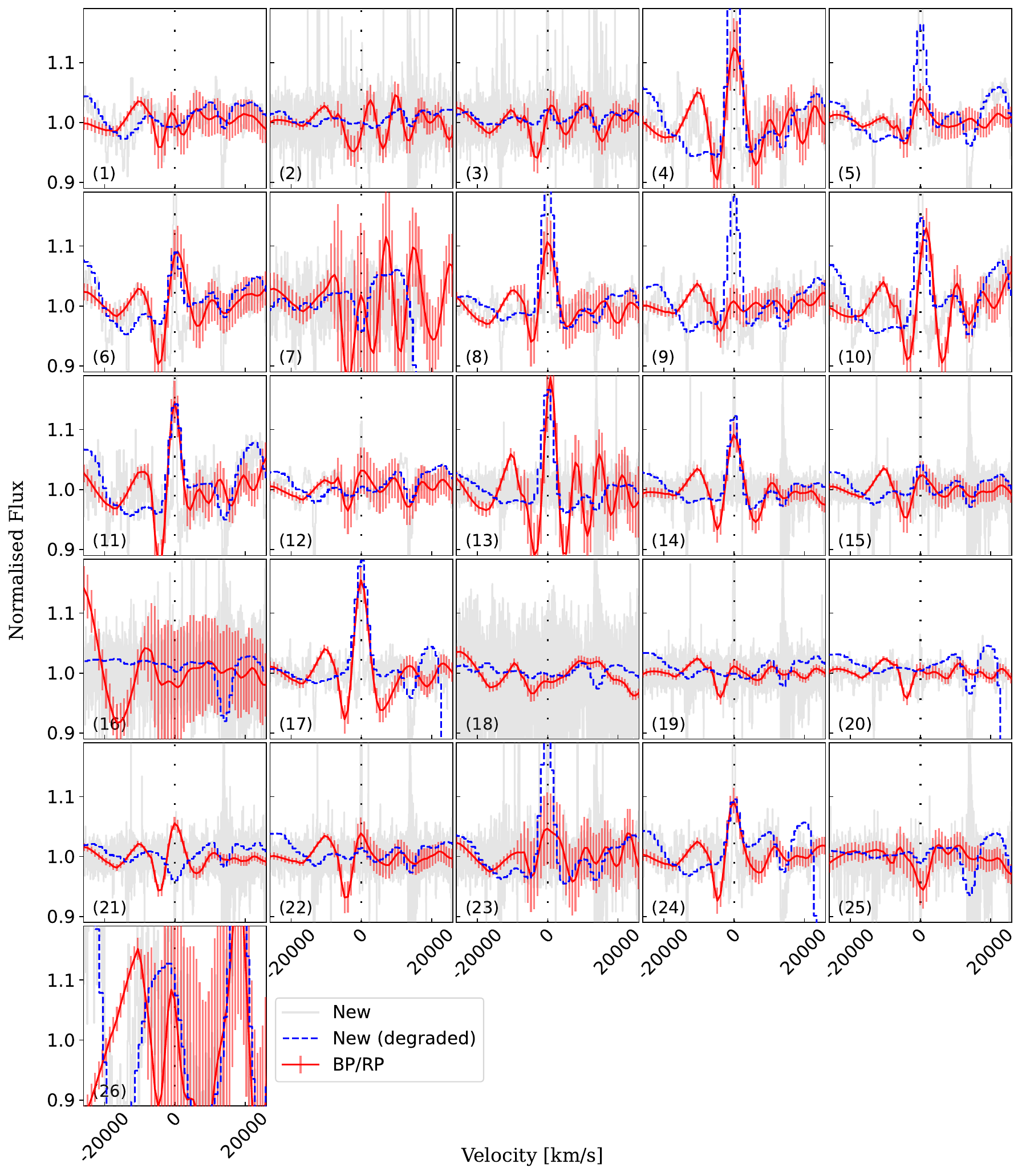}
   \caption{Flux normalised H$\alpha$ velocity profiles for the subsample of observed sources (similar to Fig. \ref{h_prof}). We compare the \textit{Gaia} low-resolution BP/RP spectrum (red solid line) with our new follow-up spectrum, degraded to match the BP/RP resolution (blue dashed line). The original-resolution follow-up spectrum is plotted in grey for comparison. All sources are shown on the same scale. The index corresponding to each source (see Table \ref{table:spec}) is shown in the bottom left corner of each spectrum.}
              \label{h_prof_gaia_xp}
\end{figure*}


\end{appendix}

\end{document}